\documentclass[twocolumn, preprint]{aastex631}

\usepackage{graphicx}	
\usepackage{amsmath}	
\usepackage{amssymb}	

\usepackage{xspace}
\usepackage{bibentry}
\usepackage{tabularx}
\usepackage{array}
\usepackage{multirow,bigdelim}

\usepackage{newtxtext,newtxmath}
\newcolumntype{C}[1]{>{\centering\let\newline\\\arraybackslash\hspace{0pt}}m{#1}}

\newcommand{\SvH}{[S~II]/H$\alpha$\xspace}
\newcommand{\OvH}{[O~I]/H$\alpha$\xspace}
\newcommand{\NvH}{[N~II]/H$\alpha$\xspace}
\newcommand{\OvHb}{[O~III]/H$\beta$\xspace}
\newcommand{\zsun}{Z$_\odot$\xspace}
\newcommand{\msun}{M$_\odot$\xspace}
\newcommand{\mstar}{M$_*$\xspace}
\newcommand{\mgas}{M$_{gas}$\xspace}
\newcommand{\mbary}{M$_{bary}$\xspace}
\newcommand{\mhalo}{M$_{halo}$\xspace}
\newcommand{\hb}{H$\beta$\xspace}
\newcommand{\ha}{H$\alpha$\xspace}

\newcommand{\zzero}{z$\sim$0\xspace}

\newcommand{\wonetwo}{(W1-W2)$\rm _{r90}$\xspace}
\newcommand{\wtwothree}{(W2-W3)$\rm _{r90}$\xspace}
\defcitealias{Kewley01}{Ke01}
\defcitealias{Kauffmann03}{Ka03}
\defcitealias{Stasinska06}{S06}
\defcitealias{Polimera22}{P22}
\defcitealias{Richardson22}{R22}
\defcitealias{Richardson25}{R25}
\defcitealias{Jarrett11}{J11}
\defcitealias{Stern12}{St12}
\defcitealias{Satyapal18}{Sa18}
\defcitealias{Kewley2006}{K06}
\defcitealias{Eckert15}{E15}
\defcitealias{Moffett15}{M15}
\defcitealias{Eckert16}{E16}
\defcitealias{Brinchmann08}{B08}
\defcitealias{Salim16}{S16}

\shorttitle{Comprehensive AGN Inventory for the RESOLVE and ECO Surveys}
\shortauthors{Polimera et al.}

\accepted{August 19, 2026}
\submitjournal{ApJ}
\begin{document}

\title{A Comprehensive AGN Inventory for the Dwarf-Dominated RESOLVE and ECO Surveys: Mass-Dependent AGN Types and Occupation Fractions}

\correspondingauthor{Mugdha S. Polimera}
\email{mugdha.polimera@cfa.harvard.edu}

\author[0000-0001-6162-3963]{Mugdha S. Polimera}
\affil{Center for Astrophysics | Harvard \& Smithsonian, 60 Garden Street, Cambridge, MA 02138, USA}
\affil{Department of Physics and Astronomy, University of North Carolina at Chapel Hill, Phillips Hall, 120 East Cameron Ave., Chapel Hill, NC 27514, USA}

\author[0000-0002-3378-6551]{Sheila J. Kannappan}
\affil{Department of Physics and Astronomy, University of North Carolina at Chapel Hill, Phillips Hall, 120 East Cameron Ave., Chapel Hill, NC 27514, USA}

\author[0000-0002-3703-0719]{Chris T. Richardson}
\affil{Physics Department, Elon University, 100 Campus Drive, Elon, NC 27244, USA}

\author{David V. Stark}
\affil{Space Telescope Science Institute, 3700 San Martin Dr., Baltimore, MD 21218, USA}

\author[0000-0002-1407-4700]{Kathleen D. Eckert}
\affil{Department of Physics and Astronomy, University of North Carolina at Chapel Hill, Phillips Hall, 120 East Cameron Ave., Chapel Hill, NC 27514, USA}

\author[0000-0002-4939-734X]{Thomas H. Jarrett}
\affil{Department of Astronomy, University of Cape Town, Rondebosch, South Africa}

\author[0009-0007-5354-3817]{Derrick S. Carr}
\affil{Department of Physics and Astronomy, University of North Carolina at Chapel Hill, Phillips Hall, 120 East Cameron Ave., Chapel Hill, NC 27514, USA}

\author[0000-0002-8574-5495]{Zackary L. Hutchens}
\affil{Physics Department, Elon University, 100 Campus Drive, Elon, NC 27244, USA}
\affil{Department of Physics and Astronomy, University of North Carolina Asheville, 1 University Heights, Asheville, NC 28804, USA}
\affil{Department of Physics and Astronomy, University of North Carolina at Chapel Hill, Phillips Hall, 120 East Cameron Ave., Chapel Hill, NC 27514, USA}

\author[0000-0001-7596-8372]{Jillian M. Bellovary}
\affil{Astrophysics Program, Queensborough Community College, City University of New York, 222-05 56th Ave, Bayside, NY, 11364}
\affil{Department of Astrophysics, American Museum of Natural History, Central Park West at 79th Street, New York, NY 10024, USA}
\affil{Department of Physics, Graduate Center, City University of New York, New York, NY 10016, USA}

\author[0000-0002-7001-805X]{Mark A. Norris}
\affil{Jeremiah Horrocks Institute, University of Lancashire, Preston, Lancashire, PR1 2HE, United Kingdom}

\begin{abstract}
We catalog Active Galactic Nuclei (AGN) in the volume- and mass-limited RESOLVE and ECO surveys using (i) the BPT plot with standard and alternate demarcations, (ii) the BPT plot and VO plots (using [O~I] or [S~II]) to find ``SF-AGN'' with mixed star-forming and AGN classifications, and (iii) mid-IR WISE color selection. Dividing RESOLVE+ECO into dwarf, transitional, and giant galaxies by baryonic mass, we find AGN occupation fractions of 2--5\%, 16--27\%, and 20--48\%, respectively, {indicating a steep $\sim$5$\times$ increase across baryonic mass $\sim$10$^{9.7}$\msun.} {These percentages are computed relative to all galaxies in each mass division, including those not searchable for AGN, with numerical ranges representing} large systematics between crossmatched emission-line measurements from different SDSS catalogs. Stellar-mass divisions yield similar results. Adding (incomplete) archival X-ray/broadline AGN detections, the dwarf AGN occupation fraction reaches 3--6\%. WI
SE AGN are detected in $<$1\% of galaxies at all masses with optimized WISE photometry. We show that standard WISE catalog photometry yields high rates of both false positives and false negatives. AGN demographics shift with mass due to correlations with gas content, star formation, and group environment, forming a continuum: SF-AGN at the high-SF end, then ``Bonus AGN'' (defined by alternate literature BPT demarcations), then Composites, then Conventional AGN at the low-SF end. Typical (metal-poor, star-forming) dwarfs most often host SF-AGN, whereas typical transitional galaxies most often host Bonus AGN and Composites. We release homogenized SDSS line measurements, reprocessed WISE data, and UV+mid-IR star formation rates for RESOLVE and ECO.
\end{abstract}
\setcounter{footnote}{10}

\section{Introduction} 
\setcounter{footnote}{11}
\label{sec:ch3intro}
Different strategies for detecting active galactic nuclei (AGN) generally identify hosts with complementary properties. An effort to build a multi-method inventory across a wide range in galaxy mass is needed to take stock of all flavours of \zzero AGN. For this, we can use the local, dwarf-dominated, mass and volume-limited REsolved Spectroscopy Of a Local VolumE (RESOLVE) and Environmental COntext (ECO) surveys, whose statistical completeness enables robust measurement of AGN occupation fractions for dwarfs and giants alike. \\

Historically, most AGN detection methods have been biased toward finding AGN in giants, but many recent studies have focused on detecting dwarf AGN, using a variety of methods based on X-ray, optical, mid-IR, and radio data (see \citealt{Polimera22} for a detailed review; hereafter \citetalias{Polimera22}). However, even dwarf AGN detection methods often have selection biases against AGN detection in typical star-forming and metal-poor dwarfs (e.g., \citealt{Wasleske24}), so use of multiple detection strategies is essential. \\   

While a key motivation for building a complete dwarf-to-giant AGN inventory is obtaining black hole (BH) occupation statistics that constrain BH growth scenarios, in practice the fueling that causes a BH to light up as an AGN depends on mass. A fundamental difference between dwarf and giant galaxies is their cold gas content and ability to form stars \citep{Dekel86, Kauffmann03mass, Kannappan04, Geha12}. Giant galaxies reside in dark matter halos with masses above the bimodality scale \mhalo $\sim 10^{12.0}$\msun (corresponding to central galaxy stellar and baryonic masses $\sim$10$^{10.5}$\msun and $\sim$10$^{10.6}$\msun, respectively, e.g., \citealt{Eckert16}), where the gas in the dark matter halo is shock heated such that gas cooling and accretion gets shut down \citep{Dekel06, Kannappan13}. Most dwarfs in the local Universe are the central galaxies of less massive dark matter halos below \mhalo $\sim 10^{11.4}$\msun, the gas-richness threshold scale in the terminology of
 \citet{Kannappan13}. Below this mass scale, theory predicts abundant inflow and accretion of cold gas \citep{Keres05, Dekel06}. Most dwarfs are extremely gas-rich, with a majority having 1-10x as much cold gas as stars \citep{Kannappan04, Kannappan13, Stark16, SaintongeCatinella2022}. Due to the wide variety of gas-to-stellar mass ratios, the central galaxy mass equivalent for the threshold scale is not precise; we adopt stellar mass $\sim$10$^{9.5}$\msun and baryonic mass $\sim$10$^{9.7}$\msun, near the bottom of the observed equivalence range \citep{Kannappan13}, in order to highlight the regime where gas richness becomes ubiquitous \citep{Geha12}. Between the gas-richness threshold and bimodality mass scales, dark matter halos of galaxies are partially shock heated and have reduced gas accretion. Galaxies in this intermediate-mass regime are likely transitioning from being dwarfs to giants and also developing bulges as they begin to process their gas with more efficiency \citep{Kannappan13, Hunt2020}. Thus, in line with these key astrophysical processes, for this paper we define dwarfs, transitional galaxies, and giants using the gas-richness and bimodality masses (and the survey completeness floors, Section~\ref{subsec:ch3surveys}). For example, in terms of baryonic mass, where the survey completeness limit is $10^{9.2}$ \msun, we define dwarfs by $10^{9.2} \leq $ \mbary $\leq 10^{9.7}$\msun, transitional galaxies by $10^{9.7} < $ \mbary $< 10^{10.6}$\msun, and giants by \mbary $\geq 10^{10.6}$\msun. \\

This work expands the efforts of \citetalias{Polimera22} to build a uniquely comprehensive and systematic \zzero AGN inventory for RESOLVE and ECO, using the traditional BPT diagnostic \citep*{Baldwin81} as well as three methods specifically designed to improve dwarf AGN detection. The first is a multiple optical emission-line classification system from \citetalias{Polimera22} that uses the BPT and VO \citep{Veilleux87} diagnostic diagrams to identify a new category of AGN, SF-AGN, that are star-forming in the BPT diagram alone, but are AGN in the VO plots. SF-AGN are common in metal-poor and star-forming dwarfs. The second is the redefined demarcation line for the BPT plot from \cite{Stasinska06}. This line highlights AGN whose spectral contributions are weak or diluted, as is typical of highly star-forming dwarfs. The third method is mid-IR colour selection based on the criteria proposed by \cite{Satyapal18} for WISE colors. Mid-IR selection holds promise for detecting obsc
ured and/or low-mass black holes, albeit low-mass BHs' mid-IR spectral energy distributions (SEDs) are still uncertain \citep[][hereafter R22]{Richardson22}. \textit{We focus on these three methods because optical emission line and WISE mid-IR data exist for nearly all of RESOLVE and ECO, and our goal is to compile a complete AGN census and compute meaningful statistics. We will also discuss compiled literature X-ray and optical/broadline H$\alpha$ AGN detections, but such data are incomplete.} \\ 

In Section \ref{sec:ch3data}, we describe the RESOLVE and ECO surveys, our adjustments to the catalog emission lines from the Sloan Digital Sky Survey (SDSS), our custom WISE mid-IR photometry, and our star formation rate estimation. In Section \ref{sec:ch3midiragn} we discuss our choice of WISE color selection method. In Section \ref{sec:ch3opticalagnmethod} we describe the three optical AGN detection methods that use the BPT and/or VO diagrams. In Section \ref{sec:ch3agninventory}, we compare the statistics of the different AGN types, highlighting shifts in characteristic AGN type between the dwarf, transitional, and giant mass regimes. In Section \ref{sec:ch3discussion}, we compare with literature AGN results and discuss how AGN statistics change with host galaxy and group halo properties. In Section \ref{sec:conclusions}, we summarize our results.  \\

Throughout this paper, we assume a Hubble constant of $H_0=70$ $km s^{-1} /Mpc$.\\

Also, for brevity, we will generally use the term \textit{AGN} with the understanding that absent multi-wavelength, high-resolution confirmation, we consider all AGN to be AGN candidates.

\section{Data and Observations}\label{sec:ch3data}
\subsection{Parent Surveys: RESOLVE and ECO}\label{subsec:ch3surveys}
We used the RESOLVE DR4 and ECO DR3 catalogs as our parent samples \citep[][hereafter H23]{Hutchens23}. RESOLVE is a statistically-complete volume-limited galaxy survey \citep{Kannappan08} within two equatorial footprints and between $cz_{\rm{grp}}$=4500--7000 $\rm{km/s}$. Here $cz_{\rm{grp}}$ is the recessional velocity of the galaxy's group (which can be a group of one) relative to the Local Group. ECO is an archival volume-limited catalog \citep{Moffett15} that complements RESOLVE over a 10x larger volume covering a large fraction of the northern spring sky and spanning $cz_{\rm{grp}}$=3000--7000 $\rm{km/s}$, so ECO encompasses the A semester of RESOLVE. Throughout this paper, we combine the RESOLVE and ECO surveys together in a non-duplicating manner (i.e., without double-counting the overlap). The combined RESOLVE+ECO parent surveys have 11,314 unique galaxies. Further details on the surveys are provided in \citetalias{Polimera22}. Although \citetalias{Polimera22} used e
arlier RESOLVE and ECO data releases, the survey design and footprints remain the same. The H23 data releases incorporate updated mass estimates and new group finding algorithms. The dark matter halo masses we adopt from H23 are group halo masses (where a galaxy alone in its halo can be considered a group of one) based on the G3 group-finder and the \citet{Warren06} halo mass function (see H23 for more details).  \\

For this study, as for \citetalias{Polimera22}, we have imposed a baryonic (stellar+cold atomic gas) mass selection limit of \mbary > $10^{9.2}$\msun, where the atomic gas mass is the best estimate from \citet{Hutchens23} following their algorithm that mixes 21cm detections ($\times$1.4 to account for Helium), deconfused detections, and photometric gas fraction estimates (constrained by upper limits or by confused detections when available). Though we also tabulate stellar-mass-limited AGN statistics in Section \ref{sec:ch3agninventory} for general use, this paper focuses on analyzing baryonic-mass-limited statistics. RESOLVE and ECO were originally selected using the SDSS $r$-band (which tracks baryonic mass better than stellar mass, \citealt{Kannappan13}) and also offer a state-of-the-art gas census \citep{Stark16,Hutchens23}, both of which argue for baryonic-mass selection. \\

In Section \ref{sec:ch3agninventory}, we provide AGN classifications for all galaxies with emission-line data in the RESOLVE and ECO parent surveys regardless of galaxy mass, with flags indicating which of these galaxies qualify for our preferred baryonic-mass-selected sample with \mbary > $10^{9.2}$\msun and/or for an alternative stellar-mass-selected sample with \mstar $> 10^{8.9}$\msun. These mass limits define samples within the parent surveys that are highly complete with respect to the mass metric in question \citep{Eckert15}. Our analysis focuses on mass-limited, complete samples to simplify statistical analysis (for uniformity, we do not use the 0.2 dex lower mass floor specified for RESOLVE-B by \citealt{Eckert15}, so we do not take advantage of its extra depth).
The baryonic mass limit yields 8352 unique galaxies in the RESOLVE and ECO surveys combined. Among these, 3691 are dwarfs ($10^{9.2} \leq$ \mbary $\leq 10^{9.7}$ \msun), 3863 are transitional galaxies ($10^{9.7}<$ \mbary $< 10^{10.6}$ \msun), and 798 are giants (\mbary $\geq 10^{10.6}$ \msun).
The stellar mass limit yields 7066 unique galaxies in the combined surveys. Among these, 2952 are dwarfs ($10^{8.9} \leq$ \mstar $\leq 10^{9.5}$ \msun), 3215 are transitional galaxies ($10^{9.5}<$ \mstar $< 10^{10.5}$ \msun), and 899 are giants (\mstar $\geq 10^{10.5}$ \msun). Note that these dwarf and giant definitions differ from those in \citetalias{Polimera22}, where we had a binary dwarf/giant division and it was set at \textit{stellar} (not baryonic) mass $10^{9.5}$\msun.  \\

In the rest of the paper, we mainly discuss results for our default baryonic-mass-selected sample, unless explicitly mentioned otherwise. 

\subsection{Custom Mid-IR Photometry from WISE Data}\label{subsec:ch3midirphot}

\begin{table*}[]
    \begin{center}
   \caption{Mid-IR photometry and derived quantities for RESOLVE and ECO galaxies}
    \begin{tabular}{c|p{10cm}|c}
        \hline
        Column Name &  Description & Units\\
        \hline
        resname/econame  & Galaxy Name in RESOLVE and/or ECO & - \\
        \hline
        \multicolumn{3}{c}{Mid-IR WISE Photometry}       \\
        \hline
          mw1 & Total W1 aperture magnitude (W1$_{\rm AP-total}$) & mag\\
          mw2 & Total W2 aperture  magnitude (W2$_{\rm AP-total}$) & mag\\
          mw3 & Total W3 aperture magnitude (W3$_{\rm AP-total}$) & mag\\
          mw4 & Total W4 aperture magnitude (W4$_{\rm AP-total}$) & mag\\
          emw1\_rand\ & Random error in W1 magnitude & mag\\
          emw2\_rand\ & Random error in W2 magnitude & mag\\
          emw3\_rand\ & Random error in W3 magnitude & mag\\
          emw4\_rand\ & Random error in W4 magnitude & mag\\
          emw1\_sys\ & Systematic error in W1 magnitude & mag\\
          emw2\_sys\ & Systematic error in W2 magnitude & mag\\
          emw3\_sys\ & Systematic error in W3 magnitude & mag\\
          emw4\_sys\ & Systematic error in W4 magnitude & mag\\
          emw1\ & Total error in W1 magnitude & mag\\
          emw2\ & Total error in W2 magnitude & mag\\
          emw3\ & Total error in W3 magnitude & mag\\
          emw4\ & Total error in W4 magnitude & mag\\
          mw1\_exp & W1 exponential magnitude & mag\\
          mw2\_exp & W2 exponential magnitude & mag\\
          mw3\_exp & W3 exponential magnitude & mag\\
          mw4\_exp & W4 exponential magnitude & mag\\
          emw1\_exp & Random error in W1 exponential magnitude & mag\\
          emw2\_exp & Random error in W2 exponential magnitude & mag\\
          emw3\_exp & Random error in W3 exponential magnitude & mag\\
          emw4\_exp & Random error in W4 exponential magnitude & mag\\
          w12\_r90 & W1 - W2 color in r$_{90}$, \wonetwo & mag\\
          w23\_r90 & W2 - W3 color in r$_{90}$, \wtwothree & mag\\
          w12\_r90\_err & Random error in \wonetwo & mag\\
          w23\_r90\_err & Random error in \wtwothree & mag\\
          peak\_snr\_w1 & Peak S/N ratio for W1 within r$_{90}$ & - \\
          peak\_snr\_w2 & Peak S/N ratio for W2 within r$_{90}$ & - \\
         \hline
        \multicolumn{3}{c}{Star Formation Rates}       \\
        \hline
          sfr\_uv & $\rm SFR_{UV}$ for galaxies with custom photometry of deep GALEX NUV data & \msun/yr \\
          sfr\_rfuv & $\rm SFR_{UV}$ for galaxies with random forest NUV mags from C25   & \msun/yr \\
          sfr\_uv\_err & Error in $\rm SFR_{UV}$ & \msun/yr \\
          sfr\_rfuv\_err & Error in $\rm SFR_{RFUV}$   & \msun/yr \\
          sfr\_ir & $\rm SFR_{IR}$ using WISE W3 and/or W4 luminosities & \msun/yr \\
          sfr\_ir\_err & Error in $\rm SFR_{IR}$ & \msun/yr \\
          sfr / sfr\_rf & Total SFR = $\rm SFR_{UV}$ + $\rm SFR_{IR}$ OR $\rm SFR_{RFUV}$ + $\rm SFR_{IR}$ & \msun/yr \\
          sfr\_err / sfr\_rf\_err & Error in SFR or $\rm SFR_{RF}$  & \msun/yr \\
        \hline
        \hline
    \end{tabular}\label{tab:irdatatable}
    \end{center}
\end{table*}

We reprocessed mid-IR imaging from WISE in the W1, W2, W3, and W4 bands with a custom photometry pipeline described below, designed to match the galaxy profiles with the custom optical photometry of \cite{Eckert15}. \\

\subsubsection{Mid-IR Photometric Pipeline}\label{subsec:ch3midirphot_qc}

We downloaded primary and uncertainty images in all four WISE bands for all galaxies in the RESOLVE and ECO parent surveys from the WISE archive\footnote{https://irsa.ipac.caltech.edu/Missions/wise.html} \citep{Cutri12}, served by the NASA/IPAC Infrared Science Archive (IRSA). We then cropped each frame to cover only the region within $15\times r_{90}$ where $r_{90}$ is the 90\% light radius from optical photometry \citep{Eckert15}. \\

We masked background sources in multiple steps, first constructing preliminary masks from raw images for the purpose of sky subtraction, then constructing final masks from the sky-subtracted images in two additional steps for the purpose of photometry. We performed sky subtraction by running \texttt{Source-Extractor} \citep{Bertin96} on each frame to identify background sources and create preliminary masks. For the preliminary masks, we chose Source-Extractor settings based on the recommendations in the WISE Explanatory Supplement\footnote{https://wise2.ipac.caltech.edu/staff/fmasci/SEx\_WPhot.html} and used the WISE uncertainty images as \texttt{weight images} to compute source detection thresholds for detecting and masking both bright and faint sources effectively. The region within $2\times r_{90}$ of the galaxy center was also masked. We created a sky model by fitting a two-dimensional third-order polynomial to the masked image and then subtracted the sky model. For galax
ies found in or split between multiple frames, we stitched the sky-subtracted frames together using \texttt{SWarp} \citep{Bertin02} and averaged overlapping pixels. We also corrected the uncertainty images to account for the improved statistics in the overlapping pixels.  \\

After sky subtraction, we created the final photometric masks in two steps, this time using the sky-subtracted WISE images in each band. Since the mid-IR background is full of faint sources speckled over the entire image, we optimized our masking algorithm to aggressively mask out faint background sources in addition to the bright objects. First, we created segmentation maps (\texttt{CHECKIMAGE\_TYPE} keyword set to \texttt{SEGMENTATION}) using only the primary images to identify and mask the bright sources. Then, similar to the process for sky subtraction, we used the uncertainty images as weight maps (using \texttt{WEIGHT\_IMAGE} keyword) to identify and mask the faint objects (\texttt{CHECKIMAGE\_TYPE} keyword set to \texttt{OBJECTS}; see WISE documentation). This setting enabled the identification of small and faint background objects and was the key to our aggressive masking strategy. Lastly, we combined the two masks to create the final photometric mask where both brigh
t and faint sources were effectively masked. We found that this two-step technique proved particularly important while performing photometry of faint and/or diffuse objects like dwarf galaxies, which make up $\sim$2/3 of our volume-limited surveys. We visually inspected a subset of masked images and confirmed that our masking technique does well at removing the speckled structure in the mid-IR background. For some galaxies with very close bright neighboring objects, our automated process failed to mask the neighbors correctly. In such cases, we manually fixed the masks (details given later in this Section). \\ 

We used the images and their final photometric masks to create elliptical photometric profiles and models using \texttt{IRAF} \citep{Tody86}. For consistency with our previously published custom RESOLVE/ECO optical photometry, we resampled the images to the \textit{r}-band pixel scale to enable use of the \textit{r}-band profiles from \cite{Eckert15} and \cite{Moffett15}. In addition, we forced the \texttt{ellipse} task in \texttt{IRAF} to use the same central coordinates, position angle, and axial ratio used by \cite{Eckert15} and \cite{Moffett15} for the optical images. Flux values in masked regions of the WISE images were replaced with the corresponding values from the 2D model generated by the \texttt{ellipse} task. \\
    
We used two methods to estimate the total magnitudes from these model-corrected images in the four WISE bands: 
\begin{itemize}
       \item[1)] Large aperture method (AP): We defined a large aperture ellipse with the semi-major axis equal to the convolution of 2$\times$FWHM of each band's PSF with 2$\times r_{90}$ to ensure a generous boundary.\footnote{We note that we set the maximum aperture to be extremely large to ensure all the flux of the galaxy is captured, especially given the poor resolution of WISE. However, in reality, the final image on which the photometry is performed does not have much flux all the way out to $2\times r_{90}$, because most of the pixels beyond a radius approximately between 1$\times-$1.5$\times$ $r_{90}$ are masked (due to the crowded background and large PSF) and filled in with very small values from the models.} We calculated the magnitude inside this ellipse from the final model-corrected images. Throughout this paper, we use the subscript `AP-total' to refer to these magnitudes (e.g., W1$_{\rm AP-total}$).    \\
       \item[2)] Exponential disk method (EXP): Following \cite{Eckert15}'s approach to the optical photometry, we defined an outer disk fitting region where the flux inside the annulus is between 1 and 5 times the 1$\sigma$ sky noise level. For galaxies with heavy masking, we chose a region with flux 3 to 5 times the 1$\sigma$ sky noise level. Then we fit an exponential function (Sersic index n=1) to the galaxy's surface brightness profile in the chosen disk fitting region. We computed an inner flux by simply summing the total flux within the outermost radius of the disk fitting region. We then computed an outer flux by extrapolating the exponential fit to infinity beyond the outermost radius to determine the extra flux below the noise level. The total exponential fit magnitude was computed using the sum of the measured inner flux and extrapolated outer flux (see \citealt{Eckert15} for more details). \\

\end{itemize}

\begin{figure*}[t!]
    \centering
    \includegraphics[width = \linewidth]{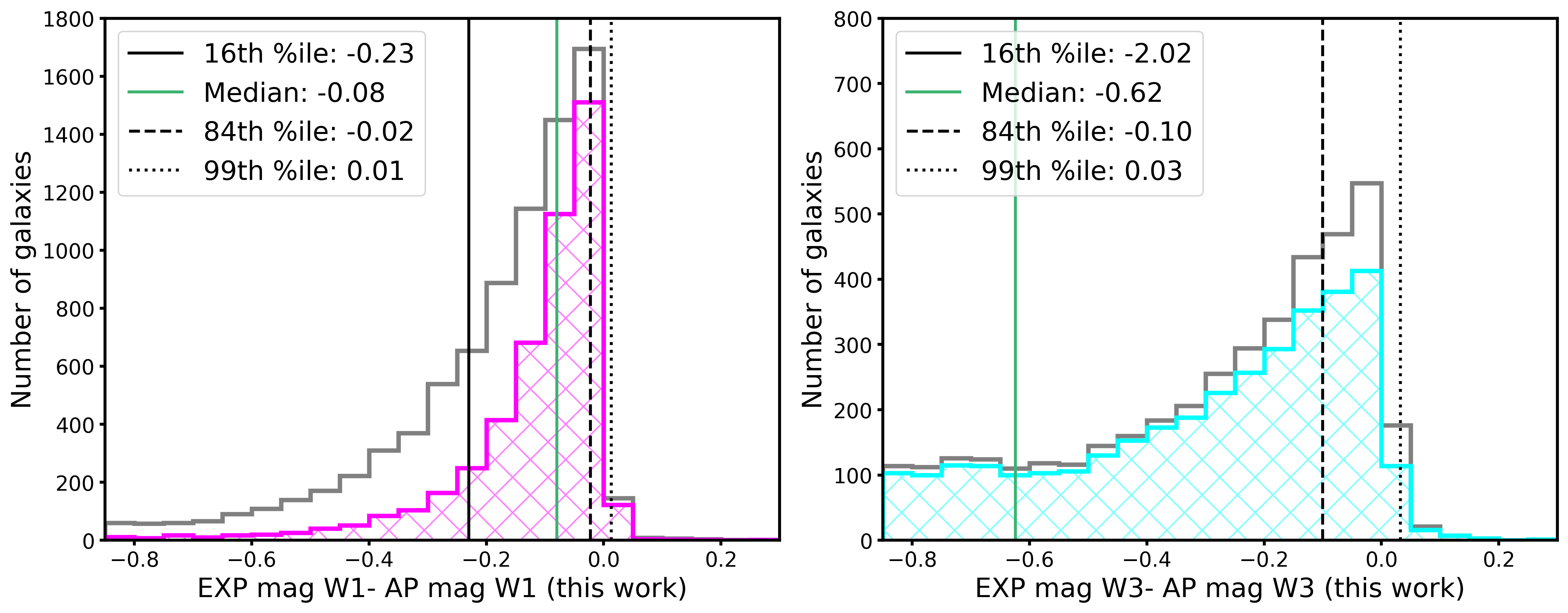}
    \caption{Comparison of mid-IR magnitudes for RESOLVE and ECO galaxies estimated by different methods. The AP method generally yields fainter magnitudes than EXP method in our custom pipeline and is used as the default total magnitude in this work (see Section \ref{subsec:ch3midirphot}). The behavior seen here for W1 and W3 magnitudes also applies to W2 and W4 magnitudes, respectively (not shown). \textit{Left:} Distribution of differences between total W1 magnitudes estimated using the exponential disk method (EXP) and using the large aperture method (AP). We show both the AGN sample (magenta crosshatch) and the parent sample (gray; see Sections \ref{subsec:ch3midirphot} and \ref{sec:ch3midiragn}). The green line shows the median of the AGN sample, and solid/dashed/dotted lines show the 16th/84th/99th percentiles, respectively, of the AGN distribution. In both the AGN and parent samples, less than 1\% of galaxies lie outside the x-axis limits of this panel.   
    \textit{Right: } Same as left panel but for W3 magnitudes for the $\rm SFR_{IR}$ sample (blue crosshatch) compared to the parent sample (gray). W3 shows worse agreement between EXP and AP magnitudes, with the 16th percentile being at $\sim$-2.02 magnitudes (not shown). In both the SFR and parent samples, $\sim$40\% of galaxies lie outside the x-axis range of this panel. The AP magnitudes still appear to be reliable for this 40\% (see Section \ref{subsec:ch3midirphot} and Figure \ref{fig:wxsc_compare}).
    }\label{fig:expap_compare}
\end{figure*} 

For RESOLVE and ECO galaxies, we have also computed $\rm W1-W2$ and $\rm W2-W3$ colors using smaller apertures defined by $r_{90}$. For each color, we blurred the band with the higher resolution (i.e., for $\rm W1-W2$ color, W1 image, and for $ \rm W2-W3$ color, W2 image) to the lower resolution of the other band. We then computed an aperture magnitude in each band with a fixed aperture computed as $r_{90}$ convolved with the FWHM of the band with the lower resolution. In the rest of this paper, we refer to these colors as \wonetwo and \wtwothree. For AGN detection, we also compute the peak S/N to ensure that there is a bright central source inside $r_{90}$. For all magnitudes and colors, we compute random statistical errors from the uncertainty images (which include a Poisson noise component). Table \ref{tab:irdatatable} reports all of our mid-IR photometry and errors. In case of failed photometry, magnitudes and errors are given as -999. {We have successfully measured W4 photometry for $\sim$52\% of the combined ECO+RESOLVE catalog. These data are included in the published catalog, though we defer analysis of the W4 band to future studies.} \\ 

\subsubsection{Mid-IR Quality Control}

We performed quality control on our photometry by using a combination of color plots to identify outliers and then manually reprocessing the outliers. (i) Galaxies with $r - \rm W1_{\rm AP-total}$ optical-WISE colors outside the 1st-99th percentile range of the color distribution were flagged as outliers. (ii) We also used a purely WISE ($\rm W1_{\rm AP-total} - \rm W2_{\rm AP-total}$) vs. ($\rm W2_{\rm AP-total} - \rm W3 _{\rm AP-total}$) color plot to select outliers with unusual WISE colors. Based on past work (e.g., \citealt{Jarrett19}), most galaxies have WISE colors $-1 <  (\rm W1_{AP-total} - W2_{AP-total}) < 1.5$ and $-1 < (\rm W2_{AP-total} - W3_{AP-total}) < 4.5$. We flagged all galaxies outside these ranges as outliers. (iii) Additionally, since it is rare to find galaxies to the left of the WISE AGN selection box from \cite{Jarrett11}, we flagged galaxies in that region as outliers (i.e., $(\rm W1_{AP-total} - W2_{AP-total}) > 0.52$ and $(\rm W2_{AP-total} - W3_{A
P-total}) < 2.2$).\\

Using these criteria, we initially identified $\sim$10\% of all RESOLVE and ECO galaxies to be outliers. However, in the end, all the outliers that were identified solely using WISE colors could be fixed by using $r_{90}$ aperture colors instead of AP-total colors, so we could remove their outlier flags. Thus, only the 310 galaxies that were identified as outliers in $r - \rm W1_{\rm AP-total}$ optical-WISE colors were further investigated; these make up only $\sim$2\% of RESOLVE/ECO galaxies.\\

We manually inspected this smaller outlier sample and found that the galaxies displayed one of the following problems: (i) poor masking of bright nearby objects, (ii) diffraction spike going through the galaxy, (iii) possible merger or interaction with a neighboring galaxy, (iv) extremely large radius or irregular shape, or (v) extremely low surface brightness. Upon further inspection, we found the primary issue for these cases to be inappropriate masking. We were able to manually fix the masks and successfully rerun the photometry for 12 of the outliers, and we rejected the photometry for the other 298 outlier galaxies. In our table, we use -999 values to indicate rejected mid-IR photometry. \\

\begin{figure*}[]
    \centering
    \includegraphics[width = 0.75\linewidth]{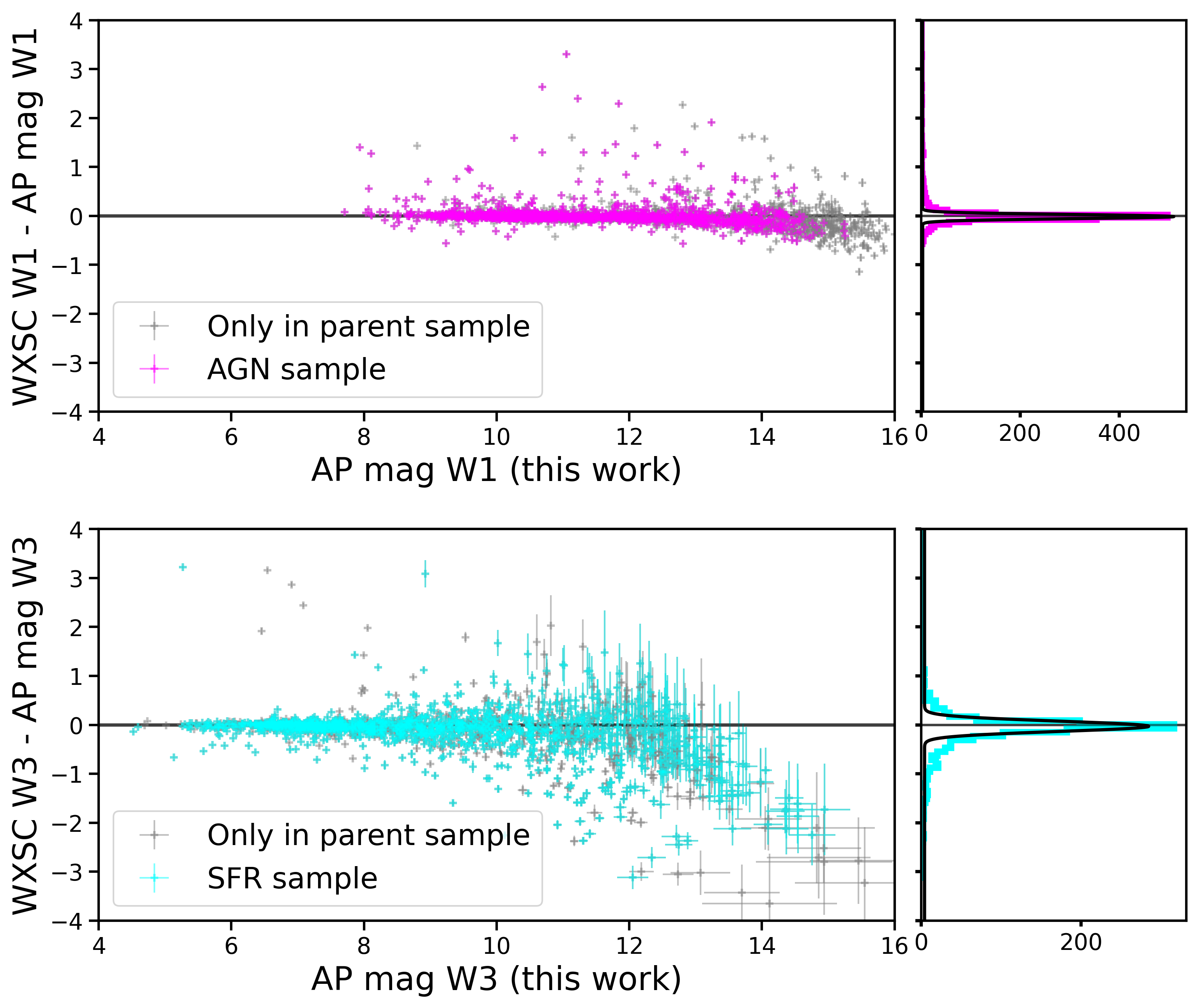}
    \caption{Comparison of mid-IR magnitude estimates for RESOLVE and ECO galaxies with the WXSC catalog (T.\ H.\ Jarrett, 2024, priv.\ comm.) for a crossmatched sample. \textit{Top left:} Differences between W1 total magnitudes from the WXSC and from this work (using the AP method) as a function of W1 magnitudes. We highlight the AGN sample (magenta; see Sections \ref{subsec:ch3midirphot} and \ref{sec:ch3midiragn}) within the rest of the parent sample (gray). The differences are centered around zero but get larger for fainter apparent magnitudes. 
    \textit{Top right:} Histogram of W1 differences (magenta) and fit with a Gaussian distribution (mean $\sim$ 0.02; $\sigma$ $\sim$ 0.03). The fit reveals a tail of outliers towards negative numbers, implying that our large aperture magnitudes tend to be fainter than those in the WXSC for the faintest sources.  
    \textit{Bottom left:} Same as top left panel but comparing W3 magnitudes and highlighting the $\rm SFR_{IR}$ sample (cyan) within the rest of the parent sample (gray). 
    \textit{Bottom right:} Histogram of W3 differences and fit with a Gaussian distribution (mean $\sim$ 0.03; $\sigma$ $\sim$ 0.1). The tail of outliers towards negative differences is more extreme for W3 than for W1.
    }
    \label{fig:wxsc_compare}
\end{figure*}

Figure~\ref{fig:expap_compare} compares the AP and EXP methods for W1 and W3 photometry. In this work, the W1 and W2 bands are mainly used for AGN selection, and the W3 and W4 bands are mainly used for estimating mid-IR star formation rates (SFRs). We have confirmed that W2 and W4 photometry follow the same trends as seen in Figure~\ref{fig:expap_compare} for W1 and W3, respectively, so we do not show them for brevity. For W1, $\sim$90\% of galaxies with non-null measurements have EXP and AP magnitudes that agree within a range of $\pm$0.25 mag, with a median difference of only $-$0.08. However, the distribution is clearly not Gaussian; for $\sim$85\% of the sample, AP magnitudes are fainter than EXP magnitudes. There exists a much longer tail towards more negative differences (not shown in plot). The right panel of Figure \ref{fig:expap_compare} shows similar trends for W3 magnitudes, but with much worse agreement between the two methods -- the 16th percentile value (i.e., -
1$\sigma$ level) is -2.02 mag.\footnote{For W1 and W2, the AGN subsample shows comparable or tighter agreement between EXP and AP magnitudes than the parent sample, with no evidence of a disproportionately long negative tail. For W3 and W4, the SFR$_\mathrm{IR}$ and parent samples show similarly broad EXP$-$AP differences, consistent with the known limitations of the exponential profile method at these bands as discussed above.} \\

We found that these large discrepancies between the AP and EXP methods occur mostly for galaxies with large errors for faint EXP magnitudes (typically W3$>$10 mag). Manual examination of these cases showed that the EXP method had failed, with unrealistic magnitudes or errors. \textit{Thus in this work, we have used AP magnitudes as the default total magnitudes for our custom mid-IR photometry.} We believe that our AP magnitudes are reliable, even when our EXP magnitudes are not, for two reasons. First, all non-null AP magnitudes fall in the expected range of magnitudes for nearby galaxies (e.g., 8-16 mag for W1, and 4-16 mag for W3). Second, in the following paragraphs, we show good agreement between our AP magnitudes and those from other WISE catalogs, especially the state-of-the-art Wise Extended Source Catalog \citep[WXSC, T.\ H.\ Jarrett, 2024, priv.\ comm.\, an expansion of the WXSC in][]{Jarrett19}. \\

{We first compared our mid-IR photometry with catalog photometry from AllWISE. For 8352 RESOLVE+ECO galaxies crossmatched to the AllWISE catalog\footnote{https://wise2.ipac.caltech.edu/docs/release/allwise/}, our W1 AP magnitudes are systematically brighter than the AllWISE profile-fitted W1 magnitudes (``w1mpro'') by $\sim$0.8~mag. These AllWISE magnitudes come from extrapolated profile fits, making them more analogous to our EXP magnitudes than to our AP magnitudes. However, since our EXP magnitudes are mostly brighter than our AP magnitudes, comparing to our EXP magnitudes would only exacerbate the discrepancy. \\

Notwithstanding this discrepancy, we trust our custom AP magnitudes for two reasons. First, \cite{Cluver14} has argued that the standard WISE catalog photometry typically underestimates the extended flux of nearby, well-resolved sources. Second, our magnitudes agree well with two other WISE catalogs with custom photometry, the GAMA and WXSC catalogs. For a crossmatched sample of 160 RESOLVE+ECO galaxies in the GAMA DR3 catalog \citep{Baldry18}, we found that our W1 AP magnitudes agree with the recommended GAMA magnitudes (``MAG\_W1''), with a slight systematic offset toward brighter values ($\sim$0.06 mag) after conversion between the AB and Vega conventions. For these 160 galaxies, $\sim$70\% of the GAMA recommended magnitudes are isophotal measurements and the rest are aperture measurements \citep{Cluver14}, meaning these magnitudes are not extrapolated, similar to our AP magnitudes. For a subset of 3231 galaxies in common with the new version of the WXSC (T.\ H.\ Jarrett, 
2024, priv.\ comm.), we found even better agreement. The WXSC has high levels of quality control applied to each galaxy in the catalog, including careful masking of neighboring objects, and WXSC ``total'' magnitudes represent a best effort to use the algorithm most appropriate for a given galaxy (extrapolated or not). Figure \ref{fig:wxsc_compare} plots the WXSC magnitudes minus our AP magnitudes as a function of our AP magnitudes, creating a sparse tail where our faintest magnitudes deviate from brighter WXSC counterpart magnitudes due to large error bars and axis covariance. Otherwise, the magnitude differences scatter close to zero, with a small population of outliers that appear to be associated with difficulty masking bright neighbors, though these outliers were not flagged by our algorithm. The magnitude difference distributions (histograms in right panels) are well fit by Gaussians: mean $\sim -0.01$ and $\sigma \sim 0.04$ for W1, and mean $\sim -0.03$ and $\sigma \sim 0.1$ fo
 r W3. As these $\sigma$ values are $\sim$30$\times$ higher than the median random error computed for AP magnitudes, we have adopted the absolute values of $\sigma$ as our systematic uncertainties. Thus, for W1 and W3 the systematic errors are 0.04 and 0.1, respectively. An analogous analysis of W2 and W4 yields systematic errors of 0.06 and 0.23, respectively. These systematic errors are also provided in Table \ref{tab:irdatatable}. \\}

For the crossmatched sample, we also compared \wonetwo from RESOLVE/ECO to $\rm W1-W2$ from the WXSC, because this color is our main mid-IR tool for AGN detection. There is much less discrepancy in colors with $\sim$99\% of galaxies agreeing within $\pm$0.25. The distribution of color differences is well fit by a Gaussian with mean $\sim$ $-$0.01 and $\sigma \sim 0.03$. Based on this and an analogous analysis of $\rm W2-W3$, we estimate the systematic uncertainties in $\rm W1-W2$ and $\rm W2-W3$ colors as $\sim$0.03 and $\sim$0.11 respectively. We note that \cite{Norris14} has independently found that $\rm W1-W2$ shows a scatter of 0.026 due to various systematic effects in excess of photometric errors.\\

\subsection{Optical emission line fluxes from SDSS}\label{subsec:ch3emlinedata}
\begin{table*}[]
    \begin{center}
    \caption{Fully corrected optical emission line fluxes and errors based on MPA-JHU, Portsmouth, and NSA Catalogs\tablenotemark{1}}
    \begin{tabular}{c|c}
        \hline
        Column Name &  Description \\
        \hline
        resname/econame  & Galaxy Name in RESOLVE and/or ECO \\
         oiii\_5007\_flux\_$<$sdsscat$>$\tablenotemark{2} & [O~III]  $\lambda$5007 flux \\ 
         hbeta\_4861\_flux\_$<$sdsscat$>$ &  \hb  $\lambda$4861 flux \\ 
         oi\_6300\_flux\_$<$sdsscat$>$ &  [O~I]  $\lambda$6300 flux \\ 
         nii\_6548\_flux\_$<$sdsscat$>$ &  [N~II]  $\lambda$6548 flux \\ 
         halpha\_6563\_flux\_$<$sdsscat$>$ &  \ha  $\lambda$6563 flux \\ 
         nii\_6584\_flux\_$<$sdsscat$>$ &  [N~II]  $\lambda$6584 flux \\ 
         sii\_6717\_flux\_$<$sdsscat$>$ &  [S~II]  $\lambda$6717 flux \\ 
         sii\_6731\_flux\_$<$sdsscat$>$ &  [S~II]  $\lambda$6731 flux \\ 
         err\_oiii\_5007\_flux\_$<$sdsscat$>$ & Error in [O~III]  $\lambda$5007 flux \\ 
         err\_hbeta\_4861\_flux\_$<$sdsscat$>$ & Error in \hb  $\lambda$4861 flux \\ 
         err\_oi\_6300\_flux\_$<$sdsscat$>$ & Error in [O~I]  $\lambda$6300 flux \\ 
         err\_nii\_6548\_flux\_$<$sdsscat$>$ & Error in [N~II]  $\lambda$6548 flux \\ 
         err\_halpha\_6563\_flux\_$<$sdsscat$>$ &  Error in \ha  $\lambda$6563 flux \\ 
         err\_nii\_6584\_flux\_$<$sdsscat$>$ & Error in [N~II]  $\lambda$6584 flux \\ 
         err\_sii\_6717\_flux\_$<$sdsscat$>$ & Error in [S~II]  $\lambda$6717 flux \\ 
         err\_sii\_6731\_flux\_$<$sdsscat$>$ & Error in [S~II]  $\lambda$6731 flux \\ 
         \hline
        \hline
    \end{tabular}
    \label{tab:optdatatable}
    \end{center}
\tablenotetext{1}{All flux and error values are corrected for internal and foreground extinction as described in Section \ref{subsec:ch3emlinedata} and \citetalias{Polimera22}. Units are 10$^{-17}$ ergs/s/cm$^2$ }
\tablenotetext{2}{$<$sdsscat$>$ = `jhu', `port', or `nsa' representing the MPA-JHU, Portsmouth, or NSA catalog, respectively}
\tablenotetext{}{Note: Full machine-readable table will be provided online.}

\end{table*}

To get optical emission line fluxes, we cross-matched galaxies in the RESOLVE and ECO parent surveys with the MPA-JHU \citep{Tremonti04}, NSA \citep{Blanton11}, and Portsmouth \citep{Thomas13} catalogs, all based on SDSS. We processed the data as described in \citetalias{Polimera22}. To account for galaxy internal extinction corrections, we performed emission line dereddening using Balmer decrements and a linear combination of the SMC and Milky Way extinction curves. We also corrected fluxes from the Portsmouth catalog to account for Milky Way foreground extinction and \textit{r}-band flux rescaling. For further details, see \citetalias{Polimera22}. We provide the corrected emission line measurements from all three catalogs for all RESOLVE and ECO galaxies in Table \ref{tab:optdatatable}.  \\

To use the system of multiple emission-line diagnostics from \citetalias{Polimera22} -- especially to identify SF-AGN (see Section \ref{subsec:ch3sfagn}) -- we need all three of the BPT and VO plots (\OvHb vs. the three ratios \NvH, \SvH, and \OvH), hereafter often referred to as the [N~II], [S~II], and [O~I] diagnostic plots.\footnote{Note that following \citet{Kewley2006}, we have measured [S~II] (or [S~II]6720) as the sum of the doublet [S~II]6717+[S~II]6731, and following \citetalias{Polimera22} we have also combined the signal in both [N~II] doublet lines then rescaled the signal to [N~II]6584.} In \citetalias{Polimera22}, we required reliable, positive-valued, and finite fluxes and errors, as well as integrated flux S/N $>$ 5, for \textit{all} six emission lines used in the three aforementioned diagnostic plots. The strict S/N cut on the weak [O~I] line greatly reduced the sample size, but we found [O~I] to be essential to the identification of SF-AGN, since [S~II] is m
ore sensitive to dilution by strong star formation.  \\

For this paper, we have expanded our search sample by dropping the S/N requirements for [S~II] and [O~I] except when we search for SF-AGN. We also subclassify BPT AGN with these plots but only when the data permit. This approach allows all galaxies with adequate S/N in the [N~II] plot to receive some classification {(using the [N~II] plot with traditional BPT categories revised to include ``S06 Bonus AGN'' as described in Section \ref{subsec:ch3s06bonus})}, while SF-AGN are only identified when adequate S/N is additionally available for both the [S~II] and [O~I] plots. With this more inclusive approach, our expanded BPT search sample has 5325~/~3857~/~3388 RESOLVE and ECO galaxies when defined using crossmatched data from the MPA-JHU~/~Portsmouth~/~NSA emission-line catalogs, respectively, increasing the \citetalias{Polimera22} search sample by $\sim$2-3$\times$. Among these galaxies, 2494~/~1916~/~1598 galaxies are dwarfs in the respective samples. As in \citetalias{Polimera22}, when we report one number (as when plotting points in Figures) we will default to the MPA-JHU version of our AGN catalog, which includes more galaxies than the other two versions and often, though not always, yields an intermediate value among the three results. More often, we will report ranges of values encompassing all three (and their uncertainties), since \citetalias{Polimera22} found no decisive basis for judging one catalog superior. \\

For robustness and parallelism with \citetalias{Polimera22}, we have required both [S~II] and [O~I] to have S/N $>$ 5 while searching for SF-AGN, i.e., the search sample for SF-AGN is selected in the same way as in \citetalias{Polimera22} and is equivalent to a BPT+VO sample. However, SF-AGN are defined to have AGN-like behavior in the [S~II] {\it and/or} [O~I] plots, meaning it would in principle be possible for follow-up studies to identify SF-AGN requiring adequate S/N in just one of the lines, depending on the available data. In RESOLVE and ECO, there exists a subset of galaxies for which [O~I] does not have enough S/N, but [S~II] does, though we do not include them in the SF-AGN search sample in this paper. We investigated this subsample and have low confidence in the status of these galaxies for two reasons. First, the atmospheric telluric contamination is potentially severe for [S~II] at the redshifts of our surveys \citep[see e.g.,][]{Lebouteiller25} and it is not alw
ays fully corrected. Second, the AGN status of the [S~II]-only SF-AGN is quite ambiguous in the [S~II] plot, with most points sitting right near the intersection of the LINER/Seyfert division line and the SF/AGN division line (from \citealt{Kewley2006} and \citealt{Kewley01}, respectively). Complementary to these [S~II]-only galaxies, we find a handful of galaxies that have S/N $>$ 5 for the weaker [O~I] line but not the stronger [S~II] doublet, which casts doubt on their flux data. As the [O~I] VO plot is a better identifier of SF-AGN (see Section~\ref{subsec:ch3sfagn}), these objects might merit reobservation to improve their flux data, but there are so few that including them would not change our reported SF-AGN frequencies within the uncertainties when using crossmatched MPA-JHU or NSA data, albeit there could be a larger increase when using crossmatched Portsmouth data. In any case, as stated, both sets of galaxies are rejected from our SF-AGN search sample in this work, both fo
 r robustness and for parallelism with \citetalias{Polimera22}.\\

The SF-AGN search sample (i.e., the BPT+VO sample) in this paper has 2566~/~1149~/1396 RESOLVE+ECO galaxies when selected using data crossmatched from the MPA-JHU~/~Portsmouth~/~NSA catalogs respectively, and it is a complete subset of the expanded BPT search sample. Among these galaxies, 1223~/~583~/~597 galaxies are dwarfs in the respective samples. We note that despite our use of the exact same selection criteria for the SF-AGN search sample in this work and in \citetalias{Polimera22}, the final sample numbers are slightly different due to small updates to redshifts, masses, and RA/Dec values in the H23 data releases for RESOLVE and ECO. \\

\subsection{Star Formation Rates}\label{subsec:ch3sfrs}
We estimated star formation rates (SFRs) using our mid-IR WISE photometry (discussed in Section \ref{subsec:ch3midirphot}) and deep GALEX NUV photometry or its equivalent, as discussed below\footnote{The combination of UV and IR optimally captures both unobscured and obscured star formation for the entire galaxy, not just the fiber aperture where the \ha from SDSS is measured (e.g., \citealt{Leroy12}).}. We computed mid-IR SFRs only for the 5333 galaxies for which our $\rm W3_{AP-total}$ or $\rm W4_{AP-total}$ magnitudes and errors have finite positive values and S/N $>$ 5, where the noise used is the statistical random error (see Table \ref{tab:irdatatable}). These SFRs were estimated using the prescription given in \cite{Jarrett13}:
\begin{equation}
    SFR_{IR-W3} = 4.91 \times 10^{-10} \nu_{W3} L_{W3} (M_\odot yr^{-1})
\end{equation}
\begin{equation}
    SFR_{IR-W4} = 7.5 \times 10^{-10} \nu_{W4} L_{W4} (M_\odot yr^{-1}),
\end{equation}
where $\nu_{W3}$ and $\nu_{W4}$ are the isophotal frequencies of the W3 and W4 bands \citep[see Table 1 in][]{Jarrett11}. We then averaged the two SFRs to get the total obscured star formation rate, $\rm SFR_{IR}$. \\

To compute UV SFRs, we used either directly measured or machine-learning-generated NUV magnitudes, depending on the availability of custom-processed NUV data from \citet{Eckert15} or \citet[][based on photometry from \citealt{Eckert16}]{Hutchens23}. Following \citet{Eckert15}, these studies provided custom-reprocessed GALEX NUV data only for exposure times $>$1000s (which we refer to as \emph{deep} data, comparable to the GALEX MIS survey). The goal of this exposure time restriction was to avoid unreliable NUV photometry: the distribution of GALEX image noise is bimodal reflecting an underlying bimodality in typical exposure times \citep{Leroy19}. Automated GALEX pipeline magnitudes differ systematically depending on image depth: for a subset of ECO galaxies with both deep and shallow GALEX imaging, \citet[][henceforth C25]{Carr25} compared SFRs computed from deep vs.\ shallow GALEX pipeline magnitudes and found systematic offsets with median fractional offsets, i.e., (deep $
-$ shallow)/deep values, ranging from 0.02 to 0.15 for high and low SFR bins, respectively, with 1$\sigma$ scatter out to $-$0.2 and +0.4 dex in the lowest SFR bin. Thus, we choose to report directly measured UV SFRs, $\rm SFR_{UV}$, only for galaxies with deep, custom-processed NUV data from \citet{Eckert15} or \citet{Hutchens23}. For these RESOLVE and ECO galaxies we have computed SFRs from the k-corrected and foreground and internal extinction-corrected NUV magnitude \emph{outputs} of SED fitting that used either custom-reprocessed  GALEX photometry \citep{Eckert15, Hutchens23} or equivalently high-quality Swift NUV photometry \citep{Eckert15} as inputs. For RESOLVE, $\sim$97\% of galaxies have deep NUV data, mostly from GALEX with a small fraction from Swift. Only $\sim$45\% of ECO galaxies have deep NUV data.\\

For the remaining $\sim$55\% of ECO without deep NUV data, we provide ``random forest'' star formation rates, $\rm SFR_{RFUV}$, derived using NUV magnitudes generated by random forest machine-learning in C25. The random-forest NUV magnitudes in C25 are designed to emulate the fully-corrected custom-processed NUV magnitudes of \citet{Eckert15} and \citet{Hutchens23}. These random-forest magnitudes outperform shallow GALEX pipeline magnitudes even when the shallow GALEX magnitudes are processed through SED fitting (C25). In Table~\ref{tab:irdatatable}, we provide either $\rm SFR_{UV}$ or $\rm SFR_{RFUV}$ but not both for each galaxy; null values are given as -999. Both $\rm SFR_{UV}$ and $\rm SFR_{RFUV}$ are unobscured SFRs, as both were computed from the NUV magnitude outputs of an SED fitting code that extinction-corrects using the prescription from \cite{Wilkins12}.   \\

We used the prescription from \cite{Buat11} to combine the non-dusty (UV) and dusty (IR) SFRs to infer total SFR as:
\begin{equation}
\label{eqn:sfr}
    \rm SFR = (1 - \eta) \rm SFR_{IR} + \gamma \rm SFR_{UV},
\end{equation} 
{where we adopted the constant values of $\eta = 0.17$ and $\gamma = 1$ from \cite{Buat11}}. As expected, we found that SFR$_{\rm UV}$ is typically dominant, except for dusty galaxies where SFR$_{\rm IR}$ contributes significantly. For ECO galaxies without deep NUV data, we have provided $\rm SFR_{RF}$ using the same prescription as above but replacing $\rm SFR_{UV}$ with $\rm SFR_{RFUV}$. \\

Table \ref{tab:irdatatable} provides SFR and $\rm SFR_{RF}$ for all of RESOLVE and ECO. The Appendix provides a detailed comparison of our directly measured SFRs with published values from the GSWLC-1 catalog \citep{Salim16}. We note that the new SFRs in this paper supersede the UV-only SFRs (without internal extinction corrections) previously provided for RESOLVE in \cite{Hood18}. \\

\begin{figure*}[]
    \centering
    \includegraphics[width = \linewidth]{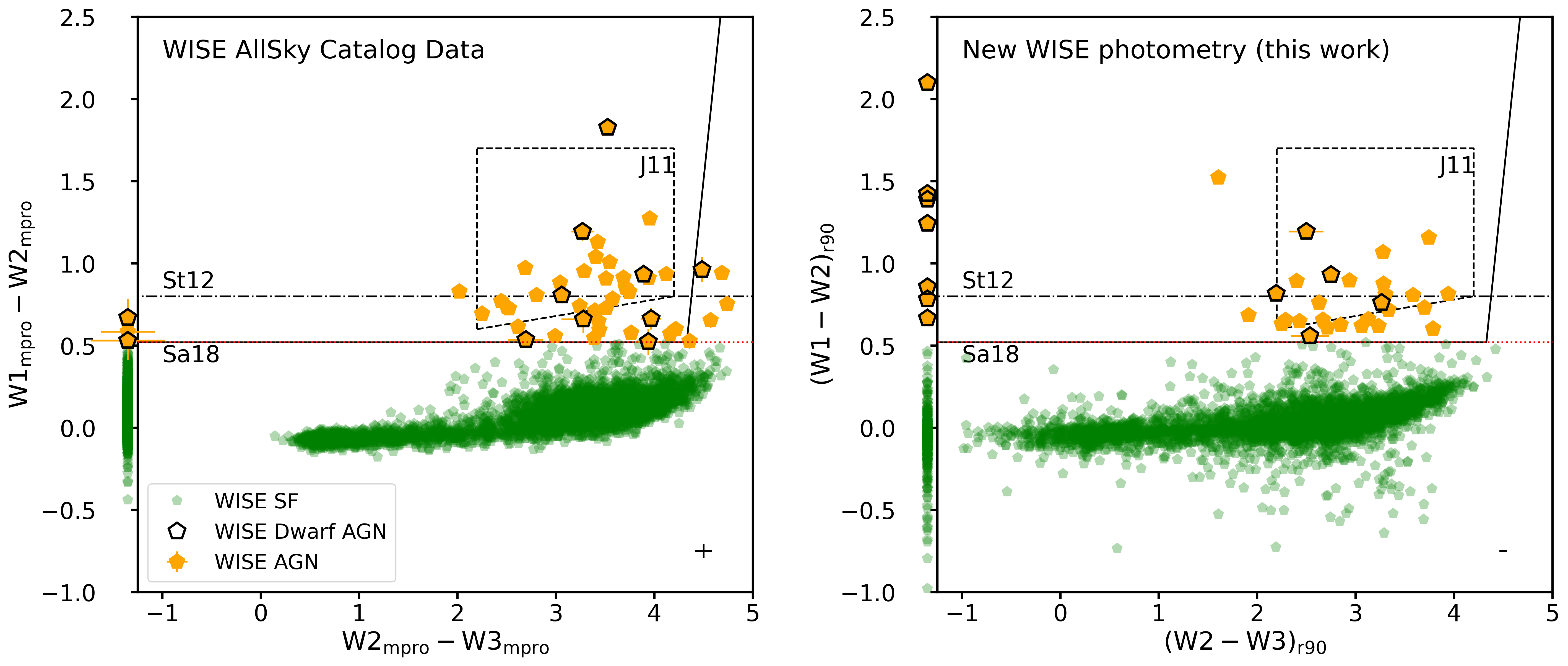}
    \caption{WISE AGN (orange points) in RESOLVE and ECO detected by mid-IR color selection for AGN search samples using AllWISE catalog data compared to our new photometry. Points with W3 S/N$<$5 are included in the AGN search sample and shown off-plot to the left of the y-axes. Dwarf AGN are shown in black outlines. Demarcation lines represent the AGN selection criteria from \citet[][dashed box]{Jarrett11}, \citet[][dot-dashed line]{Stern12}, \citet[][solid black line]{Satyapal18}, and our modified \citet[][]{Satyapal18} criterion (red dotted line). In both panels, $<$1\% of search sample galaxies fall outside the \wonetwo bounds of the plots. Typical error bars are shown in black at the bottom right corners.
      \textit{Left:} AllWISE catalog AGN search sample selected with S/N$>$5 for AllWISE W1 and W2 profile-fit magnitudes. Colors are computed as differences of profile-fit magnitudes. This sample selection is comparable to previous work (see Section \ref{sec:ch3midiragn}). \textit{Right:} Our new AGN search sample selected with peak S/N$>$5 for W1 and W2 (where the peak is measured inside \textit{r90}) from our custom WISE photometry presented in this work. Colors are computed as \wonetwo and \wtwothree, where our \textit{r90} colors use photometry convolved to the same resolution.
    }
    \label{fig:midiragn}
\end{figure*}

\section{AGN Detection using Mid-IR WISE Colors}\label{sec:ch3midiragn}

Mid-IR color selection can identify AGN because the hard radiation field from the accretion disk creates hot dust, producing redder mid-IR colors than those from only SF \citep{Jarrett11}. This method is especially successful in identifying dust-obscured AGN. In general, mid-IR color selection can potentially identify AGN in dwarfs despite their low dust content since dwarfs host smaller intermediate-mass BHs, which have accretion disks with higher temperatures \citep[][see also \citetalias{Richardson22}]{Peterson97}. \\

Figure \ref{fig:midiragn} shows WISE AGN selection criteria from \citet[][henceforth St12]{Stern12}, \citet[][henceforth Sa18]{Satyapal18}, and \citet[][henceforth J11]{Jarrett11} applied to RESOLVE and ECO galaxies using photometry from the AllWISE catalog and our custom pipeline described in Section \ref{subsec:ch3midirphot}. The left panel shows a WISE AGN search sample of 7501 galaxies drawn from the crossmatched sample of 8352 galaxies in common between RESOLVE+ECO and the AllWISE catalog. This sample is selected using AllWISE profile-fit magnitudes (e.g., w1$_{\rm mpro}$) with S/N$>$5 in W1 and W2, approximating the selection used in a number of previous studies (e.g., \citealt{Satyapal18}, \citealt{Kaviraj19}, \citealt{Lupi20}). Other past studies (e.g., \citealt{Sartori15} and \citealt{Hainline16}) have used more lenient restrictions like S/N $>$ 2 or 3. The colors in the left panel of Figure \ref{fig:midiragn} are computed by subtracting AllWISE profile-fit magnitude
s, e.g., w1$_{\rm mpro}$-w2$_{\rm mpro}$. Note that we do not restrict W3 S/N, so some galaxies lack measurements of W2-W3 and are plotted just to the left of the y-axis representing only their W1-W2 colors.\\

The right panel of Figure~\ref{fig:midiragn} shows our preferred WISE AGN search sample of 5443 galaxies, selected more stringently using our custom photometry to require that \wonetwo values must be finite and non-null and the peak S/N inside \textit{r90} must be $>$5 for both W1 and W2 (where the noise N includes only the random error). The colors used for selection and displayed in the right panel are the \textit{r90} aperture colors discussed in Section \ref{subsec:ch3midirphot}. We prefer these colors for AGN detection, because we have convolved the two bands to the same resolution to ensure that the magnitudes being subtracted are measured within the same region of the galaxy. Almost 99\% of the galaxies in the right panel (with our new photometry) are also in the left panel (with AllWISE catalog data), but only $\sim$72\% of the galaxies in the left panel are in the right panel. Our WISE AGN search sample has fewer galaxies because we used peak S/N to select our sample
, which is more restrictive than integrated S/N.\footnote{In this work, our final AGN sample is selected using only peak S/N inside \textit{r90}. However, for studies using only total magnitudes and integrated S/N cuts, our tests lead us to recommend also applying a S/N$>$2 cut for W3$_{\rm AP-total}$, regardless of whether W3 is used for AGN selection or not. \cite{Lupi20} followed a similar procedure to ensure a robust mid-IR sample.}\\

Our approach likely yields fewer false positives and false negatives compared to the more typical approach seen in the left panel of Figure~\ref{fig:midiragn}. Comparing the left and right panels, we see that our approach yields a cleaner separation between AGN and star-forming galaxies. This separation agrees with theoretical expectations for the division between the two populations (especially the extended horizontal Sa18 line in red). Using AllWISE catalog data with standard S/N cuts, 49 RESOLVE+ECO galaxies would be classified as WISE AGN using the red line, compared to the 35 WISE AGN we report. Among the 35 we report, only 21 are classified as AGN using both approaches. We infer the presence of 28 likely false positives (49-21) and 14 likely false negatives (35-21) in the AGN classifications that are based on AllWISE catalog data and standard S/N cuts. The cleaner separation of the SF/AGN regions in Figure~\ref{fig:midiragn} validates our belief that our approach (impro
ved photometry, use of peak rather than integrated S/N cuts, and use of resolution-matched \textit{r90} aperture colors) yields more secure results. \textit{Thus, using AllWISE catalog data with typical S/N cuts could result in an AGN sample in which $\sim$60\% of reported AGN are false positives, and could also lead to missing $\sim$40\% of genuine AGN \citep[similar to the findings of][]{Lupi20}.} \\

We expect some false negatives from our approach as well, because the low resolution of WISE imaging makes it difficult to isolate the mid-IR light of the central AGN from the light of the galaxy. {Approximately 93\% of our baryonic dwarfs are marginally resolved by WISE (median diameter $\sim$13 arcsec vs. $\sim$6.1 arcsec FWHM for the W1 band), with the rest unresolved. Thus, our WISE AGN tally must suffer from incompleteness due to missing AGN that have relatively weak accretion rates and/or reside in galaxies where the mid-IR color is dominated by star formation (as is common for non-satellite dwarf galaxies). The low resolution of WISE also implies potential contamination from close sources or companions, especially for faint or small sources \citep{Lupi20, Hainline16}, but we do not find evidence that masking such sources significantly affects our WISE AGN inventory. To assess whether overmasking could inadvertently remove mid-IR emission from an off-center AGN, such as
 a wandering black hole, we wrote a script to flag all dwarfs with masked pixels inside $r50$ or with $\geq$20\% of the $r90$ area masked. Among the 1901 dwarfs with good IR photometry, 163 galaxies ($\sim$9\%) were flagged. Upon manual visual inspection, every flagged case involved an identifiable contaminant seen clearly in optical imaging (foreground star or close companion); we found no examples of candidate AGN with WISE emission masked out. Conversely, among the 163 galaxies flagged for manual review, we found 21 galaxies ($\sim$1\% of mid-IR dwarf sample) where nearby contaminants were \textit{under}masked. For two of these galaxies where independent WXSC photometry with superior masking was available (from T. Jarrett, priv. comm.; see Section \ref{subsec:ch3midirphot_qc}), the W1-W2 colors differed minimally, by less than needed to alter their non-AGN classification. Another undermasked galaxy was classified as a WISE AGN (\textit{ECO01085}, a dwarf) but lacks confirmation at
  other frequencies and suffers from undermasking of a diffraction spike, so it may be a spurious AGN. None of the remaining 19 undermasked cases were classified as AGN. We conclude that both false and missed WISE AGN detections due to masking errors may exist, but are not common in our data set.} \\

As mentioned, Figure~\ref{fig:midiragn} shows WISE AGN selection criteria from \citetalias{Stern12} and \citetalias{Jarrett11} in addition to \citetalias{Satyapal18}. The \citetalias{Jarrett11} criteria form a purely empirical box in color-color space, based on a calibration sample biased toward relatively massive/high-redshift AGN. The \citetalias{Stern12} criterion is a line from semi-empirical modeling tuned to optically detected early-type galaxy AGN. The original \citetalias{Satyapal18} criteria consist of two lines derived from theoretical modeling that accounts for low metallicity and high star-formation; we use the red line as a modified \citetalias{Satyapal18} criterion (discussed below). The \citetalias{Stern12} and \citetalias{Jarrett11} cuts do not apply to typical dwarfs. Most dwarfs are not early-type galaxies. For the purposes of detecting AGN in dwarfs, the \citetalias{Satyapal18} cuts are the most directly applicable since they were defined using a suite of m
odels that explicitly included metal-poor and star-forming galaxies. The position of the nearly vertical sloped line proposed by the \citetalias{Satyapal18} (solid black line in Figure \ref{fig:midiragn}) could be debated: using SDSS optical spectroscopy, \citet{Hviding22} Figure~14 shows a decreasing emission-line AGN fraction to the right of this line in \wtwothree, so Hviding et al.\ argue for a lower upper limit in \wtwothree to enhance the purity of AGN selection. However, the numerator of their AGN fraction includes only traditional type I and II AGN, omitting the most common classes of dwarf AGN --- Composites, SF-AGN, and S06 Bonus AGN. When they consider a \wtwothree selection including Composites (their Figure~A1), their preferred vertical line position is very similar to that of \citetalias{Satyapal18}. \\ 

Assuming idealized mid-IR colors (lacking observational noise or resolution issues), \citetalias{Satyapal18} use SED modeling to conclude that purely star-forming dwarfs can contaminate their AGN selection only for non-physical ionization parameters that are not observed in typical \zzero dwarfs. These models do not account for the possibility of shocks, different interstellar medium (ISM) geometries, and lower BH masses, which could change the optimal demarcation lines. Also, though less conservative than the \citetalias{Stern12} and \citetalias{Jarrett11} cuts, the \citetalias{Satyapal18} cuts may be too conservative in that they do not identify AGN with spectral contributions $<$16\%, which may be common for star-forming dwarf AGN \citepalias{Richardson22, Polimera22}. Conversely, some observational studies have argued for a high likelihood of false AGN detections even using more conservative mid-IR color cuts due to contamination from young stellar populations \citep{Sartori15, Hainline16}. As dust heating near HMXBs and ULXs can match temperatures near AGN, the question of false positives challenges our limited understanding the frequency and distribution of stellar-mass compact objects vs.\ intermediate- and higher-mass BHs. \citet{Sturm25} go so far as to argue that mid-IR color-detected AGN without multi-wavelength corroboration are spurious, but since star formation and/or dust may hide optical or X-ray AGN signatures, we believe that pursuing AGN detection with WISE colors may offer complementary insights to other techniques.\\

Bearing these caveats in mind, the \citetalias{Satyapal18} cuts remain our default preference, since they were designed to apply to dwarf galaxies and Figure~\ref{fig:midiragn} does seem to confirm a natural division near the horizontal \citetalias{Satyapal18} line. However, the sloped, nearly vertical line in \wonetwo vs \wtwothree proposed by \citetalias{Satyapal18} does not affect any galaxy in our sample, so we have removed the $\rm W2-W3$ dependency by modifying the \citetalias{Satyapal18} cuts to use only the requirement that \wonetwo $\geq$ 0.452 (red dashed line in Figure \ref{fig:midiragn}). \textit{This single color cut is the main WISE AGN selection criterion for this paper.} We have made this choice specifically for the data set used in this paper, and future work using a different sample and/or photometry may need to reconsider this approach. The advantage of dropping the S/N requirement for W3 (due to removing the $\rm W2-W3$ dependency) is that we are able to i
ncrease our WISE AGN search sample by 6\% and increase the number of detected WISE AGN by seven galaxies in our default selection. Since Figure \ref{fig:midiragn} uses \wtwothree, the main plot shows only galaxies with usable W3 (peak S/N$>$5), and the additional galaxies that enter the search sample because of omitting any W3 S/N restriction are plotted off to the left of the y-axis. All seven AGN candidates added using the modified \citetalias{Satyapal18} selection (i.e., omitting any \wtwothree selection) are dwarf galaxies, so the effect of omitting any W3 S/N requirement is significant: our mid-IR dwarf AGN percentage of 0.6\% relative to the modified-\citetalias{Satyapal18} search sample (Section~\ref{sec:ch3agninventory}) would have been 0.3\% if computed relative to a search sample requiring W3 S/N$>$5. These seven WISE AGN should of course be considered less reliable.\\

The natural separation between SF- and AGN-hosting galaxies seen in our data (evident in the right panel of Figure~\ref{fig:midiragn}) does not align with the \citetalias{Stern12} criterion \wonetwo $\geq$ 0.8 as well as it does with the lower \citetalias{Satyapal18} cut. Nevertheless, we tabulate the \citetalias{Stern12} classifications in Section~\ref{sec:ch3agninventory}. {We do not separately tabulate the \citetalias{Jarrett11} classifications because the \citetalias{Jarrett11} color box is entirely encompassed by the \citetalias{Stern12} criterion for our sample, as seen in Figure~\ref{fig:midiragn}. Moreover, as noted above, the \citetalias{Jarrett11} criteria were calibrated on relatively massive and higher redshift AGN and are not designed for dwarf galaxies. We note that all photometric data necessary to apply the \citetalias{Jarrett11} or any other WISE color criteria are provided in Table~\ref{tab:irdatatable}.}\\

We also note that there are other WISE AGN selection criteria in the literature. For instance, \cite{Mateos12} defined a wedge to select AGN using W1, W2, and W3 magnitudes based on a complete flux-limited sample of bright X-ray-selected AGN in the ultrahard band (4.5-10 keV) of the Newton-XMM survey. However, this entire wedge would be encompassed inside the \citetalias{Satyapal18} criterion that we prefer. Additionally, a criterion defined using bright X-ray AGN could potentially miss out on identifying WISE AGN in dwarfs, which are known to be underluminous in the X-rays \citep[][see also Section~\ref{sec:ch3discussion}]{Latimer19}. \cite{Jarrett19} also defined a new criterion for so-called warm AGN that could have mixed contribution from AGN and SF \citep{Yao20}. Using this criterion could identify more WISE AGN with \wonetwo colors between 0.25-0.52, but with an increased risk of contamination from purely star-forming galaxies. \\

In the remainder of this paper, our analysis focuses on WISE AGN statistics based on our custom photometry and our modified \citetalias{Satyapal18} color cut. We provide a brief comparison to results using the \citetalias{Stern12} cut in Section \ref{sec:ch3agninventory}, where we also discuss some of the issues with using WISE colors to detect AGN. WISE color selection may yield more false positive detections compared to optical AGN detection methods. However, mid-IR detection is potentially sensitive to different host properties and evolutionary stages, so we do not wish to ignore WISE AGN candidates, rather we simply consider them less secure. \\

\section{Optical Emission-line AGN Detection}\label{sec:ch3opticalagnmethod}

Optical emission line ratio diagrams such as the BPT and VO plots serve as the most common means of AGN detection, where the shape of the incident ionizing radiation spectrum on the ISM in a galaxy controls the line ratios. Different diagrams use different forbidden lines that are more or less sensitive to ISM metallicity \citetalias{Polimera22}, while the incident spectrum is itself sensitive to dust in the ISM as well as the relative flux and geometry of SF and AGN radiation (\citetalias{Richardson22}). For this paper we have used two AGN detection methods requiring only the BPT plot (Sections \ref{subsec:ch3tradbpt} and \ref{subsec:ch3s06bonus}) as well as our SF-AGN detection method requiring the BPT and VO plots (Section \ref{subsec:ch3sfagn}). The VO plots were also used to subclassify Conventional AGN. The requirement of adequate S/N for all relevant emission lines (Section~\ref{subsec:ch3emlinedata}) yields an \textit{expanded BPT search sample} with 5325 galaxies and
 a \textit{BPT+VO search sample} with 2566 galaxies, quoting sample sizes for our preferred MPA-JHU crossmatched data set. Subtracting these numbers, 2759 galaxies are in a \textit{BPT-only search sample} for which subclassification of Conventional AGN is not possible.

\subsection{Detecting Conventional AGN Using the BPT Diagram}\label{subsec:ch3tradbpt}

\begin{figure*}[]
    \centering
    \includegraphics[width = \linewidth]{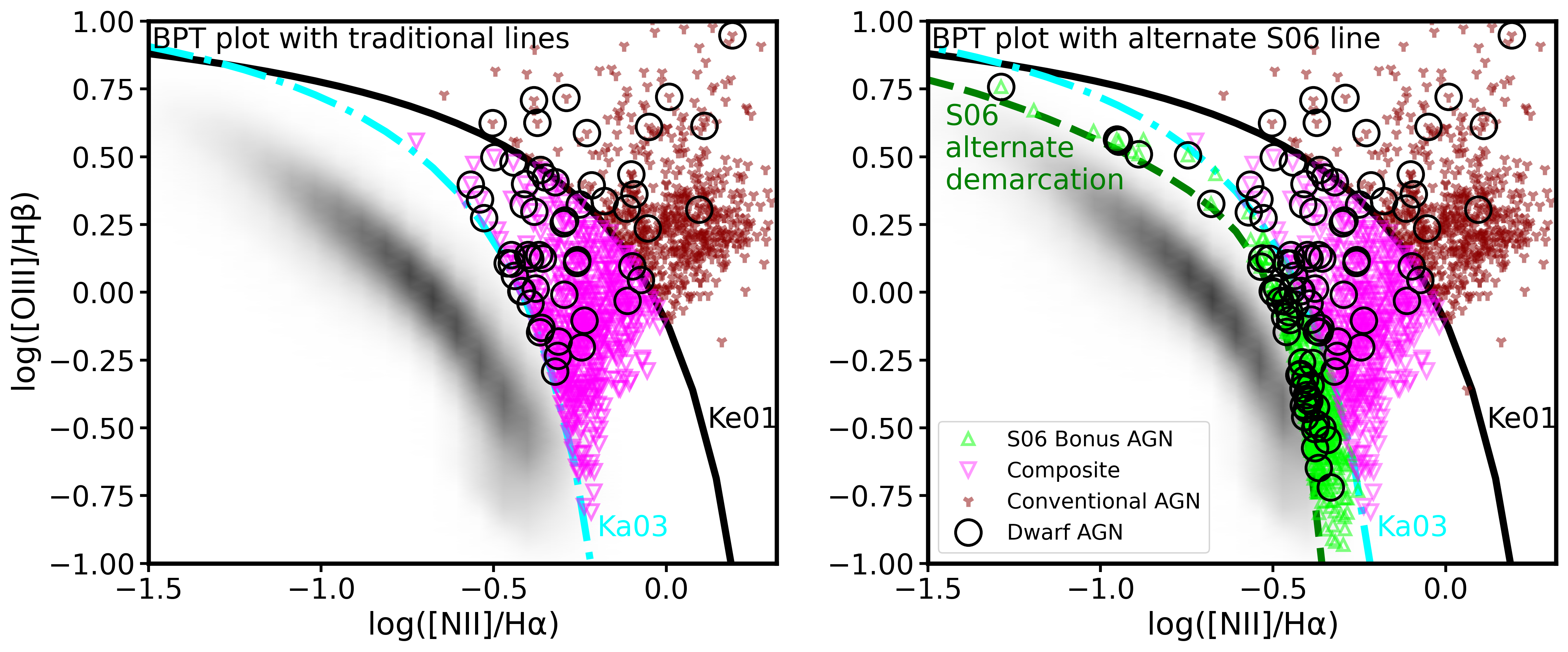}
    \caption{Comparison of traditional and alternate \citetalias{Stasinska06} demarcations in the BPT plot, used to detect optical AGN (colored symbols) and SF galaxies (grayscale) in the expanded BPT search sample obtained by crossmatching RESOLVE and ECO with the MPA-JHU emission-line catalog. \textit{Left panel:} Commonly used version of the BPT plot with the \citetalias{Kauffmann03} and \citetalias{Kewley01} demarcation lines in blue and black.  
    \textit{Right panel:} Alternate BPT demarcation line by \cite{Stasinska06} in green. 
    We refer to galaxies between the \citetalias{Kauffmann03} and \citetalias{Kewley01} demarcation lines as Composite (pink symbols), and to galaxies above the \citetalias{Kewley01} line as Conventional AGN (red symbols). We refer to the added AGN candidates between the S06 and Ka03 lines as S06 Bonus AGN (green points). The S06 Bonus AGN category includes a large number of dwarf AGN (black circles) close to the \citetalias{Stasinska06} line, likely due to relatively high spectral contributions from SF and relatively weak AGN signatures.}
    \label{fig:bptplot}
\end{figure*}

In order to separate SF and AGN activity in the BPT plot (Figure \ref{fig:bptplot} left panel), \citet[hereafter Ke01]{Kewley01} and \citet[hereafter Ka03]{Kauffmann03} provided demarcation lines from theoretical and semi-empirical modeling. We applied these demarcation lines to the expanded BPT search sample in RESOLVE and ECO (S/N $>$ 5 for all emission lines in the BPT plot), which is larger than the search sample used in \citetalias{Polimera22} (see Section \ref{subsec:ch3emlinedata}). Like in \citetalias{Polimera22}, we label the galaxies that fall between the \citetalias{Kewley01} and \citetalias{Kauffmann03} lines as \textit{Composite.} In this paper, we use \textit{Conventional AGN} as an umbrella term for all AGN categories that fall above the \citetalias{Kewley01} line in the BPT plot (i.e., the mutually exclusive Seyfert, LINER, Ambiguous-type AGN, and Low-[S~II] AGN categories from \citetalias{Polimera22} --- note that the \citetalias{Polimera22} Ambiguous-type ca
tegory is not identical to the \citetalias{Kewley01} Ambiguous category). \\

As detailed in \citetalias{Polimera22}, the BPT plot has some major drawbacks for dwarfs. Since it uses \NvH -- a metallicity proxy -- on the x-axis, it is biased against finding AGN in metal-poor galaxies. The plot is also sensitive to dilution by strong star formation. Thus, the BPT plot, as conventionally used, misses AGN in typical dwarfs, which are metal-poor and star-forming \citep{Dalcanton04, Geha12, Kannappan13}. The next two subsections describe methods that attempt to rectify some of the issues with conventional use of the BPT plot. \\

\subsection{Detecting Bonus AGN Using the BPT Diagram}\label{subsec:ch3s06bonus}

The \citetalias{Kewley01}/\citetalias{Kauffmann03} demarcation lines in the BPT plot, which were originally specified as \textit{maximum} starburst lines, were revisited using semi-empirical modelling by \citet[henceforth S06]{Stasinska06}. \citetalias{Stasinska06} provided a new \textit{typical} starburst line and gave the Ka03 line the role of the Ke01 line (Figure \ref{fig:bptplot}). Their models show that these demarcation lines can detect galaxies with AGN contributing as little as 3\% to their spectra, which is helpful because dwarf AGN may contribute relatively little to their hosts' spectra due to dilution by strong SF. \\

The right panel of Figure \ref{fig:bptplot} shows the \citetalias{Stasinska06} method applied to our expanded BPT search sample. We label the galaxies between the \citetalias{Stasinska06} and \citetalias{Kauffmann03} lines \textit{S06 Bonus AGN.} {Such galaxies were part of an expanded} Composite {category} in \citetalias{Stasinska06}, but we have {separated out the added composites as ``Bonus AGN"} to mitigate confusion with the classification categories from \cite{Kewley2006} and \citetalias{Polimera22}. Our S06 Bonus AGN were usually classified as ``Definite SF'' in \citetalias{Polimera22}, which used the \citetalias{Kauffmann03} line as the SF/AGN boundary. By adopting the lower \citetalias{Stasinska06} line, we reclassify these galaxies as AGN candidates. As in \citetalias{Polimera22}, our Composite and Conventional AGN categories (discussed in Section \ref{subsec:ch3tradbpt}) fall above the \citetalias{Kauffmann03} line, thus also above the S06 line. {If a galaxy qualifies as
 an S06 Bonus AGN we no longer classify it as SF, though following \citetalias{Polimera22} we do use the more expansive \citetalias{Kewley01}/\citetalias{Kauffmann03} SF definition as the first step in identifying SF-AGN (Section \ref{subsec:ch3sfagn}), so there is some overlap between SF-AGN and S06 Bonus AGN in our final classifications. In other words, our SF label in final classifications corresponds to the S06 SF classification.}\\  

{The Ke01/Ka03 and S06 classification systems are based on theoretical modeling, but an empirical study by \cite{deSouza17} gives us confidence in the S06 approach. \citet{deSouza17} analyzed a robust volume- and absolute-magnitude-limited sample and independently identified a population spanning the Composite and S06 Bonus AGN regions of the BPT diagram. This population is distinct from two separate SF populations (normal and starburst), supporting the interpretation that S06 Bonus AGN represent an extension of Composites rather than a contaminated tail of the SF locus (see Section~\ref{subsec:ch3sfagn} for further discussion). }\\

As S06 Bonus AGN essentially extend the Composite category to lower \NvH at values of log \OvHb below $\sim$0.25, the same caveats that apply to Composite classification apply to Bonus AGN. As defined by the BPT diagram alone, Composites and S06 Bonus AGN may include some false-positive, purely SF objects ionized by by Wolf-Rayet (WR) stars, HMXBs, and/or ULXs (\citetalias{Richardson25} and references therein). Figure~7 of \citetalias{Polimera22} shows that up to half of Composites are coincident with {Ke01/Ka03} SF galaxies in the \OvHb vs.\ [OI]/[OIII] diagram, close to though mostly above the division between purely star-forming WR galaxies and AGN-hosting WR galaxies from \citet{Brinchmann08}. However, the same figure shows that most SF-AGN, which are \textit{by definition} coincident with {Ke01/Ka03} SF galaxies in the BPT diagram, lie far above the WR plot region in the \OvHb vs.\ [OI]/[OIII] diagram. Conversely, Composites are by definition \textit{not} coincident with
 SF galaxies in the BPT diagram (see Figure~\ref{fig:bptplot}). This complexity reinforces the need for caution when using single diagnostic diagrams to draw conclusions (see Section~\ref{subsec:ch3sfagn}).\\

We note that \citetalias{Stasinska06} concluded --- using Starburst99-based photoionization models --- that none of the diagnostics we use (BPT and VO plots) are useful for identifying AGN in metal-poor galaxies. However, BPASS-based photoionization models suggest AGN in metal-poor galaxies actually can be identified with the VO plots, just not with the BPT plot (\citetalias{Richardson22,Polimera22}).\\

In practice, the S06 method uses only the BPT plot, so it does not add AGN candidates in the VO plots. The next section addresses the recovery of metal-poor dwarf AGN using the BPT and VO plots together, as introduced in \citetalias{Polimera22}.\\

{Before moving on to our definition of SF-AGN using the BPT+VO plots, we briefly consider two alternative classification approaches based on statistical and/or machine-learning methods applied to the BPT diagram combined with a second diagram. In the first approach, \citet{Ji20} projected the BPT and [S~II] diagrams into a common two-dimensional space ($P_1$, $P_2$) using photoionization modeling, then defined boundaries at approximately 10\% and 90\% AGN ionization fractions ($f_{\rm AGN}$; their Eqs.~8 and~9). In this reprojected diagram, our Conventional AGN and Composite populations fall in the intermediate zone between these curves, as expected. Our S06 Bonus AGN category falls predominantly below the 10\% $f_{\rm AGN}$ boundary, which is in principle consistent with S06's own assertion that their demarcation line can detect AGN contributions as low as $\sim$3\%\ \citep{Stasinska06}. We note however that the \citet{Ji20} demarcations are based on different stellar popula
tion synthesis models as well as different photoionization modeling choices than the S06 models. We therefore consider this comparison to indicate qualitative consistency between theoretically determined demarcations only. In a complementary, purely observational approach (already discussed earlier in this section), \citet{deSouza17} applied a Gaussian mixture model to the three-dimensional space of \NvH, \OvHb, and EW(H$\alpha$) emission-line data for a volume- and absolute-magnitude-limited ($\rm M_r < -19.88$) sample of SDSS DR7 galaxies. Except for its relatively bright selection floor, the \citet{deSouza17} sample is similar in design to RESOLVE+ECO. Within this sample, the authors identified four Gaussian components that correspond to star-forming (normal and starburst), composite, and AGN-hosting galaxies, where the \citet{deSouza17} composites span both the traditional Composite and S06 Bonus AGN populations in the BPT plane. As noted above, this purely empirical result indep
 endently supports the notion that the S06 Bonus AGN category represents a physically distinct population rather than a contaminated extension of the SF locus. Taken together, these two alternative statistical classification approaches corroborate our adopted BPT classification system.}

\subsection{Identifying SF-AGN Using BPT and VO diagrams}\label{subsec:ch3sfagn}

\begin{figure*}[]
    \centering
    \includegraphics[width = 0.8\linewidth]{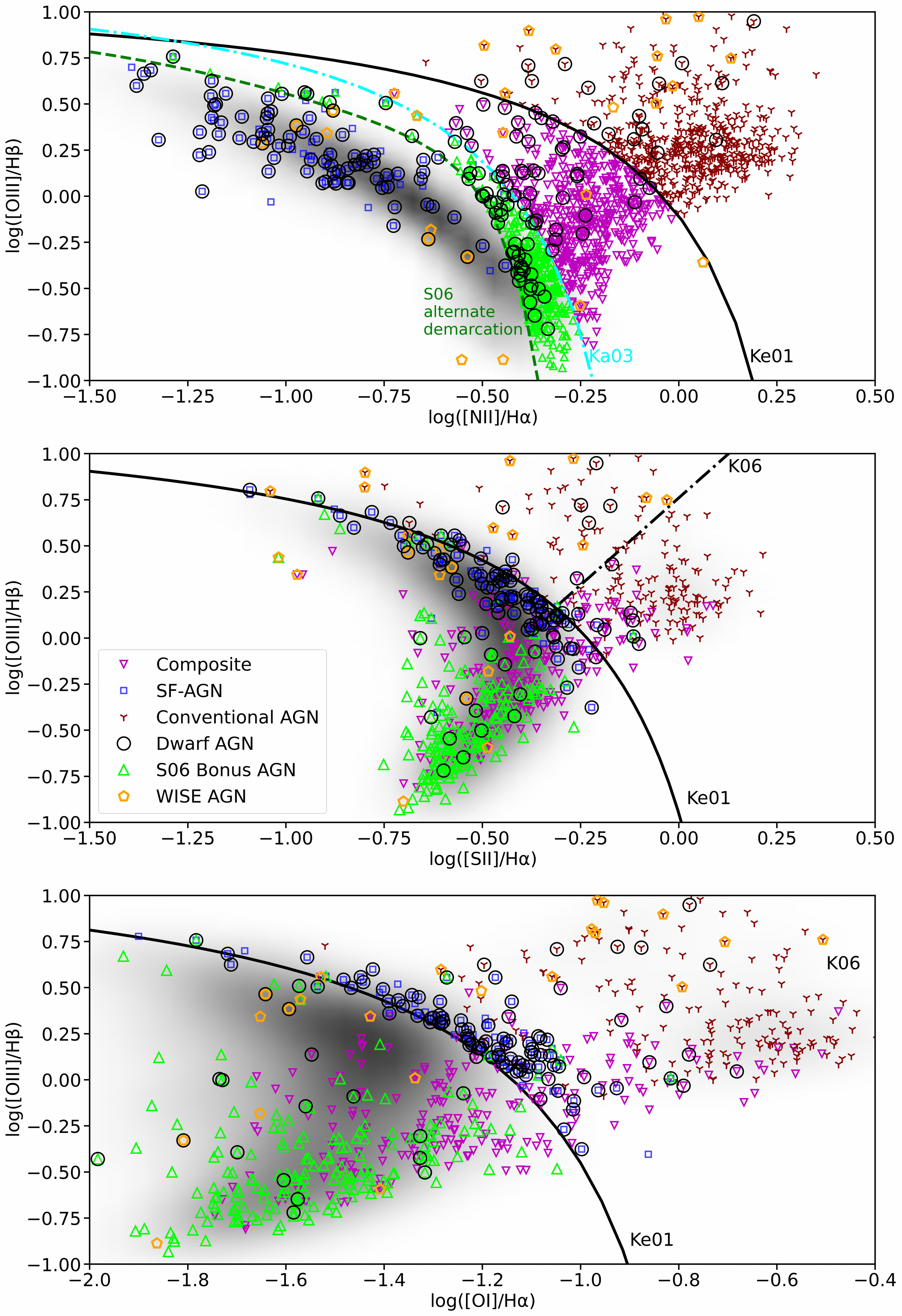}
    \caption{Optical AGN (colored symbols) and SF galaxies (grayscale) in the expanded BPT search sample defined by crossmatching RESOLVE and ECO with the MPA-JHU emission-line catalog. Red, pink, and green symbols are as in Figure \ref{fig:bptplot} and indicate traditional BPT AGN categories (Composites and Conventional AGN, Section~\ref{subsec:ch3tradbpt}) and additional AGN added by the alternate BPT demarcation of \citetalias{Stasinska06} (S06 Bonus AGN, Section~\ref{subsec:ch3s06bonus}). Blue symbols mark the new category of SF-AGN identified by the BPT+VO approach of \citetalias{Polimera22} (Section~\ref{subsec:ch3sfagn}). Only the subset of galaxies with both [O~I] and [S~II] S/N $>$ 5 (the BPT+VO search sample) appears in the two VO plots.  WISE AGN (see Section \ref{sec:ch3midiragn}) that overlap the expanded BPT search sample are marked with orange pentagons. All dwarf AGN are additionally marked with black circles as in Figure \ref{fig:bptplot}. About half of the d
warf AGN are SF-AGN, which are identified most effectively by the metallicity-insensitive [O~I] plot. Nine galaxies that are classified as both S06 Bonus AGN and SF-AGN are shown with overlapping green and blue symbols.}
    \label{fig:sfagn}
\end{figure*}

In \citetalias{Polimera22}, we used the BPT plot (\OvHb vs.\ \NvH) and VO plots (\OvHb vs.\ \SvH and \OvHb vs.\ \OvH) simultaneously to classify all galaxies in mutually exclusive categories (Figure \ref{fig:sfagn}). We defined \textit{Star-Forming AGN} or \textit{SF-AGN} as a new category of galaxies identified as AGN by the metallicity-insensitive [S~II] and/or [O~I] plots, but not by the metallicity-sensitive [N~II] plot. In Section 4 of \citetalias{Polimera22}, we provided an in-depth analysis showing that SF-AGN are a distinct AGN population that cannot be easily explained by other phenomena like extreme starbursts, diffused ionized gas, shocks, etc., and that SF-AGN hosts are missed by other dwarf AGN detection methods (though there is a small amount of overlap with S06 Bonus AGN). Furthermore, we provided photoionization modeling to show that dwarf AGN with realistic metallicities and mixed SF+AGN emission should have been expected to show up only in the VO plots and n
ot the BPT plot (\citetalias{Richardson22, Polimera22}). Thus, identifying SF-AGN allows us to recover AGN in typical dwarfs that would otherwise be missed by the conventional ways of using optical diagnostics. \\

The fact that we required S/N$>$5 for both [S~II] and [O~I] for membership in the SF-AGN search sample (Section \ref{subsec:ch3emlinedata}), but required AGN status in only one of the [S~II] or [O~I] plots for SF-AGN classification, means that we can assess the relative importance of these lines for SF-AGN detection. We will see in Section~\ref{sec:ch3agninventory} that the [O~I] plot is much more important than the [S~II] plot, consistent with theoretical models showing that the [O~I] plot is more sensitive to low fractional AGN contributions in the presence of SF (\citetalias{Richardson22, Polimera22}). This result is comforting, since as noted in Section~\ref{subsec:ch3emlinedata}, [S~II]-only SF-AGN without [O~I] data tend to fall too close to the LINER and SF regions of the [S~II] plot to warrant confident AGN classification. \\

Recently, \citet{Pucha25} argued against attempting SF-AGN detection using the [S~II] plot alone, and we would concur for the aforementioned reasons. However, the rationale put forward by \citet{Pucha25} seems to us problematic: these authors have relied on newly proposed BPT and VO plot demarcation lines from \citet{Law21} to suggest that AGN identified immediately above the \citetalias{Kewley01} line in the [S~II] plot are suspect. The problem with this argument (which could be extended to the [O~I] plot as well) is that the \citet{Law21} demarcations are (i) based on high gas velocity dispersions, which are a very indirect proxy for AGN activity and would be expected to be lower for the lower-mass BHs in dwarf galaxies, and (ii) based on a statistic sensitive to the frequency of spaxels with these high dispersions in galaxy-wide integral field spectroscopy -- in other words, based on an area-weighted rather than a flux-weighted statistic. In fact, \citet{Law21} clarified t
hat their demarcations become consistent with those of \citetalias{Kewley01} when either restricting their analysis to radii $<$3$''$ to match the SDSS fiber spectroscopy used by \citetalias{Kewley01} (and ourselves) or substituting a flux-weighted statistic, similarly favoring low radii. Also of interest, \citet{Law21} noted an abundance of moderately high velocity dispersions in non-central locations in low-metallicity galaxies. They assumed that these dispersions must reflect SF when constructing their demarcations, but both theoretical and observational studies have now established that dwarf AGN are likely to wander significantly off-center \citep[e.g.,][]{Bellovary21,Mezcua24}, so further investigation is warranted.\\

The reliability of SF-AGN as a new class of AGN is discussed extensively in \citetalias{Polimera22}. In brief, we argued that while individual SF-AGN detections might be explained away by, e.g., shocks, stellar emission (HMXBs or WRs), or DIG, the general behavior of SF-AGN as a class suggests true AGN activity. Both model fitting with NebulaBayes and analysis in the \OvHb vs.\ [OI]/[OIII] diagram, where most SF-AGN lie very far from purely star-forming WR galaxies (Figure~7 of \citetalias{Polimera22}), led \citetalias{Polimera22} to conclude that $<$10\% of SF-AGN are likely false positives. This false positive rate is negligible compared to the uncertainties caused by emission-line measurement systematics between the MPA-JHU, Portsmouth, and NSA catalogs (see Section~\ref{sec:optwisestats}). \\

Since the publication of \citetalias{Polimera22}, \citet{Garofali24} has published new theoretical models of the hard ionizing continuum from stellar-mass ULXs, and \citet[][hereafter R25]{Richardson25} has shown that for near-solar metallicity, these ULX models combined with star formation models occupy the same part of the [OI] plot as dwarf AGN + star formation models where f${_{\rm AGN}} \lesssim 0.16$ (see \citetalias{Richardson25} Figure 20). \citetalias{Richardson25} has proposed raising the \citetalias{Kewley01} line in the [OI] plot to this f${_{\rm AGN}}=0.16$ threshold to ensure a pure sample of AGN. However, we believe doing so would prioritize purity excessively over completeness. Given the body of observational evidence for AGN activity above the \citetalias{Kewley01} line and the inherently star-forming nature of SF-AGN, many genuine dwarf AGN could be missed by attempting to rule out all possibility of a ULX interpretation. Future work (Levanti et al., in prep
.) will examine ULX and AGN models for SF-AGN using all available emission lines to assess their relative contributions. \\

Figure \ref{fig:sfagn} shows the \citetalias{Polimera22} approach applied to the entire RESOLVE and ECO extended BPT search sample (see Section \ref{subsec:ch3emlinedata}), which is $\sim$2$\times$ larger than the strong emission line search sample used in \citetalias{Polimera22}. We can identify SF-AGN only within the BPT+VO subsample. The same BPT+VO subsample is also used to subclassify Conventional AGN as Seyfert, LINER, Low-[S~II], or Ambiguous-type AGN. While Figure \ref{fig:sfagn} shows that SF-AGN are mostly dwarf AGN, and we have seen that their galaxy hosts tend to be typical dwarfs in having high SF and low metallicity (\citetalias{Polimera22} and Section~\ref{sec:ch3discussion}), they are not unbiased tracers of the dwarf AGN population. One issue is that the use of the [O~I] plot to identify SF-AGN requires a cut of S/N $>$ 5 for the weak [O~I] line. \citetalias{Polimera22} accepted the sample size reduction, whereas in this paper we have partially avoided it by 
assigning only conventional BPT classifications for the BPT-only sample and assigning the SF-AGN classification only for the smaller BPT+VO sample. A second issue is that the demarcation lines used to define SF-AGN are still based on semi-empirical modeling and do not follow the locus of real galaxies optimally, as shown in \citetalias{Stasinska06}. Despite both limitations, SF-AGN still constitute the largest sub-population of dwarf AGN in our expanded sample based on two of the three SDSS emission-line catalogs (see Section~\ref{sec:optwisestats2}), and their frequency alone exceeds most previous dwarf AGN frequency estimates from \zzero optical searches (see Section 6 of \citetalias{Polimera22}), with the notable exception of \cite{Mezcua24}, who found abundant off-center AGN and SF-AGN (following the \citetalias{Polimera22} methodology) using MaNGA IFU data (but note that the definition of dwarfs in this paper has been corrected in an erratum and is now dominated by what we call 
 transitional galaxies; see Section~\ref{sec:agnfreq}). \\

\section{Optical BPT/VO and WISE AGN Inventory in RESOLVE and ECO}\label{sec:ch3agninventory}

In this section we present the RESOLVE and ECO AGN inventory based on the two data sources that provide uniform coverage across the surveys, SDSS optical spectroscopy and WISE mid-IR imaging. We consider only AGN detected using the BPT/VO plots and/or WISE colors, deferring discussion of AGN detected using other techniques to Section~\ref{sec:ch3discussion}, where we will consider previously published AGN catalogs that use other techniques as part of comparing our results with the literature. Because such archival AGN data sources are nonuniform across RESOLVE and ECO, they are excluded from the statistical discussion below.\\

{We remind the reader that our ``SF'' classification label refers to the S06 SF category (galaxies below the S06 line in the BPT diagram; see Section~\ref{subsec:ch3s06bonus}) except when referring to identification of SF-AGN based on mixed VO-plot and BPT-plot classifications using Ke01/Ka03 demarcations (following \citetalias{Polimera22}).}

\subsection{AGN Categories and Search Samples}\label{sec:flags}

\begin{figure*}[]
    \centering
    \includegraphics[width=\linewidth]{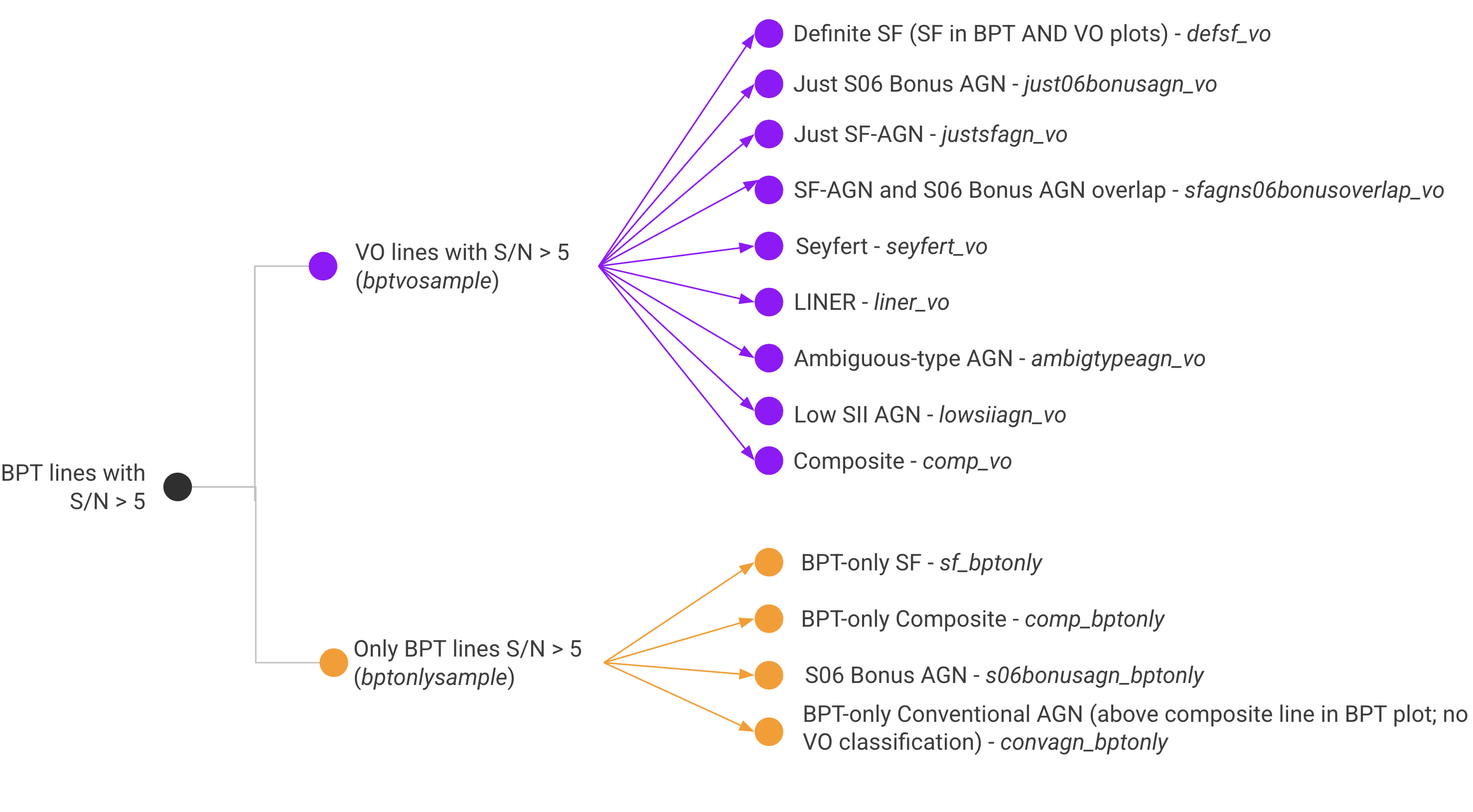}
    \caption{Tree diagram showing the optical emission-line classification categories used in Table~\ref{tab:agntable}. The BPT-only search sample (S/N $>$ 5 for \hb, [O~III], [N~II], and \ha, but \textit{not} for [S~II] and [O~I]) and the BPT+VO search sample (S/N $>$ 5 for all BPT and VO lines, as required by \citetalias{Polimera22}) are mutually exclusive and combine to make up this paper's expanded BPT search sample. All categories within each of these search samples are also mutually exclusive. A set of these category flags exists for each of the three optical emission-line catalogs crossmatched to RESOLVE and ECO, i.e., MPA-JHU, Portsmouth, and NSA. {Note that the ``Definite SF'' (\textit{defsf\_vo}) and ``SF'' (\textit{sf\_bptonly}) categories use the \citetalias{Stasinska06} line as the SF/AGN boundary in the BPT diagram, which reclassifies some of the \citetalias{Kauffmann03} SF category used by \citetalias{Polimera22} as AGN candidates (what we here call ``S06 Bonus
 AGN''; see Section~\ref{subsec:ch3s06bonus}). However, we still define SF-AGN as in \citetalias{Polimera22} using the larger \citetalias{Kauffmann03} SF category, so the (small) overlap of SF-AGN and S06 Bonus AGN is given the category flag \textit{sfagns06bonusoverlap\_vo}.} }
    \label{fig:agnflowchart}
\end{figure*}

\begin{table*}[]
    \begin{center}
    \caption{AGN Search Sample and Category Table for RESOLVE and ECO galaxies\tablenotemark{1}}
    \begin{tabular}{c|p{12cm}}
        \hline
        Column Name &  Description \\
        \hline
        name & Galaxy Name \\
        inbarysample & Flag for baryonic-mass-selected sample (\mbary $> 10^{9.2}$\msun)  \\
        instellarsample & Flag for stellar-mass-selected sample (\mstar $> 10^{8.9}$\msun)  \\
        \hline
        \multicolumn{2}{c}{Optical AGN\tablenotemark{2}}          \\
        \hline
        $<$sdsscat$>$\_bptvosample & BPT+VO AGN search sample \\
        $<$sdsscat$>$\_defsf\_vo & Definite SF in BPT+VO sample, i.e., below the \citetalias{Stasinska06} line in BPT and not identified as AGN in either VO plot \\
        $<$sdsscat$>$\_sfagndetxn\_vo & VO Plot(s) that indicate SF-AGN detection\tablenotemark{3} \\ 
        $<$sdsscat$>$\_justs06bonusagn\_vo & S06 Bonus AGN but not SF-AGN in BPT+VO sample \\
        $<$sdsscat$>$\_justsfagn\_vo & SF-AGN but not S06 Bonus AGN in BPT+VO sample \\
        $<$sdsscat$>$\_sfagns06bonusoverlap\_vo & Both S06 Bonus AGN and SF-AGN in BPT+VO sample \\
        $<$sdsscat$>$\_seyfert\_vo & Seyferts in BPT+VO sample\\
        $<$sdsscat$>$\_liner\_vo & LINERs in BPT+VO sample\\
        $<$sdsscat$>$\_ambigtypeagn\_vo & Ambiguous-type AGN in BPT+VO sample\\
        $<$sdsscat$>$\_lowsii\_vo & Low-[S~II] AGN in BPT+VO sample\\
        $<$sdsscat$>$\_comp\_vo & Composites in BPT+VO sample\\

        $<$sdsscat$>$\_bptonlysample & BPT-only AGN search sample (inadequate VO data) \\
        $<$sdsscat$>$\_sf\_bptonly & {SF in BPT-only sample, i.e., below the \citetalias{Stasinska06} line in BPT but lacking adequate VO data to rule out SF-AGN status}  \\
        $<$sdsscat$>$\_comp\_bptonly & Composites in BPT-only sample (inadequate VO data) \\
        $<$sdsscat$>$\_s06bonusagn\_bptonly & S06 Bonus AGN in BPT-only sample (inadequate VO data) \\ 
        $<$sdsscat$>$\_convagn\_bptonly & Conventional AGN in BPT-only sample (inadequate VO data)  \\

        \hline
        \multicolumn{2}{c}{Mid-IR AGN}       \\
        \hline
        inirsample & WISE AGN search sample \\
        stern12agn & AGN as per \cite{Stern12} \\
        satyapal18agn & AGN as per modified \cite{Satyapal18} criteria used in this paper\\
         \hline
        \multicolumn{2}{c}{Archival AGN}       \\
        \hline
        broadlineagn & Crossmatches in \ha broadline AGN catalog from \cite{Liu19} \\
        opticalcatalogagn & Crossmatches in optical AGN catalogs from \cite{Veron06} and \cite{Flesch15} \\
        xrayagn\tablenotemark{4} & {AGN candidates based on the $L_{\rm X,obs} \geq 3 L_{\rm exp}$ criterion of \citet{Birchall20} (preferred; see Section~\ref{sec:archival})} \\
        xrayagn\_ranalli\tablenotemark{4} & {AGN candidates based on $L_{\rm X-ray}$-SFR relation of \citet{Ranalli03} (see Section~\ref{sec:archival})} \\
        \hline
    \end{tabular}
    \label{tab:agntable}
    \end{center}
\tablenotetext{1}{Except for $<$sdsscat$>$\_sfagndetxn\_vo (see note 3 below), all category flags have values 0, 1, or -999, indicating False, True, or N/A, respectively.}
\tablenotetext{2}{To select all optical emission line AGN found in this work, define $<$sdsscat$>$\_opticalagn = (($<$sdsscat$>$\_defsf\_vo = 0) AND ($<$sdsscat$>$\_bptvosample = 1)) OR (($<$sdsscat$>$\_sf\_bptonly = 0) AND ($<$sdsscat$>$\_bptonlysample = 1)). To select all SF-AGN found in this work, define $<$sdsscat$>$\_sfagn = ($<$sdsscat$>$justsfagn\_vo OR $<$sdsscat$>$sfagns06bonusoverlap\_vo). The notation $<$sdsscat$>$ = `jhu', `port', or `nsa' for the MPA-JHU, Portsmouth, or NSA catalogs.}
\tablenotetext{3}{Not an AGN category: takes ascii values `OI', `SII', `both', or `NA' to reflect how a given SF-AGN was detected. Note that $<$sdsscat$>$\_sfagndetxn\_vo $\neq$ `NA' whenever ($<$sdsscat$>$justsfagn\_vo OR $<$sdsscat$>$sfagns06bonusoverlap\_vo) = True.}
\tablenotetext{4}{Based on X-ray luminosities from archival X-ray data from the Chandra and XMM telescopes (see Section~\ref{sec:archival}).}
\end{table*}

Table \ref{tab:agntable} provides a comprehensive AGN inventory for the full RESOLVE and ECO parent surveys. The flags inbarysample = 1 and instellarsample = 1 select the baryonic-mass-limited sample (\mbary $> 10^{9.2}$\msun) and the stellar-mass-limited sample (\mstar $> 10^{8.9}$\msun), respectively; see Section \ref{subsec:ch3surveys}.\\

Figure \ref{fig:agnflowchart} shows a tree diagram of the optical emission line classification categories used in Table~\ref{tab:agntable}. Each category has a flag with a value of 0, 1, or -999, indicating False, True, or N/A, respectively. The \textit{bptonlysample = 1} flag defines the BPT-only search sample (S/N $>$ 5 for \hb, [O~III], [N~II], and \ha), while the \textit{bptvosample = 1} flag defines the BPT+VO search sample (S/N $>$ 5 for all BPT and VO lines including [S~II] and [O~I]; see Section~\ref{subsec:ch3emlinedata}). The BPT+VO search sample serves as the SF-AGN search sample, with identical selection criteria as in \citetalias{Polimera22}. The BPT-only and BPT+VO search samples are mutually exclusive, so for Conventional AGN with a subclassification from the VO plots, \textit{bptvosample} is set to 1 along with one of \textit{seyfert\_vo},  \textit{liner\_vo}, \textit{ambigtypeagn\_vo}, or \textit{lowsiiagn\_vo}, while for Conventional AGN lacking VO data, \textit{bptonlysample} and \textit{convagn\_bptonly} are both set to 1. Each category on the right side of this diagram is mutually exclusive, so a galaxy can have a category flag set to 1 for only one category. \\

When we defined the optical AGN categories in Section~\ref{sec:ch3opticalagnmethod}, two categories were not automatically mutually exclusive: S06 Bonus AGN and SF-AGN (see Section \ref{subsec:ch3s06bonus} and \ref{subsec:ch3sfagn} for definitions). Thus, to avoid double-counting, we split the galaxies in these two categories into four subcategories. The flag \textit{justs06bonusagn\_vo = 1} selects galaxies in the \textit{bptvosample} that are \textit{only} classified as S06 Bonus AGN. In parallel, the flag \textit{justsfagn\_vo = 1} selects galaxies in the \textit{bptvosample} that are \textit{only} classified as SF-AGN. The flag the flag \textit{sfagns06bonusoverlap\_vo = 1} selects galaxies that are both SF-AGN and S06 Bonus AGN in the \textit{bptvosample}; such overlap cases are relatively few (e.g., $<$10\% of SF-AGN identified with MPA-JHU data, see Section~\ref{sec:optwisestats1}). The subset of S06 Bonus AGN that are in the \textit{bptonlysample} can be selected with
 the flag \textit{s06bonusagn\_bptonly = 1}. Table~\ref{tab:agntable} also includes the parameter \textit{sfagndetxn} with values `OI', `SII', `both', or `NA' to reflect how a given SF-AGN was detected (see discussion in Section~\ref{subsec:ch3s06bonus}).  {We distinguish ``Definite SF'' (\textit{defsf\_vo}) from SF in the BPT-only sample (\textit{sf\_bptonly}) because the former have been confirmed as non-AGN in both VO plots, whereas the latter lack adequate VO data and could include unidentified SF-AGN. Within the BPT+VO sample, our Definite SF and S06 Bonus AGN categories together correspond approximately to the ``Definite SF'' category of \citetalias{Polimera22}, which used the \citetalias{Kauffmann03} line rather than the \citetalias{Stasinska06} line as the SF boundary. The correspondence is approximate because a few S06 Bonus AGN with VO data also qualify as SF-AGN; this is the only classification category overlap allowed in our system and it is rare, as SF-AGN and S06 Bonus 
 AGN are largely complementary.}\\

Table~\ref{tab:agntable} also provides WISE AGN classification flags based on both our preferred (modified, see Section \ref{sec:ch3midiragn}) \citetalias{Satyapal18} criterion and the more restrictive \citetalias{Stern12} criterion, though we default to the former for our analysis. For both WISE AGN categories, the WISE search sample, i.e., galaxies having finite and non-null \wonetwo values and peak S/N inside $r_{90}$ greater than 5 in W1 and W2, has flag \textit{inirsample = 1}. \\

\subsection{AGN Statistics}\label{sec:optwisestats}

Table \ref{tab:agnstats} lists AGN statistics derived from the classifications in Table~\ref{tab:agntable} separately for dwarfs, giants, and transitional galaxies in the baryonic-mass selected version of the combined ECO and RESOLVE surveys. We will focus our discussion on these baryonic-mass selected results, but stellar-mass selected results are also provided in a separate table in Section~\ref{sec:optwisestats5}. All these tables also include information on archival AGN, to be discussed in Section~\ref{sec:ch3discussion}. \\

Before diving into statistics, we need to explain our reporting conventions. Our study is unusual in reporting AGN frequencies relative to both the \textit{search sample} --- the sample with adequate S/N in crossmatched emission-line data (or WISE data for BPT/VO plot analysis (or WISE mid-IR color analysis) --- and the \textit{full sample} --- the parent sample, i.e., the entire galaxy population including galaxies with inadequate data for BPT/VO analysis (or WISE color analysis). Search-sample statistics are more typical in the literature, but full-population statistics represent the natural frequency of AGN.\\

For our BPT/VO-identified AGN, we tabulate results based on crossmatched data from all three of the MPA-JHU, Portsmouth, and NSA catalogs separately, because AGN frequencies derived from optical diagnostics are sensitive to the underlying emission-line measurements. \textit{Our discussion below summarizes these results by quoting a range encompassing the three values from the three catalogs and their error bars, whereas for visualization, we default to the MPA-JHU crossmatched sample for the reasons given in Section~\ref{subsec:ch3emlinedata}.} \\

In Section 4.3 of \citetalias{Polimera22}, we provided a detailed discussion of how AGN statistics are affected by the choice between these catalogs, which have differences in their measurement pipelines. We discussed that the variation in AGN percentage for optical narrow-line methods is primarily due to irreducible systematics between the source catalogs arising from differences in spectral modeling methodologies and choices. The main drivers of the discrepancies are differences in the choice of stellar population models for spectral decomposition, differences in flux calibrations, and differences in estimates of errors on fluxes. The same effects are seen in this work, for example in the percentage of Conventional AGN in giants (and consequently the percentage of all AGN in giants), for which our estimate using MPA-JHU crossmatched data is more than 2$\times$ our Portsmouth- or NSA-based estimates. Both individual object classifications and sample definitions can disagree:
 for example, of the 79 SF-AGN found in dwarfs using MPA-JHU crossmatched data, only 16 are classified as AGN using NSA crossmatched data (15 as SF-AGN, 1 as Composite), while 30 are not in the NSA analysis sample at all (either due to missing measurements or S/N cuts) and 33 are in the NSA BPT sample but fail to qualify for VO analysis (due to S/N cuts). Using Portsmouth crossmatched data yields even greater discrepancies. All three catalogs are nominally based on the same SDSS spectroscopy, and all are considered state-of-the-art. Thus, we stress that differences in methodological choices for different data sources should always be considered when interpreting AGN statistics (see \citetalias{Polimera22} for more details). \\

For our WISE-identified AGN, we tabulate results based on two different WISE color cuts, the \citetalias{Stern12} and \citetalias{Satyapal18} criteria (the latter modified to omit the \wtwothree color cut) as described in Section~\ref{sec:ch3midiragn}, but we do not tabulate results based on AllWISE catalog data. As shown in Section \ref{sec:ch3midiragn}, AGN hosts identified using AllWISE catalog colors could potentially have $\sim$60\% false positives while also missing $\sim$40\% of the AGN we detect using our optimized WISE photometry with resolution-matched \textit{r90} fixed-aperture colors. Moreover, as discussed in Section \ref{sec:ch3midiragn}, we prefer the modified \citetalias{Satyapal18} cut because it better aligns with the natural separation between the AGN and SF sequences seen in our optimized photometry (Figure~\ref{fig:midiragn}), and also because the underlying models are better suited to find dwarf AGN (Section~\ref{sec:ch3midiragn}). Substituting the \citetalias{Stern12} cut would reject about half of the AGN in dwarf and transitional galaxies and all six of the AGN in giants found using the modified \citetalias{Satyapal18} cut (Table \ref{tab:agnstats}). As we believe our optimized photometry and the modified \citetalias{Satyapal18} are demonstrably better than the alternatives, our discussion below will quote only the WISE AGN frequencies derived using these choices. \\

\startlongtable
\begin{deluxetable*}{|l|c|c|c|c|c|}
\tablecaption{Statistics of AGN in RESOLVE and ECO Galaxies in the baryonic-mass-limited sample with \mbary $\geq 10^{9.2}$\msun \label{tab:agnstats}}
\tablehead{ \multicolumn{1}{|l|}{{AGN Category}} & \multicolumn{1}{c|}{{$\rm N_{gal,parent}$}} & \multicolumn{1}{c|}{{$\rm N_{gal,search}$}} & \multicolumn{1}{c|}{{AGN}} & \multicolumn{1}{c|}{{AGN\% in}} & \multicolumn{1}{c|}{{AGN\% in}}\\
\multicolumn{1}{|c|}{ }&\multicolumn{1}{c|}{ } &\multicolumn{1}{|c|}{ } &\multicolumn{1}{c|}{ } & \multicolumn{1}{c|}{{$\rm N_{gal,parent}$}} & \multicolumn{1}{c|}{{$\rm N_{gal,search}$}}
}
\startdata
\multicolumn{6}{|c|}{Dwarfs ($10^{9.2} \leq$ \mbary $\leq 10^{9.7}$\msun)} \\ \hline
\multicolumn{1}{|l|}{WISE AGN - St12} & \multicolumn{1}{c|}{3691} & \multicolumn{1}{c|}{1901} & \multicolumn{1}{c|}{8} & \multicolumn{1}{c|}{$0.2^{+0.1}_{-0.1}\%$} & \multicolumn{1}{c|}{$0.4^{+0.2}_{-0.1}\%$} \\
\multicolumn{1}{|l|}{WISE AGN - modified Sa18} & \multicolumn{1}{c|}{3691} & \multicolumn{1}{c|}{1901} & \multicolumn{1}{c|}{12} & \multicolumn{1}{c|}{$0.3^{+0.1}_{-0.1}\%$} & \multicolumn{1}{c|}{$0.6^{+0.2}_{-0.2}\%$} \\
\hline
\multicolumn{1}{|l|}{Composite - MPA-JHU} & \multicolumn{1}{c|}{3691} & \multicolumn{1}{c|}{2494} & \multicolumn{1}{c|}{39} & \multicolumn{1}{c|}{$1.1^{+0.2}_{-0.2}\%$} & \multicolumn{1}{c|}{$1.6^{+0.3}_{-0.2}\%$} \\
\multicolumn{1}{|l|}{Composite - Portsmouth} & \multicolumn{1}{c|}{3691} & \multicolumn{1}{c|}{1916} & \multicolumn{1}{c|}{23} & \multicolumn{1}{c|}{$0.6^{+0.1}_{-0.1}\%$} & \multicolumn{1}{c|}{$1.2^{+0.3}_{-0.2}\%$} \\
\multicolumn{1}{|l|}{Composite - NSA} & \multicolumn{1}{c|}{3691} & \multicolumn{1}{c|}{1598} & \multicolumn{1}{c|}{13} & \multicolumn{1}{c|}{$0.4^{+0.1}_{-0.1}\%$} & \multicolumn{1}{c|}{$0.8^{+0.3}_{-0.2}\%$} \\
\hline
\multicolumn{1}{|l|}{Conventional AGN - MPA-JHU} & \multicolumn{1}{c|}{3691} & \multicolumn{1}{c|}{2494} & \multicolumn{1}{c|}{16} & \multicolumn{1}{c|}{$0.4^{+0.1}_{-0.1}\%$} & \multicolumn{1}{c|}{$0.6^{+0.2}_{-0.1}\%$} \\
\multicolumn{1}{|l|}{Conventional AGN - Portsmouth} & \multicolumn{1}{c|}{3691} & \multicolumn{1}{c|}{1916} & \multicolumn{1}{c|}{8} & \multicolumn{1}{c|}{$0.2^{+0.1}_{-0.1}\%$} & \multicolumn{1}{c|}{$0.4^{+0.2}_{-0.1}\%$} \\
\multicolumn{1}{|l|}{Conventional AGN - NSA} & \multicolumn{1}{c|}{3691} & \multicolumn{1}{c|}{1598} & \multicolumn{1}{c|}{8} & \multicolumn{1}{c|}{$0.2^{+0.1}_{-0.1}\%$} & \multicolumn{1}{c|}{$0.5^{+0.2}_{-0.1}\%$} \\
\hline
\multicolumn{1}{|l|}{S06 Bonus AGN - MPA-JHU} & \multicolumn{1}{c|}{3691} & \multicolumn{1}{c|}{2494} & \multicolumn{1}{c|}{44} & \multicolumn{1}{c|}{$1.2^{+0.2}_{-0.2}\%$} & \multicolumn{1}{c|}{$1.8^{+0.3}_{-0.2}\%$} \\
\multicolumn{1}{|l|}{S06 Bonus AGN - Portsmouth} & \multicolumn{1}{c|}{3691} & \multicolumn{1}{c|}{1916} & \multicolumn{1}{c|}{28} & \multicolumn{1}{c|}{$0.8^{+0.2}_{-0.1}\%$} & \multicolumn{1}{c|}{$1.5^{+0.3}_{-0.2}\%$} \\
\multicolumn{1}{|l|}{S06 Bonus AGN - NSA} & \multicolumn{1}{c|}{3691} & \multicolumn{1}{c|}{1598} & \multicolumn{1}{c|}{57} & \multicolumn{1}{c|}{$1.5^{+0.2}_{-0.2}\%$} & \multicolumn{1}{c|}{$3.6^{+0.5}_{-0.4}\%$} \\
\hline
\multicolumn{1}{|l|}{SF-AGN - MPA-JHU} & \multicolumn{1}{c|}{3691} & \multicolumn{1}{c|}{1223} & \multicolumn{1}{c|}{79} & \multicolumn{1}{c|}{$2.1^{+0.3}_{-0.2}\%$} & \multicolumn{1}{c|}{$6.5^{+0.7}_{-0.7}\%$} \\
\multicolumn{1}{|l|}{SF-AGN - Portsmouth} & \multicolumn{1}{c|}{3691} & \multicolumn{1}{c|}{583} & \multicolumn{1}{c|}{22} & \multicolumn{1}{c|}{$0.6^{+0.1}_{-0.1}\%$} & \multicolumn{1}{c|}{$3.8^{+0.9}_{-0.7}\%$} \\
\multicolumn{1}{|l|}{SF-AGN - NSA} & \multicolumn{1}{c|}{3691} & \multicolumn{1}{c|}{597} & \multicolumn{1}{c|}{89} & \multicolumn{1}{c|}{$2.4^{+0.3}_{-0.2}\%$} & \multicolumn{1}{c|}{$14.9^{+1.5}_{-1.4}\%$} \\
\hline
\multicolumn{1}{|l|}{Total Unique AGN (optical using MPA-JHU)} & \multicolumn{1}{c|}{3691} & \multicolumn{1}{c|}{n/a} & \multicolumn{1}{c|}{186} & \multicolumn{1}{c|}{$5.0^{+0.4}_{-0.3}\%$} & \multicolumn{1}{c|}{n/a} \\
\multicolumn{1}{|l|}{Total Unique AGN (optical using Portsmouth)} & \multicolumn{1}{c|}{3691} & \multicolumn{1}{c|}{n/a} & \multicolumn{1}{c|}{91} & \multicolumn{1}{c|}{$2.5^{+0.3}_{-0.2}\%$} & \multicolumn{1}{c|}{n/a} \\
\multicolumn{1}{|l|}{Total Unique AGN (optical using NSA)} & \multicolumn{1}{c|}{3691} & \multicolumn{1}{c|}{n/a} & \multicolumn{1}{c|}{172} & \multicolumn{1}{c|}{$4.7^{+0.4}_{-0.3}\%$} & \multicolumn{1}{c|}{n/a} \\
\hline
\multicolumn{1}{|l|}{\ha Broadline AGN (archival)} & \multicolumn{1}{c|}{3691} & \multicolumn{1}{c|}{n/a} & \multicolumn{1}{c|}{1} & \multicolumn{1}{c|}{$0.03^{+0.04}_{-0.02}\%$} & \multicolumn{1}{c|}{n/a} \\
\multicolumn{1}{|l|}{Optical AGN Catalogs (archival)} & \multicolumn{1}{c|}{3691} & \multicolumn{1}{c|}{n/a} & \multicolumn{1}{c|}{19} & \multicolumn{1}{c|}{$0.5^{+0.1}_{-0.1}\%$} & \multicolumn{1}{c|}{n/a} \\
\multicolumn{1}{|l|}{X-ray AGN (Birchall et al. 2020 method, used for total unique AGN stats)} & \multicolumn{1}{c|}{3691} & \multicolumn{1}{c|}{n/a} & \multicolumn{1}{c|}{5} & \multicolumn{1}{c|}{$0.14^{+0.08}_{-0.05}\%$} & \multicolumn{1}{c|}{n/a} \\
\multicolumn{1}{|l|}{X-ray AGN (Ranalli et al. 2003 method)} & \multicolumn{1}{c|}{3691} & \multicolumn{1}{c|}{n/a} & \multicolumn{1}{c|}{4} & \multicolumn{1}{c|}{$0.11^{+0.07}_{-0.04}\%$} & \multicolumn{1}{c|}{n/a} \\
\hline
\multicolumn{1}{|l|}{Total Unique AGN including Archival AGN (optical using MPA-JHU)} & \multicolumn{1}{c|}{3691} & \multicolumn{1}{c|}{n/a} & \multicolumn{1}{c|}{199} & \multicolumn{1}{c|}{$5.4^{+0.4}_{-0.4}\%$} & \multicolumn{1}{c|}{n/a} \\
\multicolumn{1}{|l|}{Total Unique AGN including Archival AGN (optical using Portsmouth)} & \multicolumn{1}{c|}{3691} & \multicolumn{1}{c|}{n/a} & \multicolumn{1}{c|}{107} & \multicolumn{1}{c|}{$2.9^{+0.3}_{-0.3}\%$} & \multicolumn{1}{c|}{n/a} \\
\multicolumn{1}{|l|}{Total Unique AGN including Archival AGN (optical using NSA)} & \multicolumn{1}{c|}{3691} & \multicolumn{1}{c|}{n/a} & \multicolumn{1}{c|}{189} & \multicolumn{1}{c|}{$5.1^{+0.4}_{-0.4}\%$} & \multicolumn{1}{c|}{n/a} \\
\hline
\hline
\multicolumn{6}{|c|}{Transitional galaxies ($10^{9.7} < $\mbary $< 10^{10.6}$\msun)} \\ \hline
\multicolumn{1}{|l|}{WISE AGN - St12} & \multicolumn{1}{c|}{3863} & \multicolumn{1}{c|}{2868} & \multicolumn{1}{c|}{9} & \multicolumn{1}{c|}{$0.2^{+0.1}_{-0.1}\%$} & \multicolumn{1}{c|}{$0.3^{+0.1}_{-0.1}\%$} \\
\multicolumn{1}{|l|}{WISE AGN - modified Sa18} & \multicolumn{1}{c|}{3863} & \multicolumn{1}{c|}{2868} & \multicolumn{1}{c|}{17} & \multicolumn{1}{c|}{$0.4^{+0.1}_{-0.1}\%$} & \multicolumn{1}{c|}{$0.6^{+0.2}_{-0.1}\%$} \\
\hline
\multicolumn{1}{|l|}{Composite - MPA-JHU} & \multicolumn{1}{c|}{3863} & \multicolumn{1}{c|}{2436} & \multicolumn{1}{c|}{367} & \multicolumn{1}{c|}{$9.5^{+0.5}_{-0.5}\%$} & \multicolumn{1}{c|}{$15.1^{+0.7}_{-0.7}\%$} \\
\multicolumn{1}{|l|}{Composite - Portsmouth} & \multicolumn{1}{c|}{3863} & \multicolumn{1}{c|}{1760} & \multicolumn{1}{c|}{263} & \multicolumn{1}{c|}{$6.8^{+0.4}_{-0.4}\%$} & \multicolumn{1}{c|}{$14.9^{+0.9}_{-0.8}\%$} \\
\multicolumn{1}{|l|}{Composite - NSA} & \multicolumn{1}{c|}{3863} & \multicolumn{1}{c|}{1607} & \multicolumn{1}{c|}{280} & \multicolumn{1}{c|}{$7.2^{+0.4}_{-0.4}\%$} & \multicolumn{1}{c|}{$17.4^{+1.0}_{-0.9}\%$} \\
\hline
\multicolumn{1}{|l|}{Conventional AGN - MPA-JHU} & \multicolumn{1}{c|}{3863} & \multicolumn{1}{c|}{2436} & \multicolumn{1}{c|}{290} & \multicolumn{1}{c|}{$7.5^{+0.4}_{-0.4}\%$} & \multicolumn{1}{c|}{$11.9^{+0.7}_{-0.6}\%$} \\
\multicolumn{1}{|l|}{Conventional AGN - Portsmouth} & \multicolumn{1}{c|}{3863} & \multicolumn{1}{c|}{1760} & \multicolumn{1}{c|}{119} & \multicolumn{1}{c|}{$3.1^{+0.3}_{-0.3}\%$} & \multicolumn{1}{c|}{$6.8^{+0.6}_{-0.6}\%$} \\
\multicolumn{1}{|l|}{Conventional AGN - NSA} & \multicolumn{1}{c|}{3863} & \multicolumn{1}{c|}{1607} & \multicolumn{1}{c|}{120} & \multicolumn{1}{c|}{$3.1^{+0.3}_{-0.3}\%$} & \multicolumn{1}{c|}{$7.5^{+0.7}_{-0.6}\%$} \\
\hline
\multicolumn{1}{|l|}{S06 Bonus AGN - MPA-JHU} & \multicolumn{1}{c|}{3863} & \multicolumn{1}{c|}{2436} & \multicolumn{1}{c|}{308} & \multicolumn{1}{c|}{$8.0^{+0.4}_{-0.4}\%$} & \multicolumn{1}{c|}{$12.6^{+0.7}_{-0.7}\%$} \\
\multicolumn{1}{|l|}{S06 Bonus AGN - Portsmouth} & \multicolumn{1}{c|}{3863} & \multicolumn{1}{c|}{1760} & \multicolumn{1}{c|}{222} & \multicolumn{1}{c|}{$5.7^{+0.4}_{-0.4}\%$} & \multicolumn{1}{c|}{$12.6^{+0.8}_{-0.8}\%$} \\
\multicolumn{1}{|l|}{S06 Bonus AGN - NSA} & \multicolumn{1}{c|}{3863} & \multicolumn{1}{c|}{1607} & \multicolumn{1}{c|}{284} & \multicolumn{1}{c|}{$7.4^{+0.4}_{-0.4}\%$} & \multicolumn{1}{c|}{$17.7^{+1.0}_{-0.9}\%$} \\
\hline
\multicolumn{1}{|l|}{SF-AGN - MPA-JHU} & \multicolumn{1}{c|}{3863} & \multicolumn{1}{c|}{1207} & \multicolumn{1}{c|}{30} & \multicolumn{1}{c|}{$0.8^{+0.2}_{-0.1}\%$} & \multicolumn{1}{c|}{$2.5^{+0.5}_{-0.4}\%$} \\
\multicolumn{1}{|l|}{SF-AGN - Portsmouth} & \multicolumn{1}{c|}{3863} & \multicolumn{1}{c|}{514} & \multicolumn{1}{c|}{13} & \multicolumn{1}{c|}{$0.3^{+0.1}_{-0.1}\%$} & \multicolumn{1}{c|}{$2.5^{+0.8}_{-0.6}\%$} \\
\multicolumn{1}{|l|}{SF-AGN - NSA} & \multicolumn{1}{c|}{3863} & \multicolumn{1}{c|}{703} & \multicolumn{1}{c|}{31} & \multicolumn{1}{c|}{$0.8^{+0.2}_{-0.1}\%$} & \multicolumn{1}{c|}{$4.4^{+0.8}_{-0.7}\%$} \\
\hline
\multicolumn{1}{|l|}{Total Unique AGN (optical using MPA-JHU)} & \multicolumn{1}{c|}{3863} & \multicolumn{1}{c|}{n/a} & \multicolumn{1}{c|}{995} & \multicolumn{1}{c|}{$25.8^{+0.7}_{-0.7}\%$} & \multicolumn{1}{c|}{n/a} \\
\multicolumn{1}{|l|}{Total Unique AGN (optical using Portsmouth)} & \multicolumn{1}{c|}{3863} & \multicolumn{1}{c|}{n/a} & \multicolumn{1}{c|}{627} & \multicolumn{1}{c|}{$16.2^{+0.6}_{-0.6}\%$} & \multicolumn{1}{c|}{n/a} \\
\multicolumn{1}{|l|}{Total Unique AGN (optical using NSA)} & \multicolumn{1}{c|}{3863} & \multicolumn{1}{c|}{n/a} & \multicolumn{1}{c|}{713} & \multicolumn{1}{c|}{$18.5^{+0.6}_{-0.6}\%$} & \multicolumn{1}{c|}{n/a} \\
\hline
\multicolumn{1}{|l|}{\ha Broadline AGN (archival)} & \multicolumn{1}{c|}{3863} & \multicolumn{1}{c|}{n/a} & \multicolumn{1}{c|}{17} & \multicolumn{1}{c|}{$0.4^{+0.1}_{-0.1}\%$} & \multicolumn{1}{c|}{n/a} \\
\multicolumn{1}{|l|}{Optical AGN Catalogs (archival)} & \multicolumn{1}{c|}{3863} & \multicolumn{1}{c|}{n/a} & \multicolumn{1}{c|}{173} & \multicolumn{1}{c|}{$4.5^{+0.3}_{-0.3}\%$} & \multicolumn{1}{c|}{n/a} \\
\multicolumn{1}{|l|}{X-ray AGN (Birchall et al. 2020 method, used for total unique AGN stats)} & \multicolumn{1}{c|}{3863} & \multicolumn{1}{c|}{n/a} & \multicolumn{1}{c|}{22} & \multicolumn{1}{c|}{$0.6^{+0.1}_{-0.1}\%$} & \multicolumn{1}{c|}{n/a} \\
\multicolumn{1}{|l|}{X-ray AGN (Ranalli et al. 2003 method)} & \multicolumn{1}{c|}{3863} & \multicolumn{1}{c|}{n/a} & \multicolumn{1}{c|}{40} & \multicolumn{1}{c|}{$1.0^{+0.2}_{-0.2}\%$} & \multicolumn{1}{c|}{n/a} \\
\hline
\multicolumn{1}{|l|}{Total Unique AGN including Archival AGN (optical using MPA-JHU)} & \multicolumn{1}{c|}{3863} & \multicolumn{1}{c|}{n/a} & \multicolumn{1}{c|}{1018} & \multicolumn{1}{c|}{$26.4^{+0.7}_{-0.7}\%$} & \multicolumn{1}{c|}{n/a} \\
\multicolumn{1}{|l|}{Total Unique AGN including Archival AGN (optical using Portsmouth)} & \multicolumn{1}{c|}{3863} & \multicolumn{1}{c|}{n/a} & \multicolumn{1}{c|}{674} & \multicolumn{1}{c|}{$17.4^{+0.6}_{-0.6}\%$} & \multicolumn{1}{c|}{n/a} \\
\multicolumn{1}{|l|}{Total Unique AGN including Archival AGN (optical using NSA)} & \multicolumn{1}{c|}{3863} & \multicolumn{1}{c|}{n/a} & \multicolumn{1}{c|}{767} & \multicolumn{1}{c|}{$19.9^{+0.6}_{-0.6}\%$} & \multicolumn{1}{c|}{n/a} \\
\hline
\hline
\multicolumn{6}{|c|}{Giants (\mbary $\geq 10^{10.6}$\msun)} \\ \hline
\multicolumn{1}{|l|}{WISE AGN - St12 cut} & \multicolumn{1}{c|}{798} & \multicolumn{1}{c|}{674} & \multicolumn{1}{c|}{0} & \multicolumn{1}{c|}{$0.00^{+0.13}_{-0.00}\%$} & \multicolumn{1}{c|}{$0.00^{+0.15}_{-0.00}\%$} \\
\multicolumn{1}{|l|}{WISE AGN - modified Sa18 cut} & \multicolumn{1}{c|}{798} & \multicolumn{1}{c|}{674} & \multicolumn{1}{c|}{6} & \multicolumn{1}{c|}{$0.8^{+0.4}_{-0.3}\%$} & \multicolumn{1}{c|}{$0.9^{+0.4}_{-0.3}\%$} \\
\hline
\multicolumn{1}{|l|}{Composite - MPA-JHU} & \multicolumn{1}{c|}{798} & \multicolumn{1}{c|}{395} & \multicolumn{1}{c|}{106} & \multicolumn{1}{c|}{$13.3^{+1.2}_{-1.2}\%$} & \multicolumn{1}{c|}{$26.8^{+2.3}_{-2.2}\%$} \\
\multicolumn{1}{|l|}{Composite - Portsmouth} & \multicolumn{1}{c|}{798} & \multicolumn{1}{c|}{181} & \multicolumn{1}{c|}{56} & \multicolumn{1}{c|}{$7.0^{+1.0}_{-0.9}\%$} & \multicolumn{1}{c|}{$30.9^{+3.5}_{-3.3}\%$} \\
\multicolumn{1}{|l|}{Composite - NSA} & \multicolumn{1}{c|}{798} & \multicolumn{1}{c|}{183} & \multicolumn{1}{c|}{60} & \multicolumn{1}{c|}{$7.5^{+1.0}_{-0.9}\%$} & \multicolumn{1}{c|}{$32.8^{+3.6}_{-3.4}\%$} \\
\hline
\multicolumn{1}{|l|}{Conventional AGN - MPA-JHU} & \multicolumn{1}{c|}{798} & \multicolumn{1}{c|}{395} & \multicolumn{1}{c|}{233} & \multicolumn{1}{c|}{$29.2^{+1.6}_{-1.6}\%$} & \multicolumn{1}{c|}{$59.0^{+2.4}_{-2.5}\%$} \\
\multicolumn{1}{|l|}{Conventional AGN - Portsmouth} & \multicolumn{1}{c|}{798} & \multicolumn{1}{c|}{181} & \multicolumn{1}{c|}{91} & \multicolumn{1}{c|}{$11.4^{+1.2}_{-1.1}\%$} & \multicolumn{1}{c|}{$50.3^{+3.7}_{-3.7}\%$} \\
\multicolumn{1}{|l|}{Conventional AGN - NSA} & \multicolumn{1}{c|}{798} & \multicolumn{1}{c|}{183} & \multicolumn{1}{c|}{95} & \multicolumn{1}{c|}{$11.9^{+1.2}_{-1.1}\%$} & \multicolumn{1}{c|}{$51.9^{+3.7}_{-3.7}\%$} \\
\hline
\multicolumn{1}{|l|}{S06 Bonus AGN - MPA-JHU} & \multicolumn{1}{c|}{798} & \multicolumn{1}{c|}{395} & \multicolumn{1}{c|}{31} & \multicolumn{1}{c|}{$3.9^{+0.7}_{-0.6}\%$} & \multicolumn{1}{c|}{$7.8^{+1.5}_{-1.2}\%$} \\
\multicolumn{1}{|l|}{S06 Bonus AGN - Portsmouth} & \multicolumn{1}{c|}{798} & \multicolumn{1}{c|}{181} & \multicolumn{1}{c|}{20} & \multicolumn{1}{c|}{$2.5^{+0.6}_{-0.5}\%$} & \multicolumn{1}{c|}{$11.0^{+2.5}_{-2.1}\%$} \\
\multicolumn{1}{|l|}{S06 Bonus AGN - NSA} & \multicolumn{1}{c|}{798} & \multicolumn{1}{c|}{183} & \multicolumn{1}{c|}{19} & \multicolumn{1}{c|}{$2.4^{+0.6}_{-0.5}\%$} & \multicolumn{1}{c|}{$10.4^{+2.5}_{-2.0}\%$} \\
\hline
\multicolumn{1}{|l|}{SF-AGN - MPA-JHU} & \multicolumn{1}{c|}{798} & \multicolumn{1}{c|}{136} & \multicolumn{1}{c|}{0} & \multicolumn{1}{c|}{$0.00^{+0.13}_{-0.00}\%$} & \multicolumn{1}{c|}{$0.00^{+0.73}_{-0.00}\%$} \\
\multicolumn{1}{|l|}{SF-AGN - Portsmouth} & \multicolumn{1}{c|}{798} & \multicolumn{1}{c|}{52} & \multicolumn{1}{c|}{0} & \multicolumn{1}{c|}{$0.00^{+0.13}_{-0.00}\%$} & \multicolumn{1}{c|}{$0.00^{+1.89}_{-0.00}\%$} \\
\multicolumn{1}{|l|}{SF-AGN - NSA} & \multicolumn{1}{c|}{798} & \multicolumn{1}{c|}{96} & \multicolumn{1}{c|}{0} & \multicolumn{1}{c|}{$0.00^{+0.13}_{-0.00}\%$} & \multicolumn{1}{c|}{$0.00^{+1.03}_{-0.00}\%$} \\
\hline
\multicolumn{1}{|l|}{Total Unique AGN (optical using MPA-JHU)} & \multicolumn{1}{c|}{798} & \multicolumn{1}{c|}{n/a} & \multicolumn{1}{c|}{372} & \multicolumn{1}{c|}{$46.6^{+1.8}_{-1.8}\%$} & \multicolumn{1}{c|}{n/a} \\
\multicolumn{1}{|l|}{Total Unique AGN (optical using Portsmouth)} & \multicolumn{1}{c|}{798} & \multicolumn{1}{c|}{n/a} & \multicolumn{1}{c|}{172} & \multicolumn{1}{c|}{$21.6^{+1.5}_{-1.4}\%$} & \multicolumn{1}{c|}{n/a} \\
\multicolumn{1}{|l|}{Total Unique AGN (optical using NSA)} & \multicolumn{1}{c|}{798} & \multicolumn{1}{c|}{n/a} & \multicolumn{1}{c|}{176} & \multicolumn{1}{c|}{$22.1^{+1.5}_{-1.4}\%$} & \multicolumn{1}{c|}{n/a} \\
\hline
\multicolumn{1}{|l|}{\ha Broadline AGN (archival)} & \multicolumn{1}{c|}{798} & \multicolumn{1}{c|}{n/a} & \multicolumn{1}{c|}{7} & \multicolumn{1}{c|}{$0.9^{+0.4}_{-0.3}\%$} & \multicolumn{1}{c|}{n/a} \\
\multicolumn{1}{|l|}{Optical AGN Catalogs (archival)} & \multicolumn{1}{c|}{798} & \multicolumn{1}{c|}{n/a} & \multicolumn{1}{c|}{92} & \multicolumn{1}{c|}{$11.5^{+1.2}_{-1.1}\%$} & \multicolumn{1}{c|}{n/a} \\
\multicolumn{1}{|l|}{X-ray AGN (Birchall et al. 2020 method, used for total unique AGN stats)} & \multicolumn{1}{c|}{798} & \multicolumn{1}{c|}{n/a} & \multicolumn{1}{c|}{25} & \multicolumn{1}{c|}{$3.1^{+0.7}_{-0.6}\%$} & \multicolumn{1}{c|}{n/a} \\
\multicolumn{1}{|l|}{X-ray AGN (Ranalli et al. 2003 method)} & \multicolumn{1}{c|}{798} & \multicolumn{1}{c|}{n/a} & \multicolumn{1}{c|}{40} & \multicolumn{1}{c|}{$5.0^{+0.8}_{-0.7}\%$} & \multicolumn{1}{c|}{n/a} \\
\hline
\multicolumn{1}{|l|}{Total Unique AGN including Archival AGN (optical using MPA-JHU)} & \multicolumn{1}{c|}{798} & \multicolumn{1}{c|}{n/a} & \multicolumn{1}{c|}{385} & \multicolumn{1}{c|}{$48.2^{+1.8}_{-1.8}\%$} & \multicolumn{1}{c|}{n/a} \\
\multicolumn{1}{|l|}{Total Unique AGN including Archival AGN (optical using Portsmouth)} & \multicolumn{1}{c|}{798} & \multicolumn{1}{c|}{n/a} & \multicolumn{1}{c|}{205} & \multicolumn{1}{c|}{$25.7^{+1.6}_{-1.5}\%$} & \multicolumn{1}{c|}{n/a} \\
\multicolumn{1}{|l|}{Total Unique AGN including Archival AGN (optical using NSA)} & \multicolumn{1}{c|}{798} & \multicolumn{1}{c|}{n/a} & \multicolumn{1}{c|}{206} & \multicolumn{1}{c|}{$25.8^{+1.6}_{-1.5}\%$} & \multicolumn{1}{c|}{n/a} \\
\hline
\enddata
\tablecomments{Note: All error bars are computed using binomial
confidence intervals. Total unique AGN numbers count duplicate AGN
detections only once -- overlapping detections are possible between our S06
Bonus AGN and SF-AGN categories within BPT/VO AGN, between any BPT/VO AGN
category and WISE AGN, between any of our AGN categories and archival AGN
detections, or between different archival AGN detections. {Total Unique AGN
using only the \cite{Birchall20} X-ray AGN classifications (see Section
\ref{sec:archival}).} See Section~\ref{sec:ch3agninventory} for information
on our BPT/VO and WISE AGN categories and Section~\ref{sec:ch3discussion}
for information on archival AGN.}
\end{deluxetable*}

\subsubsection{AGN Statistics across the Galaxy Population}\label{sec:optwisestats1}

\begin{figure*}[]
    \centering
    \includegraphics[width=\linewidth]{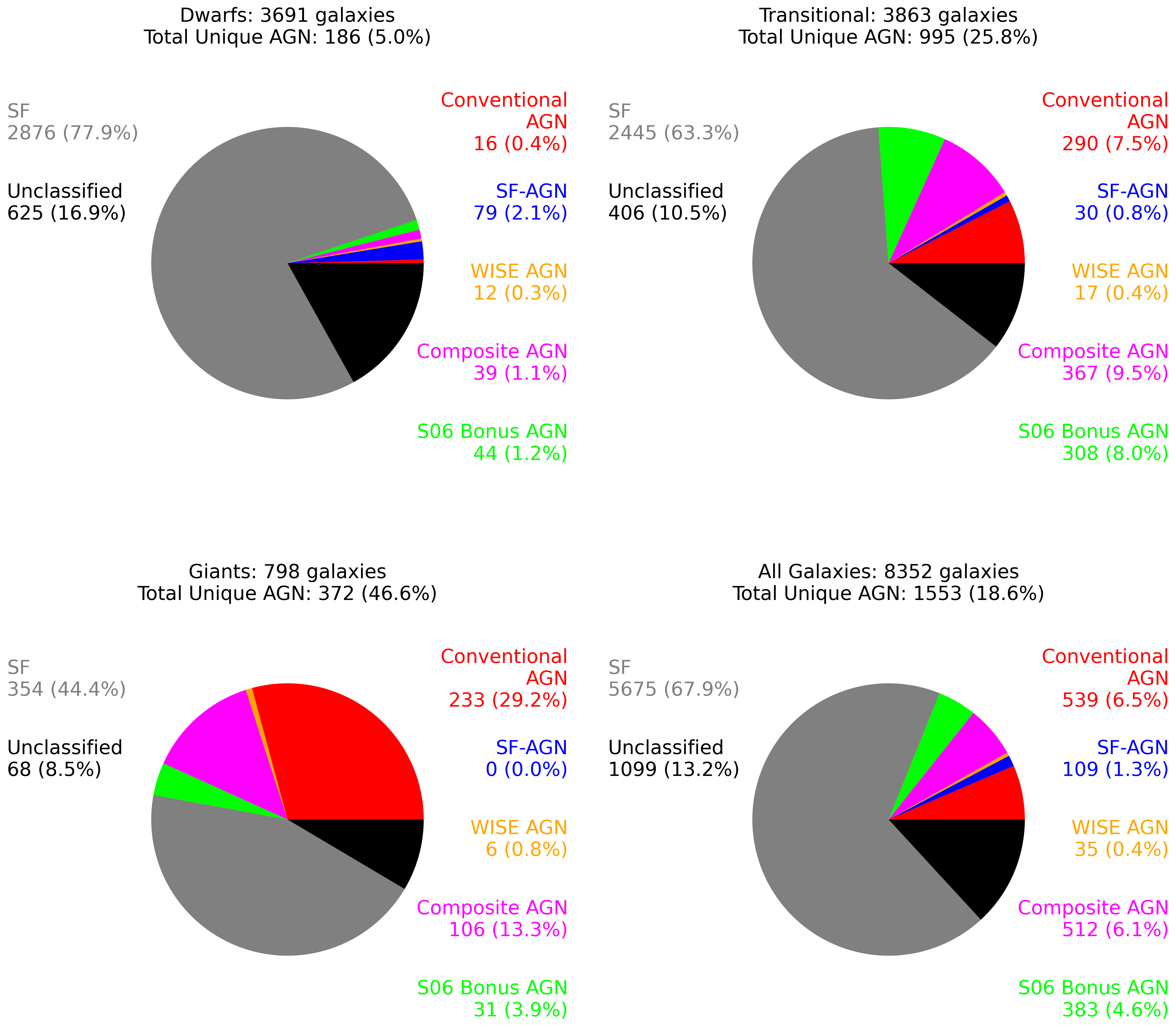}
    \caption{Breakdown of optical and mid-IR WISE AGN categories in all dwarf (top left), transitional (top right), and giant (bottom left) galaxies and in the full survey (bottom right) as percentages of the full RESOLVE and ECO surveys (including galaxies lacking adequate emission line and/or WISE data). Optical AGN statistics shown are based on MPA-JHU catalog emission-line fluxes crossmatched to the combined RESOLVE and ECO surveys; results based on crossmatches to the Portmouth and NSA catalogs are also given in Table~\ref{tab:agnstats}. No archival AGN are included in this figure. {The ``SF'' category refers to the S06 SF classification (below the \citetalias{Stasinska06} line in the BPT diagram; see Section~\ref{subsec:ch3s06bonus}).} \textit{\textbf{Note:}} Among all dwarfs, 0.1\% are double-counted in the pie chart (four galaxies are both S06 Bonus AGN and SF-AGN). Among all transitional galaxies, 0.4\% are double-counted in the pie chart (seven galaxies are both Conventional 
AGN and WISE AGN, four are both Composites and WISE AGN, one is both an S06 Bonus AGN and WISE AGN, and five are both S06 Bonus AGN and SF-AGN). Among all giants, 0.5\% are double-counted in the pie chart (four galaxies are both Conventional AGN and WISE AGN). The total AGN counts and percentages report only unique AGN. }
    \label{fig:agnpiechart}
\end{figure*}

For the full, baryonic-mass selected RESOLVE and ECO parent sample of 8352 galaxies, we measure an overall BPT/VO + WISE AGN frequency of $\sim$10.3--19.0\%, where the range reflects variation based on which optical emission line catalog was crossmatched and also includes the error bars, as explained above (see preamble to Section~\ref{sec:optwisestats}). This frequency may seem low partly because it is normalized relative to the full population rather than to the separate SDSS/WISE data search samples (see Table~\ref{tab:agnstats} to compare statistics normalized both ways side by side), but an even larger effect is the strong falloff in AGN detection with decreasing mass.\\

Figure~\ref{fig:agnpiechart} provides pie charts illustrating the complete RESOLVE and ECO BPT/VO + WISE AGN inventory using MPA-JHU crossmatched data, showing full-population AGN frequencies for the dwarf, transitional, and giant mass regimes as well as for galaxies overall. The overall AGN frequency of 18.6\% seen in this figure is substantially lower than the giant-only AGN frequency of 46.6\%, which is expected given that $<$10\% of the galaxies in our complete, volume-limited surveys are giants. As we will detail in the subsections below, dwarf and transitional galaxies differ from giants not only in having lower AGN frequencies, but also in having different typical AGN flavours. Based on the MPA-JHU crossmatched data shown: (1) giants most often host Conventional AGN, with Composites being about half as common; (2) transitional galaxies host Composites, S06 Bonus AGN, and Conventional AGN in roughly equal numbers; and (3) dwarfs host most often host SF-AGN, plus a signi
ficant number of Composites and S06 Bonus AGN in roughly 2:1:1 proportions. In the next three subsections, we consider each mass regime in more detail as well as results based on crossmatching to the Portsmouth and NSA catalogs. \\

For representation in the pie chart, we do not single-count duplicate AGN detections as in the Total Unique AGN entries in Table~\ref{tab:agnstats} but instead repeatedly count AGN that are both WISE AGN and BPT/VO AGN, or both S06 Bonus AGN and SF-AGN, in all relevant pie segments. Such duplicates represent $\sim$8\% of SF-AGN (which duplicate $\sim$2\% of S06 Bonus AGN) and $\sim$46\% of WISE AGN (which duplicate $\sim$2\% of Conventional AGN and $\sim$1\% of Composites and S06 Bonus AGN).\\

We note that while WISE AGN comprise $<$1\% of galaxies at all masses, the rate of WISE AGN overlapping BPT/VO AGN seems higher for transitional and giant galaxies ($\sim$50\% using MPA-JHU or NSA crossmatched data and $\sim$20\% using Portsmouth data) than for dwarfs (zero using MPA-JHU or Portsmouth data and just one galaxy or $\sim$8\% using NSA data). That said, 7/12 dwarf WISE AGN were allowed into the sample by removing W3 S/N requirements (see Section~\ref{sec:ch3midiragn}) implying less reliable detection, so it seems premature to conclude that the non-overlap of dwarf WISE AGN with BPT/VO AGN implies anything physical, especially given the small number statistics and discrepant overlap rates depending on emission-line data source. 

\subsubsection{Dwarfs}\label{sec:optwisestats2}

Among dwarfs ($10^{9.2} \leq$ \mbary $\leq 10^{9.7}$\msun), SF-AGN have the highest search-sample detection frequency, comprising 3.1-16.4\% of dwarfs in the SF-AGN (i.e., BPT+VO) search sample. In the expanded BPT search sample (i.e., the BPT-only and BPT+VO samples combined), S06 Bonus AGN comprise 1.2--4.1\%, Composites comprise 0.6--1.8\%, and Conventional AGN comprise 0.3--0.8\% of search sample dwarfs. WISE AGN are detected in $\sim$0.6\% of the WISE search sample {(12 galaxies, none of which overlap our BPT/VO AGN using MPA-JHU data)}. \\

If we take advantage of the census-style design of RESOLVE and ECO to instead compute statistics with respect to the full dwarf galaxy population, including non-search-sample dwarfs, we necessarily obtain smaller AGN frequencies but the various AGN types can be compared on a common basis. For the full dwarf population in the baryonic-mass-selected sample (3691 galaxies), Table~\ref{tab:agnstats} shows that 0.5--2.7\% of dwarfs host SF-AGN, 0.6--1.8\% host S06 Bonus AGN, 0.3--1.2\% host Composites, 0.2--0.6\% host Conventional AGN, and $\sim$0.3\% host WISE AGN. The overall frequency of BPT/VO and WISE AGN is 2.2--5.4\% among all dwarfs in RESOLVE and ECO.\\

The top left panel of Figure \ref{fig:agnpiechart} shows that when normalizing to the full galaxy population, \textit{SF-AGN and to a lesser extent Composites and S06 Bonus AGN dominate the AGN inventory in the dwarf galaxy population.} For MPA-JHU crossmatched data, including SF-AGN nearly doubles the number of dwarf AGN found using the BPT plot alone (Composite, Conventional, and S06 Bonus AGN). A similar result holds for NSA crossmatched data, whereas for Portmouth crossmatched data, SF-AGN are proportionally less common, with S06 Bonus AGN being the most common type.\\

These results are {roughly} consistent with those of \citetalias{Polimera22} taking into account the change in sample definition {(baryonic-mass-selection rather than the stellar-mass-selection used in \citetalias{Polimera22}) and the adoption of the \citetalias{Stasinska06} line as the SF/AGN boundary. Across all masses, \citetalias{Polimera22} reported that $\sim$82\% of ``strong emission line" galaxies -- equivalent to the BPT+VO sample in this work -- are Definite SF using the \citetalias{Kauffmann03} boundary (Table~2 of \citetalias{Polimera22}). Applying the same \citetalias{Kauffmann03} definition to this paper's BPT+VO sample yields $\sim$80\%, confirming that the change in sample definition has minimal effect. Under the \citetalias{Stasinska06} boundary adopted here, the overall Definite SF fraction drops to $\sim$73\%, with the difference reflecting the reclassification of galaxies between the \citetalias{Kauffmann03} and \citetalias{Stasinska06} lines as S06 Bonus 
AGN. Within this paper's baryonic mass framework, the \citetalias{Stasinska06} reclassification effect grows with mass -- the \citetalias{Kauffmann03}-based SF fraction of $\sim$92\% drops only to $\sim$91\% using the \citetalias{Stasinska06} line for dwarfs, but from $\sim$76\% to $\sim$63\% for transitional galaxies and from $\sim$18\% to $\sim$7\% for giants (using BPT+VO sample from MPA-JHU catalog).}.

\subsubsection{Transitional Galaxies}\label{sec:optwisestats3}

Transitional galaxies ($10^{9.7}< $ \mbary $< 10^{10.6}$\msun) display a shift in AGN types and abundances. Within the transitional galaxy BPT+VO sample (SF-AGN search sample), SF-AGN are $\sim$2$\times$ less common than in the analogous dwarf sample, comprising 1.9--5.3\% of search sample galaxies. In contrast, within the expanded BPT sample for transitional galaxies, S06 Bonus AGN, Composites, and Conventional AGN are collectively $\sim$10$\times$ \textit{more} common than in the analogous dwarf sample, comprising 11.8--18.6\%, 14.1--18.4\%, and 6.8--12.6\%, respectively, of search sample galaxies. WISE AGN are detected in only $\sim$0.6\% of the transitional-galaxy WISE search sample {(17 galaxies, of which 12 are also BPT/VO AGN using MPA-JHU data)}.\\

Recomputing these frequencies relative to the full transitional galaxy population (3863 galaxies including non-search-sample galaxies), we find 0.3--1.0\% of galaxies host SF-AGN, 5.4--8.4\% host S06 Bonus AGN, 6.4--10.0\% host Composites, 2.8--7.9\% host Conventional AGN, and $\sim$0.4\% host WISE AGN. The overall frequency of BPT/VO and WISE AGN is 15.6--26.5\% among all transitional galaxies in RESOLVE and ECO, roughly 5$\times$ the overall frequency for dwarfs.\\

The top right panel of Figure \ref{fig:agnpiechart} shows that when normalizing to the full galaxy population, \textit{Composites, S06 Bonus AGN, and to a lesser extent Conventional AGN dominate the AGN inventory in the transitional galaxy population.} When using the MPA-JHU crossmatched data as shown, these types occur in almost equal proportions, whereas Conventional AGN are up to $\sim$2$\times$ less common than Composites or S06 Bonus AGN when using NSA or Portsmouth crossmatched data, for which the overall frequency of AGN is also substantially lower.\\

Comparing dwarf and transitional galaxies, we see that in both mass regimes the two AGN types associated with moderate star formation and metallicity (Composites and S06 Bonus AGN) are typical. Where the results differ is that in dwarfs there is an even larger population of SF-AGN associated with strong star formation and low metallicity, whereas in transitional galaxies there is an increasingly important Conventional AGN population associated with weak star formation and high metallicity.

\subsubsection{Giants}\label{sec:optwisestats4}

In giant galaxies (\mbary $> 10^{10.6}$\msun) the AGN population is more traditional. We find no SF-AGN using any optical catalog. Conventional AGN and Composites dominate, comprising 46.6--61.4\% and 24.7--36.3\%, respectively, of giants in the expanded BPT search sample. Combining these two traditional categories, the AGN frequency for giants in the expanded BPT search sample reaches $>$80\%. This high percentage reflects the fact that we have defined giants based on an expectation of quenching (i.e., baryonic mass above the bimodality scale; see Section \ref{sec:ch3intro}), so star formation should be moderate at most (and distributed outside the central region probed by SDSS spectroscopy), implying that strong emission lines generally indicate AGN activity. The other AGN category associated with moderate star formation, S06 Bonus AGN, comprises another 6.6--13.6\% of giants in this expanded BPT search sample. WISE AGN are detected in $\sim$0.9\% of the giant-galaxy WISE s
earch sample {(6 galaxies, of which 4 are also BPT/VO AGN using MPA-JHU data)}. \\

Recomputing these frequencies relative to the full giant-galaxy population (798 galaxies including non-search-sample galaxies), we find 0\% host SF-AGN, 1.9--4.6\% host S06 Bonus AGN, 6.2--14.5\% host Composites, 10.3--30.8\% host Conventional AGN, and $\sim$0.8\% host WISE AGN. The overall frequency of BPT/VO and WISE AGN is 20.1--48.4\% among all giant galaxies in RESOLVE and ECO, roughly 10$\times$ and 2$\times$ the overall frequencies for dwarfs and transitional galaxies, respectively. \\

The bottom left panel of Figure \ref{fig:agnpiechart} shows that when normalizing to the full galaxy population, \textit{Conventional AGN and to a lesser extent Composites dominate the AGN inventory in the giant galaxy population.} When using the MPA-JHU crossmatched data as shown, Conventional AGN are over twice as common as Composites, whereas when using NSA or Portsmouth crossmatched data, Conventional AGN are only $\sim$50\% more common and the overall frequency of AGN is $\sim$2$\times$ lower.\\

Comparing transitional and giant galaxies, the increased proportion of Conventional AGN for giants makes sense given the association of such AGN with higher metallicity and lower star formation. It is worth noting, however, that the discrepancies between results based on crossmatched MPA-JHU, Portsmouth, and NSA measurements are most noticeable in the giant regime, especially for Conventional AGN, {which make up $\sim$29\% of the giant parent sample using MPA-JHU measurements but only $\sim$11--12\% using Portsmouth or NSA measurements (see Section~\ref{sec:optwisestats1} for discussion of the catalog systematics driving these differences).}

\subsubsection{Stellar-mass Selected Results}\label{sec:optwisestats5}

As discussed in Section~\ref{subsec:ch3surveys}, we prefer our baryonic-mass-selected sample for scientific analysis, in line with the design of the RESOLVE and ECO surveys. However, to facilitate comparison to previous work, Table~\ref{tab:agnstatsstellar} provides analogous statistics to Table~\ref{tab:agnstats} for a complete stellar-mass-limited sample (7066 galaxies with \mstar $> 10^{8.9}$\mstar; see Section \ref{subsec:ch3surveys}). In this redefined parent sample, the combined full-population BPT/VO + WISE AGN percentages are 11.9--21.6\% for all galaxies, 2.1--4.9\% for dwarfs (now defined by $10^{8.9} \leq$ \mstar $\leq 10^{9.5}$\msun), 18.1--30.2\% for transitional galaxies (now defined by $10^{9.5} < $ \mstar $< 10^{10.5}$\msun), and 20.8--47.6\% for giants (now defined by \mstar $\geq 10^{10.5}$\msun). Unsurprisingly, the AGN percentage among giants is virtually unchanged between the stellar-mass and baryonic-mass selected samples because the baryonic mass of gia
nts is heavily dominated by their stellar mass. Among transitional galaxies, the combined BPT/VO + WISE AGN percentage is higher in the stellar-mass-selected sample than in the baryonic-mass-selected sample by $\sim$2--3\%, reflecting the difficulty of defining an exact correspondence between baryonic and stellar mass ranges as gas content changes. For dwarfs, the overall AGN frequency is very similar, but with fewer SF-AGN (0.3--1.7\%) and more S06 Bonus AGN (0.9--2.8\%). These shifts reflect both the inexact correspondence at the upper mass boundary and the fact that the stellar mass floor at $10^{8.9}$\msun, being a factor of two lower than the baryonic mass floor at $10^{9.2}$\msun, excludes dwarfs with \mgas $>$ \mstar. Such gas-dominated dwarfs are abundant in volume-limited surveys \citep{Kannappan13} and are the typical hosts of most SF-AGN (see Figure \ref{fig:agnprop}).

\startlongtable
\begin{deluxetable*}{|l|c|c|c|c|c|}
\tablecaption{Statistics of AGN in RESOLVE and ECO Galaxies in the stellar-mass-limited sample with \mstar $\geq 10^{8.9}$\msun \label{tab:agnstatsstellar}}

\tablehead{ \multicolumn{1}{|l|}{{AGN Category}} & \multicolumn{1}{c|}{{$\rm N_{gal,parent}$}} & \multicolumn{1}{c|}{{$\rm N_{gal,search}$}} & \multicolumn{1}{c|}{{AGN}} & \multicolumn{1}{c|}{{AGN\% in}} & \multicolumn{1}{c|}{{AGN\% in}}\\
\multicolumn{1}{|c|}{ }&\multicolumn{1}{c|}{ } &\multicolumn{1}{|c|}{ } &\multicolumn{1}{c|}{ } & \multicolumn{1}{c|}{{$\rm N_{gal,parent}$}} & \multicolumn{1}{c|}{{$\rm N_{gal,search}$}}}
\startdata
\multicolumn{6}{|c|}{Dwarfs ($10^{8.9} \leq$ \mstar $\leq 10^{9.5}$\msun)} \\ \hline
\multicolumn{1}{|l|}{WISE AGN - St12} & \multicolumn{1}{c|}{2952} & \multicolumn{1}{c|}{1749} & \multicolumn{1}{c|}{5} & \multicolumn{1}{c|}{$0.2^{+0.1}_{-0.1}\%$} & \multicolumn{1}{c|}{$0.3^{+0.2}_{-0.1}\%$} \\
\multicolumn{1}{|l|}{WISE AGN - modified Sa18} & \multicolumn{1}{c|}{2952} & \multicolumn{1}{c|}{1749} & \multicolumn{1}{c|}{8} & \multicolumn{1}{c|}{$0.3^{+0.1}_{-0.1}\%$} & \multicolumn{1}{c|}{$0.5^{+0.2}_{-0.1}\%$} \\
\hline
\multicolumn{1}{|l|}{Composite - MPA-JHU} & \multicolumn{1}{c|}{2952} & \multicolumn{1}{c|}{1968} & \multicolumn{1}{c|}{26} & \multicolumn{1}{c|}{$0.9^{+0.2}_{-0.2}\%$} & \multicolumn{1}{c|}{$1.3^{+0.3}_{-0.2}\%$} \\
\multicolumn{1}{|l|}{Composite - Portsmouth} & \multicolumn{1}{c|}{2952} & \multicolumn{1}{c|}{1636} & \multicolumn{1}{c|}{15} & \multicolumn{1}{c|}{$0.5^{+0.1}_{-0.1}\%$} & \multicolumn{1}{c|}{$0.9^{+0.3}_{-0.2}\%$} \\
\multicolumn{1}{|l|}{Composite - NSA} & \multicolumn{1}{c|}{2952} & \multicolumn{1}{c|}{1388} & \multicolumn{1}{c|}{9} & \multicolumn{1}{c|}{$0.3^{+0.1}_{-0.1}\%$} & \multicolumn{1}{c|}{$0.6^{+0.3}_{-0.2}\%$} \\
\hline
\multicolumn{1}{|l|}{Conventional AGN - MPA-JHU} & \multicolumn{1}{c|}{2952} & \multicolumn{1}{c|}{1968} & \multicolumn{1}{c|}{9} & \multicolumn{1}{c|}{$0.3^{+0.1}_{-0.1}\%$} & \multicolumn{1}{c|}{$0.5^{+0.2}_{-0.1}\%$} \\
\multicolumn{1}{|l|}{Conventional AGN - Portsmouth} & \multicolumn{1}{c|}{2952} & \multicolumn{1}{c|}{1636} & \multicolumn{1}{c|}{5} & \multicolumn{1}{c|}{$0.2^{+0.1}_{-0.1}\%$} & \multicolumn{1}{c|}{$0.3^{+0.2}_{-0.1}\%$} \\
\multicolumn{1}{|l|}{Conventional AGN - NSA} & \multicolumn{1}{c|}{2952} & \multicolumn{1}{c|}{1388} & \multicolumn{1}{c|}{5} & \multicolumn{1}{c|}{$0.2^{+0.1}_{-0.1}\%$} & \multicolumn{1}{c|}{$0.4^{+0.2}_{-0.1}\%$} \\
\hline
\multicolumn{1}{|l|}{S06 Bonus AGN - MPA-JHU} & \multicolumn{1}{c|}{2952} & \multicolumn{1}{c|}{1968} & \multicolumn{1}{c|}{52} & \multicolumn{1}{c|}{$1.8^{+0.3}_{-0.2}\%$} & \multicolumn{1}{c|}{$2.6^{+0.4}_{-0.3}\%$} \\
\multicolumn{1}{|l|}{S06 Bonus AGN - Portsmouth} & \multicolumn{1}{c|}{2952} & \multicolumn{1}{c|}{1636} & \multicolumn{1}{c|}{33} & \multicolumn{1}{c|}{$1.1^{+0.2}_{-0.2}\%$} & \multicolumn{1}{c|}{$2.0^{+0.4}_{-0.3}\%$} \\
\multicolumn{1}{|l|}{S06 Bonus AGN - NSA} & \multicolumn{1}{c|}{2952} & \multicolumn{1}{c|}{1388} & \multicolumn{1}{c|}{74} & \multicolumn{1}{c|}{$2.5^{+0.3}_{-0.3}\%$} & \multicolumn{1}{c|}{$5.3^{+0.6}_{-0.6}\%$} \\
\hline
\multicolumn{1}{|l|}{SF-AGN - MPA-JHU} & \multicolumn{1}{c|}{2952} & \multicolumn{1}{c|}{1017} & \multicolumn{1}{c|}{41} & \multicolumn{1}{c|}{$1.4^{+0.2}_{-0.2}\%$} & \multicolumn{1}{c|}{$4.0^{+0.7}_{-0.6}\%$} \\
\multicolumn{1}{|l|}{SF-AGN - Portsmouth} & \multicolumn{1}{c|}{2952} & \multicolumn{1}{c|}{482} & \multicolumn{1}{c|}{11} & \multicolumn{1}{c|}{$0.4^{+0.1}_{-0.1}\%$} & \multicolumn{1}{c|}{$2.3^{+0.8}_{-0.6}\%$} \\
\multicolumn{1}{|l|}{SF-AGN - NSA} & \multicolumn{1}{c|}{2952} & \multicolumn{1}{c|}{507} & \multicolumn{1}{c|}{43} & \multicolumn{1}{c|}{$1.5^{+0.2}_{-0.2}\%$} & \multicolumn{1}{c|}{$8.5^{+1.3}_{-1.2}\%$} \\
\hline
\multicolumn{1}{|l|}{Total Unique AGN (optical using MPA-JHU)} & \multicolumn{1}{c|}{2952} & \multicolumn{1}{c|}{n/a} & \multicolumn{1}{c|}{132} & \multicolumn{1}{c|}{$4.5^{+0.4}_{-0.4}\%$} & \multicolumn{1}{c|}{n/a} \\
\multicolumn{1}{|l|}{Total Unique AGN (optical using Portsmouth)} & \multicolumn{1}{c|}{2952} & \multicolumn{1}{c|}{n/a} & \multicolumn{1}{c|}{70} & \multicolumn{1}{c|}{$2.4^{+0.3}_{-0.3}\%$} & \multicolumn{1}{c|}{n/a} \\
\multicolumn{1}{|l|}{Total Unique AGN (optical using NSA)} & \multicolumn{1}{c|}{2952} & \multicolumn{1}{c|}{n/a} & \multicolumn{1}{c|}{134} & \multicolumn{1}{c|}{$4.5^{+0.4}_{-0.4}\%$} & \multicolumn{1}{c|}{n/a} \\
\hline
\multicolumn{1}{|l|}{\ha Broadline AGN (archival)} & \multicolumn{1}{c|}{2952} & \multicolumn{1}{c|}{n/a} & \multicolumn{1}{c|}{1} & \multicolumn{1}{c|}{$0.03^{+0.05}_{-0.02}\%$} & \multicolumn{1}{c|}{n/a} \\
\multicolumn{1}{|l|}{Optical AGN Catalogs (archival)} & \multicolumn{1}{c|}{2952} & \multicolumn{1}{c|}{n/a} & \multicolumn{1}{c|}{11} & \multicolumn{1}{c|}{$0.4^{+0.1}_{-0.1}\%$} & \multicolumn{1}{c|}{n/a} \\
\multicolumn{1}{|l|}{X-ray AGN (Birchall et al. 2020 method, used for total unique AGN stats)} & \multicolumn{1}{c|}{2952} & \multicolumn{1}{c|}{n/a} & \multicolumn{1}{c|}{6} & \multicolumn{1}{c|}{$0.2^{+0.1}_{-0.1}\%$} & \multicolumn{1}{c|}{n/a} \\
\multicolumn{1}{|l|}{X-ray AGN (Ranalli et al. 2003 method)} & \multicolumn{1}{c|}{2952} & \multicolumn{1}{c|}{n/a} & \multicolumn{1}{c|}{5} & \multicolumn{1}{c|}{$0.2^{+0.1}_{-0.1}\%$} & \multicolumn{1}{c|}{n/a} \\
\hline
\multicolumn{1}{|l|}{Total Unique AGN including Archival AGN (optical using MPA-JHU)} & \multicolumn{1}{c|}{2952} & \multicolumn{1}{c|}{n/a} & \multicolumn{1}{c|}{141} & \multicolumn{1}{c|}{$4.8^{+0.4}_{-0.4}\%$} & \multicolumn{1}{c|}{n/a} \\
\multicolumn{1}{|l|}{Total Unique AGN including Archival AGN (optical using Portsmouth)} & \multicolumn{1}{c|}{2952} & \multicolumn{1}{c|}{n/a} & \multicolumn{1}{c|}{81} & \multicolumn{1}{c|}{$2.7^{+0.3}_{-0.3}\%$} & \multicolumn{1}{c|}{n/a} \\
\multicolumn{1}{|l|}{Total Unique AGN including Archival AGN (optical using NSA)} & \multicolumn{1}{c|}{2952} & \multicolumn{1}{c|}{n/a} & \multicolumn{1}{c|}{146} & \multicolumn{1}{c|}{$4.9^{+0.4}_{-0.4}\%$} & \multicolumn{1}{c|}{n/a} \\
\hline
\hline
\multicolumn{6}{|c|}{Transitional galaxies ($10^{9.5} < $\mstar $< 10^{10.5}$\msun)} \\ \hline
\multicolumn{1}{|l|}{WISE AGN - St12} & \multicolumn{1}{c|}{3215} & \multicolumn{1}{c|}{2452} & \multicolumn{1}{c|}{9} & \multicolumn{1}{c|}{$0.3^{+0.1}_{-0.1}\%$} & \multicolumn{1}{c|}{$0.4^{+0.1}_{-0.1}\%$} \\
\multicolumn{1}{|l|}{WISE AGN - modified Sa18} & \multicolumn{1}{c|}{3215} & \multicolumn{1}{c|}{2452} & \multicolumn{1}{c|}{15} & \multicolumn{1}{c|}{$0.5^{+0.1}_{-0.1}\%$} & \multicolumn{1}{c|}{$0.6^{+0.2}_{-0.1}\%$} \\
\hline
\multicolumn{1}{|l|}{Composite - MPA-JHU} & \multicolumn{1}{c|}{3215} & \multicolumn{1}{c|}{1802} & \multicolumn{1}{c|}{372} & \multicolumn{1}{c|}{$11.6^{+0.6}_{-0.6}\%$} & \multicolumn{1}{c|}{$20.6^{+1.0}_{-0.9}\%$} \\
\multicolumn{1}{|l|}{Composite - Portsmouth} & \multicolumn{1}{c|}{3215} & \multicolumn{1}{c|}{1244} & \multicolumn{1}{c|}{263} & \multicolumn{1}{c|}{$8.2^{+0.5}_{-0.5}\%$} & \multicolumn{1}{c|}{$21.1^{+1.2}_{-1.1}\%$} \\
\multicolumn{1}{|l|}{Composite - NSA} & \multicolumn{1}{c|}{3215} & \multicolumn{1}{c|}{1153} & \multicolumn{1}{c|}{275} & \multicolumn{1}{c|}{$8.6^{+0.5}_{-0.5}\%$} & \multicolumn{1}{c|}{$23.9^{+1.3}_{-1.2}\%$} \\
\hline
\multicolumn{1}{|l|}{Conventional AGN - MPA-JHU} & \multicolumn{1}{c|}{3215} & \multicolumn{1}{c|}{1802} & \multicolumn{1}{c|}{266} & \multicolumn{1}{c|}{$8.3^{+0.5}_{-0.5}\%$} & \multicolumn{1}{c|}{$14.8^{+0.9}_{-0.8}\%$} \\
\multicolumn{1}{|l|}{Conventional AGN - Portsmouth} & \multicolumn{1}{c|}{3215} & \multicolumn{1}{c|}{1244} & \multicolumn{1}{c|}{111} & \multicolumn{1}{c|}{$3.5^{+0.3}_{-0.3}\%$} & \multicolumn{1}{c|}{$8.9^{+0.8}_{-0.8}\%$} \\
\multicolumn{1}{|l|}{Conventional AGN - NSA} & \multicolumn{1}{c|}{3215} & \multicolumn{1}{c|}{1153} & \multicolumn{1}{c|}{109} & \multicolumn{1}{c|}{$3.4^{+0.3}_{-0.3}\%$} & \multicolumn{1}{c|}{$9.5^{+0.9}_{-0.8}\%$} \\
\hline
\multicolumn{1}{|l|}{S06 Bonus AGN - MPA-JHU} & \multicolumn{1}{c|}{3215} & \multicolumn{1}{c|}{1802} & \multicolumn{1}{c|}{298} & \multicolumn{1}{c|}{$9.3^{+0.5}_{-0.5}\%$} & \multicolumn{1}{c|}{$16.5^{+0.9}_{-0.9}\%$} \\
\multicolumn{1}{|l|}{S06 Bonus AGN - Portsmouth} & \multicolumn{1}{c|}{3215} & \multicolumn{1}{c|}{1244} & \multicolumn{1}{c|}{219} & \multicolumn{1}{c|}{$6.8^{+0.5}_{-0.4}\%$} & \multicolumn{1}{c|}{$17.6^{+1.1}_{-1.1}\%$} \\
\multicolumn{1}{|l|}{S06 Bonus AGN - NSA} & \multicolumn{1}{c|}{3215} & \multicolumn{1}{c|}{1153} & \multicolumn{1}{c|}{268} & \multicolumn{1}{c|}{$8.3^{+0.5}_{-0.5}\%$} & \multicolumn{1}{c|}{$23.2^{+1.3}_{-1.2}\%$} \\
\hline
\multicolumn{1}{|l|}{SF-AGN - MPA-JHU} & \multicolumn{1}{c|}{3215} & \multicolumn{1}{c|}{886} & \multicolumn{1}{c|}{9} & \multicolumn{1}{c|}{$0.3^{+0.1}_{-0.1}\%$} & \multicolumn{1}{c|}{$1.0^{+0.4}_{-0.3}\%$} \\
\multicolumn{1}{|l|}{SF-AGN - Portsmouth} & \multicolumn{1}{c|}{3215} & \multicolumn{1}{c|}{366} & \multicolumn{1}{c|}{3} & \multicolumn{1}{c|}{$0.09^{+0.07}_{-0.04}\%$} & \multicolumn{1}{c|}{$0.82^{+0.62}_{-0.36}\%$} \\
\multicolumn{1}{|l|}{SF-AGN - NSA} & \multicolumn{1}{c|}{3215} & \multicolumn{1}{c|}{560} & \multicolumn{1}{c|}{11} & \multicolumn{1}{c|}{$0.3^{+0.1}_{-0.1}\%$} & \multicolumn{1}{c|}{$2.0^{+0.7}_{-0.5}\%$} \\
\hline
\multicolumn{1}{|l|}{Total Unique AGN (optical using MPA-JHU)} & \multicolumn{1}{c|}{3215} & \multicolumn{1}{c|}{n/a} & \multicolumn{1}{c|}{944} & \multicolumn{1}{c|}{$29.4^{+0.8}_{-0.8}\%$} & \multicolumn{1}{c|}{n/a} \\
\multicolumn{1}{|l|}{Total Unique AGN (optical using Portsmouth)} & \multicolumn{1}{c|}{3215} & \multicolumn{1}{c|}{n/a} & \multicolumn{1}{c|}{605} & \multicolumn{1}{c|}{$18.8^{+0.7}_{-0.7}\%$} & \multicolumn{1}{c|}{n/a} \\
\multicolumn{1}{|l|}{Total Unique AGN (optical using NSA)} & \multicolumn{1}{c|}{3215} & \multicolumn{1}{c|}{n/a} & \multicolumn{1}{c|}{660} & \multicolumn{1}{c|}{$20.5^{+0.7}_{-0.7}\%$} & \multicolumn{1}{c|}{n/a} \\
\hline
\multicolumn{1}{|l|}{\ha Broadline AGN (archival)} & \multicolumn{1}{c|}{3215} & \multicolumn{1}{c|}{n/a} & \multicolumn{1}{c|}{15} & \multicolumn{1}{c|}{$0.5^{+0.1}_{-0.1}\%$} & \multicolumn{1}{c|}{n/a} \\
\multicolumn{1}{|l|}{Optical AGN Catalogs (archival)} & \multicolumn{1}{c|}{3215} & \multicolumn{1}{c|}{n/a} & \multicolumn{1}{c|}{162} & \multicolumn{1}{c|}{$5.0^{+0.4}_{-0.4}\%$} & \multicolumn{1}{c|}{n/a} \\
\multicolumn{1}{|l|}{X-ray AGN (Birchall et al. 2020 method, used for total unique AGN stats)} & \multicolumn{1}{c|}{3215} & \multicolumn{1}{c|}{n/a} & \multicolumn{1}{c|}{17} & \multicolumn{1}{c|}{$0.5^{+0.1}_{-0.1}\%$} & \multicolumn{1}{c|}{n/a} \\
\multicolumn{1}{|l|}{X-ray AGN (Ranalli et al. 2003 method)} & \multicolumn{1}{c|}{3215} & \multicolumn{1}{c|}{n/a} & \multicolumn{1}{c|}{32} & \multicolumn{1}{c|}{$1.0^{+0.2}_{-0.2}\%$} & \multicolumn{1}{c|}{n/a} \\
\hline
\multicolumn{1}{|l|}{Total Unique AGN including Archival AGN (optical using MPA-JHU)} & \multicolumn{1}{c|}{3215} & \multicolumn{1}{c|}{n/a} & \multicolumn{1}{c|}{962} & \multicolumn{1}{c|}{$29.9^{+0.8}_{-0.8}\%$} & \multicolumn{1}{c|}{n/a} \\
\multicolumn{1}{|l|}{Total Unique AGN including Archival AGN (optical using Portsmouth)} & \multicolumn{1}{c|}{3215} & \multicolumn{1}{c|}{n/a} & \multicolumn{1}{c|}{644} & \multicolumn{1}{c|}{$20.0^{+0.7}_{-0.7}\%$} & \multicolumn{1}{c|}{n/a} \\
\multicolumn{1}{|l|}{Total Unique AGN including Archival AGN (optical using NSA)} & \multicolumn{1}{c|}{3215} & \multicolumn{1}{c|}{n/a} & \multicolumn{1}{c|}{710} & \multicolumn{1}{c|}{$22.1^{+0.7}_{-0.7}\%$} & \multicolumn{1}{c|}{n/a} \\
\hline
\hline
\multicolumn{6}{|c|}{Giants (\mstar $\geq 10^{10.5}$\msun)} \\ \hline
\multicolumn{1}{|l|}{WISE AGN - St12 cut} & \multicolumn{1}{c|}{899} & \multicolumn{1}{c|}{734} & \multicolumn{1}{c|}{0} & \multicolumn{1}{c|}{$0.00^{+0.11}_{-0.00}\%$} & \multicolumn{1}{c|}{$0.00^{+0.14}_{-0.00}\%$} \\
\multicolumn{1}{|l|}{WISE AGN - modified Sa18 cut} & \multicolumn{1}{c|}{899} & \multicolumn{1}{c|}{734} & \multicolumn{1}{c|}{6} & \multicolumn{1}{c|}{$0.7^{+0.3}_{-0.2}\%$} & \multicolumn{1}{c|}{$0.8^{+0.4}_{-0.3}\%$} \\
\hline
\multicolumn{1}{|l|}{Composite - MPA-JHU} & \multicolumn{1}{c|}{899} & \multicolumn{1}{c|}{433} & \multicolumn{1}{c|}{118} & \multicolumn{1}{c|}{$13.1^{+1.2}_{-1.1}\%$} & \multicolumn{1}{c|}{$27.3^{+2.2}_{-2.1}\%$} \\
\multicolumn{1}{|l|}{Composite - Portsmouth} & \multicolumn{1}{c|}{899} & \multicolumn{1}{c|}{201} & \multicolumn{1}{c|}{66} & \multicolumn{1}{c|}{$7.3^{+0.9}_{-0.8}\%$} & \multicolumn{1}{c|}{$32.8^{+3.4}_{-3.2}\%$} \\
\multicolumn{1}{|l|}{Composite - NSA} & \multicolumn{1}{c|}{899} & \multicolumn{1}{c|}{203} & \multicolumn{1}{c|}{69} & \multicolumn{1}{c|}{$7.7^{+0.9}_{-0.8}\%$} & \multicolumn{1}{c|}{$34.0^{+3.4}_{-3.2}\%$} \\
\hline
\multicolumn{1}{|l|}{Conventional AGN - MPA-JHU} & \multicolumn{1}{c|}{899} & \multicolumn{1}{c|}{433} & \multicolumn{1}{c|}{266} & \multicolumn{1}{c|}{$29.6^{+1.5}_{-1.5}\%$} & \multicolumn{1}{c|}{$61.4^{+2.3}_{-2.4}\%$} \\
\multicolumn{1}{|l|}{Conventional AGN - Portsmouth} & \multicolumn{1}{c|}{899} & \multicolumn{1}{c|}{201} & \multicolumn{1}{c|}{103} & \multicolumn{1}{c|}{$11.5^{+1.1}_{-1.0}\%$} & \multicolumn{1}{c|}{$51.2^{+3.5}_{-3.5}\%$} \\
\multicolumn{1}{|l|}{Conventional AGN - NSA} & \multicolumn{1}{c|}{899} & \multicolumn{1}{c|}{203} & \multicolumn{1}{c|}{110} & \multicolumn{1}{c|}{$12.2^{+1.1}_{-1.1}\%$} & \multicolumn{1}{c|}{$54.2^{+3.5}_{-3.5}\%$} \\
\hline
\multicolumn{1}{|l|}{S06 Bonus AGN - MPA-JHU} & \multicolumn{1}{c|}{899} & \multicolumn{1}{c|}{433} & \multicolumn{1}{c|}{27} & \multicolumn{1}{c|}{$3.0^{+0.6}_{-0.5}\%$} & \multicolumn{1}{c|}{$6.2^{+1.3}_{-1.1}\%$} \\
\multicolumn{1}{|l|}{S06 Bonus AGN - Portsmouth} & \multicolumn{1}{c|}{899} & \multicolumn{1}{c|}{201} & \multicolumn{1}{c|}{16} & \multicolumn{1}{c|}{$1.8^{+0.5}_{-0.4}\%$} & \multicolumn{1}{c|}{$8.0^{+2.1}_{-1.7}\%$} \\
\multicolumn{1}{|l|}{S06 Bonus AGN - NSA} & \multicolumn{1}{c|}{899} & \multicolumn{1}{c|}{203} & \multicolumn{1}{c|}{16} & \multicolumn{1}{c|}{$1.8^{+0.5}_{-0.4}\%$} & \multicolumn{1}{c|}{$7.9^{+2.1}_{-1.7}\%$} \\
\hline
\multicolumn{1}{|l|}{SF-AGN - MPA-JHU} & \multicolumn{1}{c|}{899} & \multicolumn{1}{c|}{154} & \multicolumn{1}{c|}{0} & \multicolumn{1}{c|}{$0.00^{+0.11}_{-0.00}\%$} & \multicolumn{1}{c|}{$0.00^{+0.65}_{-0.00}\%$} \\
\multicolumn{1}{|l|}{SF-AGN - Portsmouth} & \multicolumn{1}{c|}{899} & \multicolumn{1}{c|}{61} & \multicolumn{1}{c|}{0} & \multicolumn{1}{c|}{$0.00^{+0.11}_{-0.00}\%$} & \multicolumn{1}{c|}{$0.00^{+1.61}_{-0.00}\%$} \\
\multicolumn{1}{|l|}{SF-AGN - NSA} & \multicolumn{1}{c|}{899} & \multicolumn{1}{c|}{112} & \multicolumn{1}{c|}{0} & \multicolumn{1}{c|}{$0.00^{+0.11}_{-0.00}\%$} & \multicolumn{1}{c|}{$0.00^{+0.88}_{-0.00}\%$} \\
\hline
\multicolumn{1}{|l|}{Total Unique AGN (optical using MPA-JHU)} & \multicolumn{1}{c|}{899} & \multicolumn{1}{c|}{n/a} & \multicolumn{1}{c|}{413} & \multicolumn{1}{c|}{$45.9^{+1.7}_{-1.7}\%$} & \multicolumn{1}{c|}{n/a} \\
\multicolumn{1}{|l|}{Total Unique AGN (optical using Portsmouth)} & \multicolumn{1}{c|}{899} & \multicolumn{1}{c|}{n/a} & \multicolumn{1}{c|}{190} & \multicolumn{1}{c|}{$21.1^{+1.4}_{-1.3}\%$} & \multicolumn{1}{c|}{n/a} \\
\multicolumn{1}{|l|}{Total Unique AGN (optical using NSA)} & \multicolumn{1}{c|}{899} & \multicolumn{1}{c|}{n/a} & \multicolumn{1}{c|}{197} & \multicolumn{1}{c|}{$21.9^{+1.4}_{-1.3}\%$} & \multicolumn{1}{c|}{n/a} \\
\hline
\multicolumn{1}{|l|}{\ha Broadline AGN (archival)} & \multicolumn{1}{c|}{899} & \multicolumn{1}{c|}{n/a} & \multicolumn{1}{c|}{9} & \multicolumn{1}{c|}{$1.0^{+0.4}_{-0.3}\%$} & \multicolumn{1}{c|}{n/a} \\
\multicolumn{1}{|l|}{Optical AGN Catalogs (archival)} & \multicolumn{1}{c|}{899} & \multicolumn{1}{c|}{n/a} & \multicolumn{1}{c|}{105} & \multicolumn{1}{c|}{$11.7^{+1.1}_{-1.0}\%$} & \multicolumn{1}{c|}{n/a} \\
\multicolumn{1}{|l|}{X-ray AGN (Birchall et al. 2020 method, used for total unique AGN stats)} & \multicolumn{1}{c|}{899} & \multicolumn{1}{c|}{n/a} & \multicolumn{1}{c|}{29} & \multicolumn{1}{c|}{$3.2^{+0.6}_{-0.5}\%$} & \multicolumn{1}{c|}{n/a} \\
\multicolumn{1}{|l|}{X-ray AGN (Ranalli et al. 2003 method)} & \multicolumn{1}{c|}{899} & \multicolumn{1}{c|}{n/a} & \multicolumn{1}{c|}{47} & \multicolumn{1}{c|}{$5.2^{+0.8}_{-0.7}\%$} & \multicolumn{1}{c|}{n/a} \\
\hline
\multicolumn{1}{|l|}{Total Unique AGN including Archival AGN (optical using MPA-JHU)} & \multicolumn{1}{c|}{899} & \multicolumn{1}{c|}{n/a} & \multicolumn{1}{c|}{429} & \multicolumn{1}{c|}{$47.7^{+1.7}_{-1.7}\%$} & \multicolumn{1}{c|}{n/a} \\
\multicolumn{1}{|l|}{Total Unique AGN including Archival AGN (optical using Portsmouth)} & \multicolumn{1}{c|}{899} & \multicolumn{1}{c|}{n/a} & \multicolumn{1}{c|}{230} & \multicolumn{1}{c|}{$25.6^{+1.5}_{-1.4}\%$} & \multicolumn{1}{c|}{n/a} \\
\multicolumn{1}{|l|}{Total Unique AGN including Archival AGN (optical using NSA)} & \multicolumn{1}{c|}{899} & \multicolumn{1}{c|}{n/a} & \multicolumn{1}{c|}{230} & \multicolumn{1}{c|}{$25.6^{+1.5}_{-1.4}\%$} & \multicolumn{1}{c|}{n/a} \\
\hline
\enddata
\tablecomments{Note: All error bars are computed using binomial confidence intervals. Total unique AGN numbers count duplicate AGN detections only once -- overlapping detections are possible between our S06 Bonus AGN and SF-AGN categories within BPT/VO AGN, between any BPT/VO AGN category and WISE AGN, between any of our AGN categories and archival AGN detections, or between different archival AGN detections. {Total Unique AGN using only the \cite{Birchall20} X-ray AGN classifications (see Section \ref{sec:archival}).} See Section~\ref{sec:ch3agninventory} for information on our BPT/VO and WISE AGN categories and Section~\ref{sec:ch3discussion} for information on archival AGN.}
\end{deluxetable*}

\section{Discussion}
\label{sec:ch3discussion}

The primary goal of this work was to create a comprehensive AGN inventory for the RESOLVE and ECO surveys, with special attention to recent advances in dwarf AGN detection. Such an inventory can provide a starting point for understanding the inventory of massive and supermassive black holes, providing statistical constraints on black hole-galaxy coevolution and black hole seeding mechanisms. The previous section summarized the statistics of our new inventory. Below, we compare our results with the literature and examine how the mass dependence of the RESOLVE+ECO AGN inventory relates to galaxy properties and group environment.\\

\subsection{Comparison with the Literature}
\label{sec:disclit}

\subsubsection{AGN Frequencies}
\label{sec:agnfreq}

Most AGN surveys are flux-limited, so dominated by giants. RESOLVE and ECO are volume- and mass-limited, so dominated by dwarfs. Thus, we stress the importance of considering the underlying parent sample when comparing AGN percentages between studies. We have provided statistics in three mass regimes with astrophysically meaningful definitions (see Section~\ref{sec:ch3intro}) to facilitate such comparison. Also, as previously mentioned, most observational studies quote AGN percentages with respect to the search sample dictated by data availability, not the full galaxy population from which the search sample is drawn. In contrast, we have tabulated percentages relative to both our search samples and our full volume- and mass-limited parent populations of dwarfs, transitional galaxies, and giants. \textit{The latter full-population statistics are directly analogous to theoretical AGN occupation fractions, because of the complete mass- and volume-limited design of RESOLVE and EC
O.}\\

To our knowledge, no prior AGN inventory has specifically separated out transitional galaxies. Rather, most prior AGN surveys have included a mix of transitional and giant galaxies. Relative to the full galaxy population (including galaxies lacking BPT/VO or WISE data) as defined by baryonic (or stellar) mass limits, we obtain AGN frequencies of 16--26\% (18--30\%) for transitional galaxies and 20--48\% (20--48\%) for giants, where the ranges as usual include the systematics between the MPA-JHU, Portsmouth, and NSA crossmatched emission-line catalogs. These numbers agree roughly with the overall frequency of AGN in the flux-limited SDSS survey --- for example, \citet{Kauffmann03} identified $\sim$18\% of their complete flux-limited SDSS sample as narrow-line AGN, where the search sample was $\sim$45\% of the complete sample. (We also included mid-IR WISE AGN, but they proved insignificant, being detected in under 1\% of our galaxies and often duplicating BPT/VO detections.) O
ur numbers do reflect additional BPT/VO-based detection strategies, but unsurprisingly, we found that for massive galaxies the traditional Composite and Conventional AGN categories are dominant. Only when isolating transitional galaxies do we see S06 Bonus AGN become comparably important, at the same time as Composites overtake Conventional AGN in abundance (see Section~\ref{sec:optwisestats3}).\\

In the dwarf regime, we see the AGN inventory shift even more strongly toward AGN categories associated with star formation: Composites, S06 Bonus AGN, and most especially SF-AGN, the new category introduced by \citetalias{Polimera22}. We should note that while SF-AGN still seem to be the most abundant dwarf AGN type in this paper's expanded search sample (based on MPA-JHU or NSA data; Portsmouth data somewhat disagree, see Section~\ref{sec:optwisestats2}), our SF-AGN percentages are lower than those reported in \citetalias{Polimera22}. This difference is not due to a reduction in SF-AGN numbers, but rather because we have expanded our search sample to enable BPT-only detections of S06 Bonus AGN and Composites (as well as Conventional AGN, rare in dwarfs). Though we have slightly changed the definition of the dwarf sample in terms of mass limits (see Section \ref{sec:ch3data}), there are no statistically significant differences between the AGN percentages relative to the smal
ler BPT+VO search sample in \citetalias{Polimera22} vs.\ this work. For example, the dwarf SF-AGN percentage, when computed relative to the BPT+VO search sample, has changed from 5--15\% in \citetalias{Polimera22} to 3--16\% in this work (using the baryonic-mass-selected sample), where our wider range mainly reflects that fact that we now report ranges including measurement error bars in addition to systematics between crossmatched emission-line catalogs, where only the latter was included by \citetalias{Polimera22}. Our results are also consistent with the SF-AGN percentage of 1--3\% relative to the full dwarf population reported in \citetalias{Polimera22}.\\

Though we also attempted to increase our dwarf AGN inventory using mid-IR WISE AGN detection, we ultimately found WISE AGN to be quite rare in dwarfs, contrary to \citet{Kaviraj19} but in agreement with other work \citep[e.g.,][]{Sartori15,Lupi20}. One issue is that due to SF contamination, the WISE AGN category potentially has more false positives among dwarfs than do optical BPT/VO AGN categories (see \citealt{Hainline16, Wasleske24}). Another issue is the low S/N of WISE data, especially standard catalog data. We found that reprocessing WISE photometry and computing resolution-matched aperture colors yields a cleaner \wonetwo separation of AGN and SF galaxies in the WISE color-color diagnostic diagram, thus better matching the theoretical predictions of \citetalias{Satyapal18}. At the same time, we had to drop the \citetalias{Satyapal18} \wtwothree color cut just to detect 7 of our 12 dwarf WISE AGN, as these seven lack usable W3 data. We have made the case for a more stri
ngent standard for usability of WISE data (see Section~\ref{sec:ch3midiragn}), i.e., peak S/N~$>$~5, noting that more typical S/N cuts applied to standard WISE catalog data could result in a $\sim$60\% rate of false positives, primarily due to scatter across the \wonetwo dividing line. By analogy, we must also caution that $\sim$60\% of our own reported WISE AGN are potential false positives (i.e., the 7 lacking usable W3 data), though we feel comfortable reporting them as AGN candidates because the \wtwothree cut is a nearly vertical line and nearly all galaxies fall on the AGN side of that cut using our reprocessed WISE colors.\\

With or without these few WISE AGN, \textit{we find a $\sim$2--5\% dwarf AGN frequency, which far exceeds other statistically complete dwarf AGN frequency measurements in the literature} though not necessarily the completeness-corrected frequencies. 
For example, \citet{Birchall20} 
obtain a raw X-ray detected AGN frequency of $\sim$1.4\% that they use to infer a completeness-corrected frequency of $\sim$4--7\% at 
\mstar\ $\sim\,10^{9}$--$10^{9.5}$ \msun\ (their Figure~10; see also \citealt{Latimer21}). 
The variability-based search of \citet{Baldassare20} yields a smaller raw percentage of 
$\sim$0.7\% near \mstar\ $\sim\,10^{9.5}$ \msun\ (their Figure~5), and their completeness corrections roughly double this number. 
Recently, \cite{Pucha25} used DESI survey early data to measure a dwarf AGN fraction of $\sim$2.1\% for the 114,496 dwarfs in their emission-line AGN search sample, using the BPT diagram alone. If we normalize their dwarf AGN percentage instead to their full population of 222,680 dwarfs (still not volume-limited), we find their full-population dwarf AGN fraction is $\sim$1.1\%, less than half of our estimated dwarf AGN percentage for the ECO and RESOLVE surveys.\footnote{These numbers are taken from the original paper, as the erratum for \citet{Pucha25}, which reduces the number of dwarf AGN candidates from 2444 to 2261 after correcting for systematically overestimated S/N ratios, does not state the size of the new search sample.} \textit{We have shown that inclusion of S06 Bonus AGN and especially SF-AGN is key to expanding the dwarf AGN inventory.} The abundant population of SF-AGN we report is in line with the recent study by \citet{Mezcua24}, who used integral field unit 
spectroscopy from MaNGA to find that SF-AGN are $\sim$2$\times$ more common than other AGN flavours among dwarfs. The shallow potential wells of dwarf galaxies imply that off-center dwarf AGN should be common \citep{Bellovary21}, and including off-center IFU detections, the \citet{Mezcua24} dwarf AGN frequency greatly exceeds our own. That said, according to their recent erratum, the statistics in \citet{Mezcua24} that were intended to apply to \mstar $<$ $10^{9.5}$ \msun\ in fact apply to \mstar $<$ $10^{10}$ \msun, so $\sim$2/3 of their nominal dwarfs are what we would call transitional galaxies, for which AGN are much more common.\\

As discussed in Sections~\ref{subsec:ch3s06bonus} and \ref{subsec:ch3sfagn}, because S06 Bonus AGN and SF-AGN are associated with strong star formation (see Section~\ref{sec:discmass}), we expect some false positives, e.g., due to ULXs \citep[][R25]{Garofali24}. We have opted to balance impurity against incompleteness in our AGN detections by including ambiguous cases that might be explained purely by ULXs/HMXBs/WRs but lie far from the observed locus of most SF galaxies. Conversely, theoretical spectral models suggest that mixed SF+AGN galaxies with AGN fractions below 0.08 are likely lurking in {the Ke01/Ka03 SF category and even the more restrictive S06 SF} category (\citetalias{Polimera22, Richardson22, Richardson25}). Breaking the theoretical degeneracy between ULXs/HMXBs/WRs and dwarf AGN will likely require more sophisticated analysis using all available emission lines. However, any such analysis will still be limited by the poor consensus between different state-of-th
e-art emission line pipelines (\citetalias{Polimera22} and Section~\ref{sec:optwisestats}) and the observational \textit{anticorrelation} between [OI] S/N and the strength of the \OvH line ratio, a counterintuitive result that implies that necessary S/N cuts on the [OI] line cause underestimation of the SF-AGN population (\citetalias{Polimera22}). \\

Compared to theory, our full-population dwarf AGN percentage of $\sim$2--5\% agrees with predicted AGN occupation fractions given by \cite{Haidar22}, though it is a lower limit based on only BPT/VO and WISE color detection (we add archival AGN detections in Section~\ref{sec:archival} to obtain a slightly higher $\sim$3--6\% estimate). Moreover, our analysis is better matched to simulations than most observational analyses, in that we quote statistics relative to the full galaxy population rather than a subpopulation (e.g., of emission-line galaxies). Nonetheless, there are important caveats to our comparison with simulations. \cite{Haidar22} note that black hole seeding and accretion mechanisms in simulations are not necessarily physically well motivated. Real accretion is almost certainly harder to detect than simulated accretion (e.g., due to wandering and obscured black holes), but black hole seeding is too poorly understood to say whether simulations are on balance overes
timating or underestimating the true black hole population. See \cite{Polimera22} for a more detailed discussion of the nuances of comparing observational and theoretical AGN percentages. \\ 

\subsubsection{Archival AGN}
\label{sec:archival}

Comparison with archival AGN detections is useful both to test and expand our AGN inventory (albeit with a poorly defined selection function). In addition to off-center AGN, we are certainly missing more AGN that cannot be detected by our chosen methods, especially among dwarfs. In \citet{Wasleske24}, dwarf AGN were selected with different spectroscopic diagnostics, WISE colors, X-ray detections, and variability, then checked for overlap between the various methods. Among the considered methods, \citet{Wasleske24} found that BPT-selected dwarf AGN and [He~II]-selected dwarf AGN had the greatest overlap, but these AGN types are intrinsically rare \citep[see also][]{Sartori15}. The authors did not consider a primarily [O~I]-selected sample of dwarf AGN like SF-AGN, but SF-AGN will naturally fail in the BPT diagram due to their metal-poor nature. In \citetalias{Polimera22}, we provided a detailed overview of past dwarf AGN searches at different wavelengths and examined the corre
spondence of SF-AGN to X-ray and/or radio-detected AGN, finding relatively little overlap. Unfortunately, with current survey data we cannot construct a uniform AGN census at X-ray or radio wavelengths for RESOLVE and ECO, which could detect complementary populations of dwarf AGN \citep{Yao22}. Nevertheless, we have compiled an incomplete inventory of archival AGN by crossmatching RESOLVE and ECO to X-ray data, {following the same crossmatch procedure to \citetalias{Polimera22} but with an updated X-ray AGN classification (see below)}, as well as to catalogs of broadline \ha AGN and BPT/VO AGN (identified independently from our own BPT/VO AGN). We have already compared to the standard approach used in WISE AGN detection in Section~\ref{sec:ch3midiragn}, where we found that our WISE AGN inventory shows approximate consistency, with differences reflecting our higher S/N standards. \\

We first crossmatched the full parent survey to two AGN catalogs from \cite{Veron06} and \cite{Flesch15}, which include both BPT/VO AGN detections and broadline H$\alpha$ AGN detections, so we collectively refer to these as \textit{optical} AGN. In the baryonic-mass-selected sample, we found 284 AGN matches (19 in dwarfs, 173 in transitional galaxies, 92 in giants). Category flags indicating these matches are included in Table~\ref{tab:agntable}. About 90\% of these matches are also identified by us, using BPT/VO and/or WISE color selection, though 8\% fail our search sample S/N requirements. Another 2\% are classified as star-forming galaxies using all three crossmatched emission-line catalogs in our work. \\

Compared to broadline AGN detections \textit{specifically}, our AGN inventory shows high overlap, though the non-overlap is still substantial. We crossmatched the full RESOLVE+ECO parent survey to a comprehensive broadline AGN catalog \citep{Liu19} and found 25 matches (1 dwarf, 17 transitional, 7 giant). Category flags indicating these matches are included in Table~\ref{tab:agntable}. Among the broadline AGN in transitional and giant galaxies, 70\% overlap our BPT/VO or WISE AGN. The one broadline AGN in a dwarf is an object we classified as a Conventional AGN. \\

Compared to archival X-ray AGN detections, our AGN inventory appears significantly complementary, based on the limited (non-uniform) X-ray coverage available for RESOLVE and ECO. We crossmatched the full RESOLVE+ECO parent survey to X-ray data from Newton-XMM and the Chandra Source Catalog. 
{In our preferred method of identifying X-ray AGN candidates, we followed the approach of \citet{Birchall20}, which improves upon the $\rm L_{X-ray}$-SFR relation of \citet{Ranalli03} used in \citetalias{Polimera22} by separately accounting for the distinct X-ray contributions from stellar-mass compact objects and hot interstellar gas. We first computed the expected non-AGN X-ray luminosity in the 0.5--8~keV band as the sum of X-ray binary and hot gas contributions:\\
\begin{equation}
L_{\rm X-ray-exp} = L_{\rm XRB} + L_{\rm gas}.
\end{equation}
The X-ray binary component follows \citet{Lehmer19}:
\begin{equation}
L_{\rm XRB} = \alpha M_{\star} + \beta \times {\rm SFR},
\end{equation}
where we used $\alpha$=$10^{29.25}$ and $\beta$=$10^{39.71}$ scaling values from \cite{Lehmer19}. The hot gas component follows \citet{Mineo12}:
\begin{equation}
L_{\rm gas} = 8.3 \times 10^{38} \times {\rm SFR} \quad {\rm erg~s^{-1}}.
\end{equation}
A galaxy is classified as an X-ray AGN candidate if $\rm L_{X,obs} \geq 3\times L_{X-ray-exp}$ \citep{Birchall20}. Since the \citet{Lehmer19} scaling relations are calibrated in the 0.5--8~keV band, we converted all observed luminosities to this common scale. For XMM-Newton luminosities, we summed EPIC bands 2--5 (spanning 0.5--12~keV) and applied a correction factor of 0.83, computed using WebPIMMS\footnote{\url{https://heasarc.gsfc.nasa.gov/cgi-bin/Tools/w3pimms/w3pimms.pl}} assuming a standard power-law spectrum with photon index $\Gamma$=1.8 \citep{Wasleske24}. For Chandra luminosities, we scaled the broad-band (0.5--7~keV) aperture flux by a factor of 1.21 under the same spectral assumption. }\\

{Using this approach based on \citet{Birchall20}, we find 52 X-ray AGN (5 in dwarfs, 22 in transitional galaxies, 25 in giants). Using the \citet{Ranalli03} criterion adopted in \citetalias{Polimera22}, we find 84 X-ray AGN (4 in dwarfs, 40 in transitional galaxies, 40 in giants). In other words, our preferred method identifies one more dwarf-galaxy X-ray AGN and many fewer transitional- and giant-galaxy X-ray AGN compared to the \citet{Ranalli03} criterion used in \citetalias{Polimera22}. We favor the \citet{Birchall20} approach because the \citet{Ranalli03} relation does not account for the stellar-mass-dependent LMXB contribution that dominates X-ray emission in massive galaxies, and that omission likely leads to systematic over-classification of non-dwarf star-forming galaxies as AGN \citep{Lehmer10, Lehmer19}. However, using our BPT/VO+WISE classifications based on the largest crossmatched SDSS catalog (MPA-JHU), we find evidence that the \citet{Birchall20} approach has 
both fewer unconfirmed X-ray AGN and fewer confirmed X-ray AGN. Specifically, we find independent confirmation for 2/4 dwarf, 17/40 transitional-galaxy, and 20/40 giant X-ray AGN using \citet{Ranalli03}, vs. independent confirmation for 2/5 dwarf, 12/22 transitional-galaxy, and 17/25 giant X-ray AGN using \citet{Birchall20}. Transitional galaxies show the largest discrepancy in confirmed X-ray AGN, with $\sim$40\% (17 vs.\ 12) more confirmed AGN based on \citet{Ranalli03}, while giants show the largest discrepancy in unconfirmed X-ray AGN, with the \citet{Ranalli03} criterion yielding 2.5$\times$ (20 vs.\ 8) as many unconfirmed as the \citet{Birchall20} criterion yields. } \\

{Given the apparent purity/completeness tradeoff between the two approaches, we include category flags for both sets of candidate X-ray AGN in Table~\ref{tab:agntable}, which also ensures reproducibility of our earlier work in \citetalias{Polimera22} and \citet{Hutchens25}. \citet{Hutchens25} used a pre-publication version of this paper’s AGN catalogue based on \citet{Ranalli03}, so we have revisited the results of that paper substituting X-ray AGN flags based on the \citet{Birchall20} approach. We found the results to be nearly unchanged, as expected given the small number of X-ray AGN and the focus of \citet{Hutchens25} on entire galaxy \textit{groups}, whose AGN-hosting status rarely depends on the presence or absence of a single X-ray AGN. (Our AGN quenching analysis in \citealt{Carr25} also used this paper’s AGN catalogue prior to publication, but used only BPT/VO+WISE AGN, exactly corresponding to the BPT/VO+WISE inventory given herein.) } \\

{We view the discrepancies between X-ray AGN identifications using \citet{Ranalli03} vs.\ using \citet{Birchall20} as a measure of systematic error, so we show statistics based on both criteria in Tables~\ref{tab:agnstats} and \ref{tab:agnstatsstellar}, but we use only our preferred approach \citep{Birchall20} for combined total unique AGN statistics. With this choice, $\sim$60/50/30\% of dwarf/transitional/giant-galaxy X-ray AGN represent new unique AGN not overlapping our BPT/VO+WISE inventory.  Among the X-ray AGN in dwarfs, two (40\%) overlap our BPT/VO AGN: one is a Conventional AGN and the other is the SF-AGN \textit{rf0477}, also identified as an X-ray AGN using the \citet{Ranalli03} $\rm L_{X-ray}$-SFR relation in \citetalias{Polimera22}. } \\

Because of the scarcity of X-ray data for SF-AGN, we cannot say how many might be detected in X-rays. We obtained an X-ray observation with the Chandra X-ray Observatory for one of our SF-AGN hosts, \textit{rs0010}, which has AGN-like \OvH ratio in the central kpc revealed by MaNGA IFU measurements (see \citetalias{Polimera22} for details). We designed the observations to emulate those for \textit{RGG118}, a dwarf AGN host with similar properties to \textit{rs0010} and with detected \ha line broadening \citep{Baldassare15}. We found that \textit{rs0010} does not have a strong X-ray signal and has an upper limit of 4$\times10^{39}$ ergs $\rm s^{-1}$ for the 0.5-7 keV band luminosity. This lack of a strong X-ray signal may not be surprising given that the overlap of X-ray and optical AGN is low, as shown in this work and in \citetalias{Polimera22}, and given literature arguments that dwarf AGN may actually be underluminous in X-rays \citep[][\citetalias{Richardson25}]{Latimer19}. \\

\textit{Including archival AGN and normalizing to the full population in RESOLVE+ECO, we obtain lower limits on the AGN occupation fractions for dwarf, transitional, and giant galaxies of $\sim$3--6\%, $\sim$17--31\%, and $\sim$24--50\%, respectively (Tables \ref{tab:agnstats} and \ref{tab:agnstatsstellar}).} To find even more AGN, we would ideally need deeper multi-wavelength surveys as well as complementary techniques like variability, mid-IR emission line diagnostics, and $\rm L_{X-ray}$-SFR or $\rm L_{radio}$-SFR relationships. Uniform coverage with such data sets is essential to understanding the statistics of AGN across the galaxy population.\\

\subsection{Mass Dependence of the AGN Inventory}
\label{sec:discmass}

\begin{figure*}[]
    \centering
    \includegraphics[width = \linewidth]{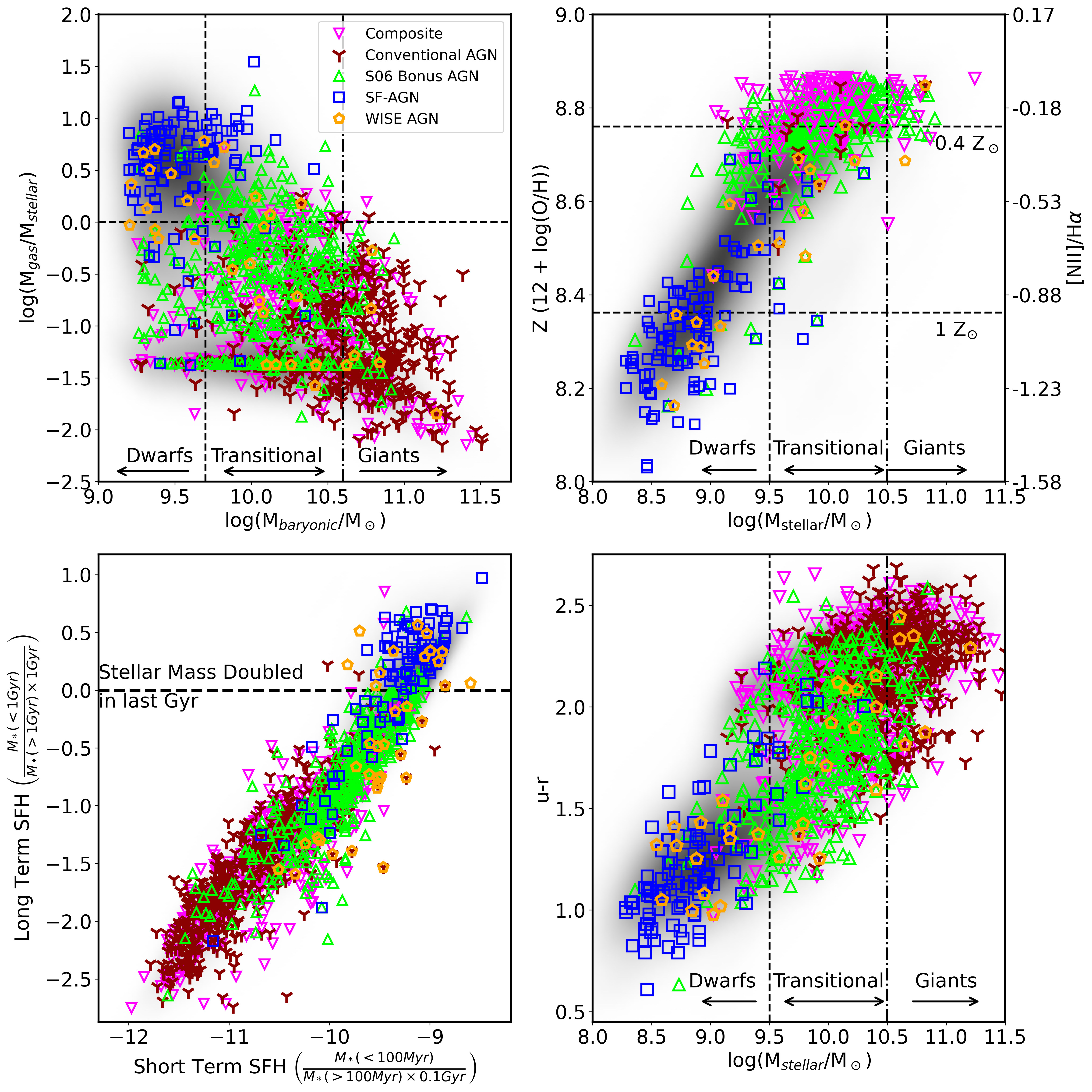}
    \caption{%
   Properties of RESOLVE and ECO AGN hosts identified by the three methods discussed in Sections \ref{sec:ch3midiragn} and \ref{sec:ch3opticalagnmethod}. 
   \textit{Top left:} Atomic gas-to-stellar-mass ratio vs.\ stellar mass using the best gas estimate of \citet{Hutchens23} as discussed in Section~\ref{subsec:ch3surveys}. 
   The concentration of points near $\log$({\mgas / \mstar}) $\sim -1.5$ represents strong upper limits taken as the gas mass value.  
   \textit{Top right:} Mass-metallicity relation. Metallicities are estimated using the \NvH calibration from \citet{Pettini04}. 
   We caution that these metallicities are only illustrative since we are using a crude proxy and not considering the AGN spectral contribution. 
   Since the calibration has an upper limit for \NvH, several Composite and Conventional AGN are not plotted in this panel.  
   \textit{Bottom left:} Long vs.\ short term star formation histories using the 1-Gyr-timescale fractional stellar mass growth rate (FSMGR) and 100-Myr-timescale FSMGR, respectively. 
   \textit{Bottom right:} Colour-stellar mass relation using extinction-corrected model colors. 
   For consistency, all panels including the color-stellar mass relation panel show our baryonic-mass selected sample. Collectively, the four panels show that SF-AGN are most common in typical dwarfs that are gas-rich, metal-poor, highly star-forming, and blue, whereas S06 Bonus AGN are most common in transitional galaxies with intermediate properties, and Conventional AGN are most common in giants that are gas-poor, metal-rich, weakly star-forming, and red. Composite hosts bridge the properties of S06 Bonus AGN hosts and Conventional AGN hosts. WISE AGN hosts span almost the entire range of properties, but few are very quenched or metal-rich. 
   \textit{Note}: We plot the handful of galaxies that are both SF-AGN and S06 Bonus AGN with both blue and green symbols. Similarly, we also double plot the galaxies that are both WISE AGN and some flavour of BPT/VO AGN. Pink symbols lie underneath the red and green symbols and may not be fully visible in crowded areas.
    }
    \label{fig:agnprop}
\end{figure*}

\begin{figure*}[]
    \centering
    \includegraphics[width = 0.9\linewidth]{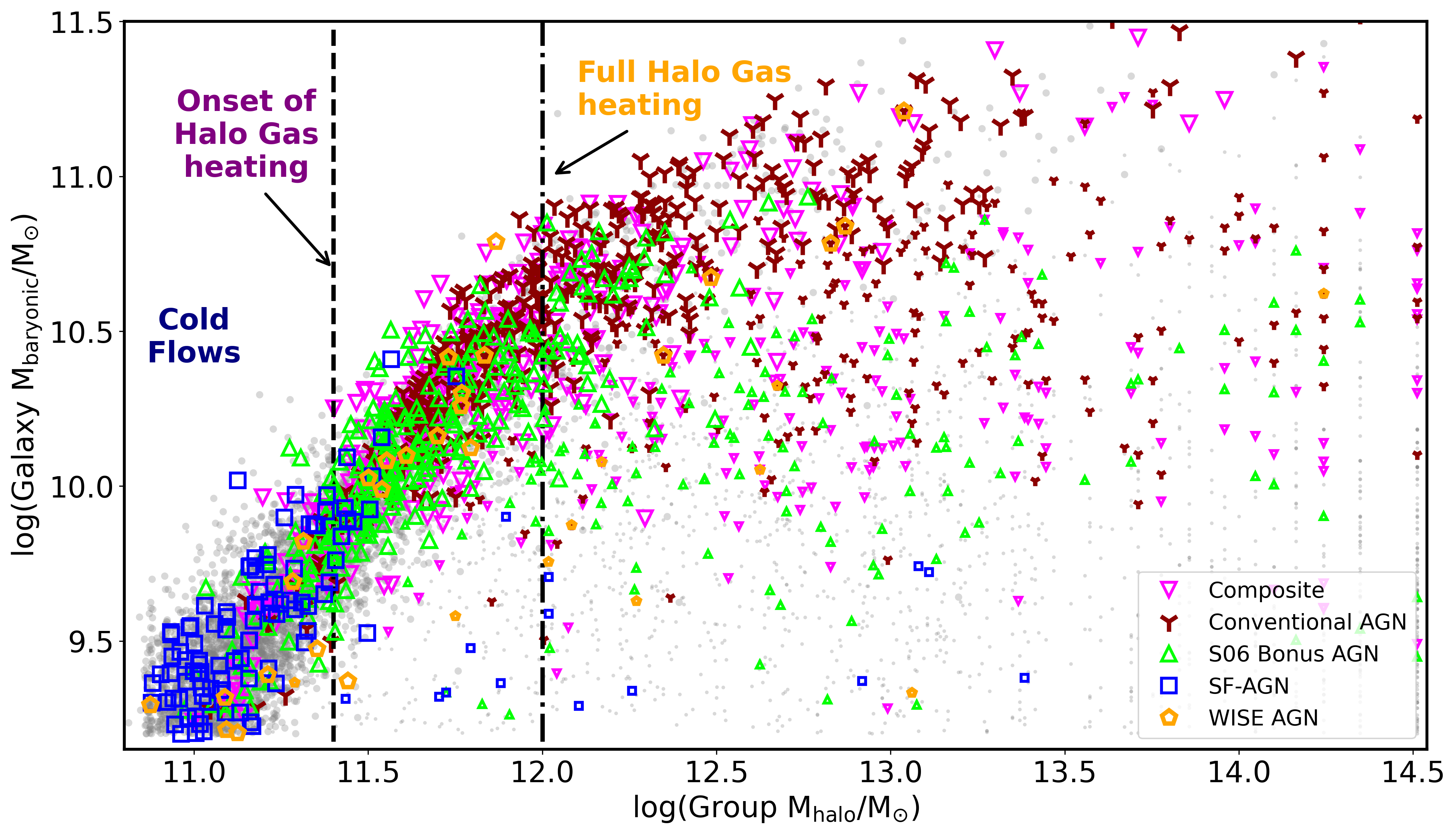}
    \caption{Galaxy baryonic mass vs.\ group halo mass relation for RESOLVE and ECO AGN hosts identified by the methods discussed in Sections \ref{sec:ch3midiragn} and \ref{sec:ch3opticalagnmethod}. Light gray background symbols show the rest of the baryonic-mass-limited combined survey. The dashed line represents the gas-richness threshold mass (\mhalo $\sim$$10^{11.4}$\msun) and the dot-dashed line represents the bimodality mass (\mhalo $\sim$$10^{12.0}$\msun). Larger and smaller symbols represent central and satellite galaxies, respectively. Different categories of AGN span different halo mass ranges representing different group environments. Some galaxies are double-plotted as in Figure~\ref{fig:agnprop}. Pink symbols lie underneath the red and green symbols and may not be fully visible in crowded areas.}
    \label{fig:halomass}
\end{figure*}

\begin{figure}[]
    \centering
    \includegraphics[width=\columnwidth]{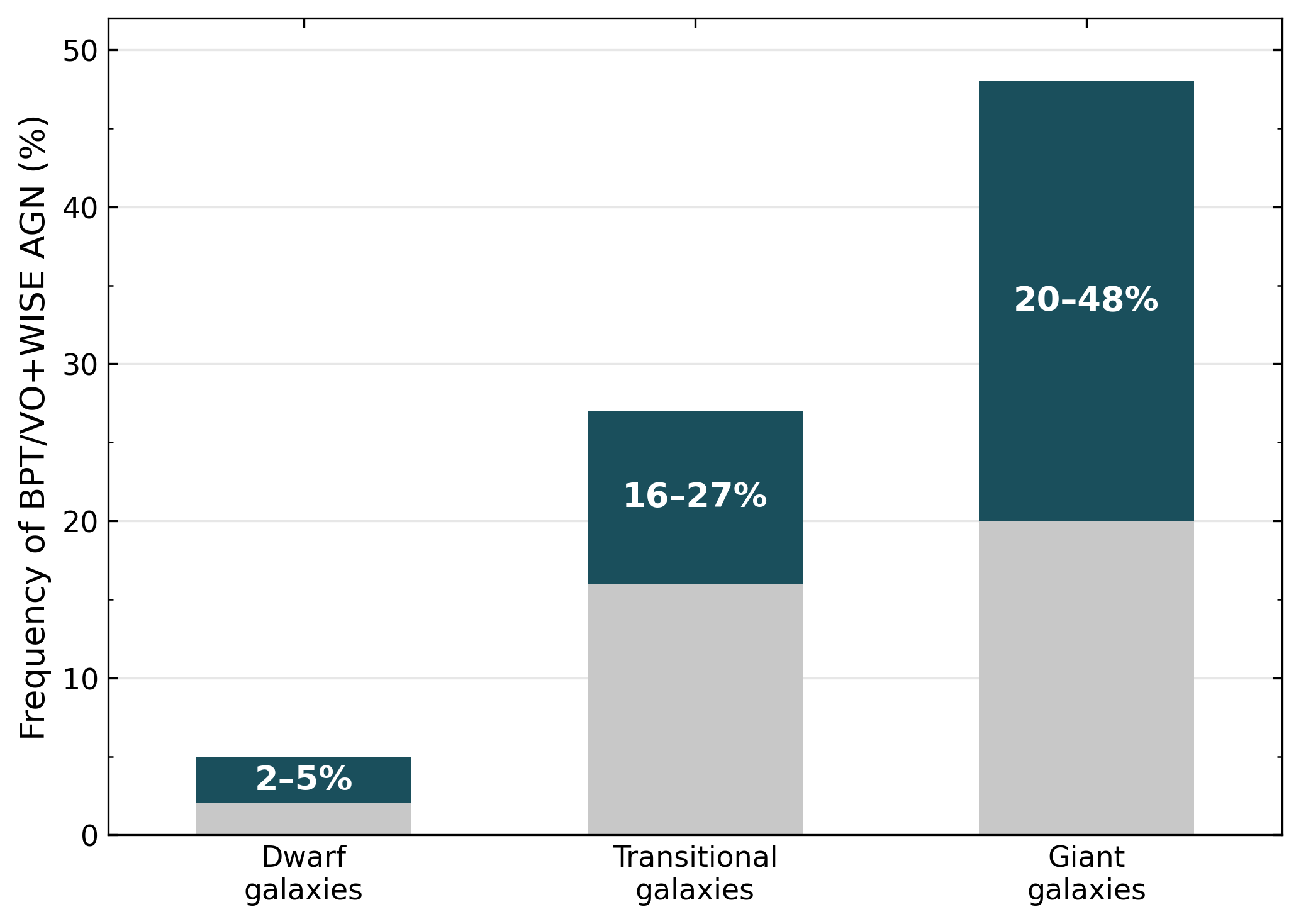}
    \caption{{Full-population AGN frequencies for dwarf, transitional, and giant galaxies in the combined RESOLVE and ECO surveys, based on BPT/VO and WISE AGN detections (excluding archival AGN to ensure uniformity). We see a steep increase from dwarfs ($\sim$2–-5\%) to transitional galaxies ($\sim$16–-27\%), a $\sim$5$\times$ jump that coincides with the onset of halo gas heating at the gas-richness threshold scale (see Section \ref{sec:discmass}). The reported frequency ranges (shown in a darker shade) span the AGN percentages obtained using the three crossmatched SDSS emission-line catalogs (MPA-JHU, Portsmouth, and NSA), reflecting irreducible systematics between catalogs. These fractions are computed relative to the full parent population in each mass bin, including galaxies without adequate data for AGN searches, making them more comparable to theoretical occupation fractions than typically reported AGN frequencies, though they remain lower limits as they exclude w
andering AGN and AGN that might be added via complementary detection strategies.}}
    \label{fig:barchart}
\end{figure}

An interesting emergent result from this work is the association of different types of AGN with characteristic mass ranges. In this study, we have divided our sample into dwarf, transitional, and giant galaxies based on their baryonic mass (see Section \ref{sec:ch3data}). Subject to some variation depending on emission-line data source, we found that: (i) in dwarfs, the most common AGN types in dwarfs are SF-AGN and to a lesser extent Composites and S06 Bonus AGN, (ii) in transitional galaxies, the most common AGN are Composites and S06 Bonus AGN and to a lesser extent Conventional AGN (iii) in giants, the most common AGN are Conventional AGN and to a lesser extent Composites. The mass dependence of these results suggests a connection to the mass dependence of star formation, with SF-AGN hosts being the most star-forming, Composite and S06 Bonus AGN hosts being moderately star-forming, and Conventional AGN hosts being the least star-forming. We validate and explore this conne
ction below. \\

Figure \ref{fig:agnprop} compares the galaxy properties of AGN hosts selected using traditional methods (Conventional and Composite AGN) and using newer approaches (S06 Bonus AGN, SF-AGN, and WISE AGN). All properties are estimated using the same methods as in \citetalias{Polimera22}. We see that Conventional, Composite, and S06 Bonus AGN hosts are separated in parameter space from SF-AGN hosts, while WISE AGN span almost the full parameter space. \\

In general, as seen in the left panels of Figure~\ref{fig:agnprop}, BPT/VO AGN flavours correlate with the gas content and recent SF history of a galaxy. The continuum runs from SF-AGN at the gas-rich/high SF end to S06 Bonus AGN, Composites, and finally Conventional AGN at the gas-poor/low SF end. \\

Figure \ref{fig:agnprop} also shows that in general, SF-AGN and WISE AGN in dwarf hosts dwell in typical dwarfs -- having high gas content (top left panel), low metallicities (top right panel), high star formation (bottom left panel) and blue colors (bottom right panel). In contrast, the redefined BPT lines from \citetalias{Stasinska06} identify AGN mostly in transitional galaxies and giants or atypical dwarfs with low gas content, high metallicities (close to 1\zsun), and low star formation activity.\\

Thus, a key conclusion of this work is that AGN hosted by \textit{typical} dwarfs are mostly in the SF-AGN category. WISE AGN also occur in dwarfs with typical properties, but we detect very few, and as discussed in Section~\ref{sec:agnfreq}, about half of the detections are not secure. Whereas WISE AGN strongly overlap BPT/VO AGN in general, we find little overlap for dwarfs, which could imply that for dwarfs SF-AGN and WISE AGN represent different evolutionary stages. However, we hesitate to speculate without a larger sample of secure WISE AGN detections.\\

Likewise, AGN hosted by \textit{typical} transitional galaxies, which tend to be bulged spirals with intermediate gas content and star formation \citep{Kannappan13}, are mostly in the S06 Bonus AGN and Composite categories. In transitional galaxies, we see a continuum in color and star formation history from mostly blue and star-forming (S06 Bonus AGN) to intermediate (Composite) to mostly red and dead (Conventional).\\

As discussed in Section~\ref{sec:ch3intro}, the underlying physics that differentiates the dwarf, transitional, and giant galaxy mass regimes is related to the dark matter halo mass of their groups (where galaxies that do not share a halo are considered to be in a group of one). Group halo mass drives accretion physics, which is a key factor in SF and AGN activity. Figure \ref{fig:halomass} shows the relation between galaxy baryonic mass and group halo mass. Note that except at high halo mass, most AGN host galaxies (like most galaxies) follow the tight locus of galaxies that are the most massive in their groups. We designate these galaxies as \textit{centrals} (large symbols) as opposed to \textit{satellites} (small symbols). The bimodality and threshold mass scales for galaxy centrals correspond to the bimodality and threshold mass scales for group halos as described in Section~\ref{sec:ch3intro}. \\

At low halo masses below the gas-richness threshold scale (\mhalo $\sim$$10^{11.4}$\msun; \citealt{Kannappan13}), theory predicts rapid halo gas cooling should dominate \citep[e.g.,][]{Dekel06, Nelson13} --- we refer to this as \textit{cold accretion}.\footnote{This terminology indicates the halo-mass dependent type of cold accretion discussed in \citep{Dekel06}, not the other type of cold accretion, a.k.a.\ cold-in-hot accretion, seen only at high redshift.} Figure \ref{fig:halomass} shows that SF-AGN dominate the low halo-mass, cold-accretion regime, with more Composites and S06 Bonus AGN appearing as halo mass increases up to the threshold mass scale. Given our surveys' baryonic-mass floor, these hosts are nearly all centrals. At the high-mass end above the bimodality mass (\mhalo $\sim$$10^{12.0}$\msun), theory predicts that the halo gas should be shock-heated out to and beyond the virial radius. In this regime, centrals are giant galaxies that tend to host Conventional A
GN and Composites, with S06 Bonus AGN confined mostly below \mhalo $\sim$$10^{12.5}$\msun. In the transitional range between the two key halo mass scales, centrals host a mix of Conventional AGN, Composites, and S06 Bonus AGN, plus a smattering of SF-AGN mostly below \mhalo $\sim$$10^{11.8}$\msun. \\

The \textit{satellites} in Figure~\ref{fig:halomass} host AGN types associated with active star formation up to higher halo masses than do the centrals. Dwarf satellites host SF-AGN up to nearly \mhalo $\sim$$10^{13.5}$\msun, while transitional galaxies (as defined by baryonic mass) host S06 bonus AGN all the way into the cluster halo-mass regime above \mhalo $\sim$$10^{14}$\msun. These results likely reflect continuing star formation in infalling subhalos and/or incomplete quenching as the primary halo heats outward (as expected from theory, \citealt{Dekel06}). WISE AGN present a slightly more complex picture, perhaps due to an association with dusty SF --- while these AGN are not more common below the threshold scale, they do span the whole range of halo masses, and they shift from central hosts to satellite hosts in group halos above the bimodality scale. \\

{Figure~\ref{fig:barchart} summarizes the steep mass dependence of the full-population BPT/VO+WISE AGN frequency across the three mass regimes. The $\sim$5$\times$ jump from dwarfs ($\sim$2--5\%) to transitional galaxies ($\sim$16--27\%) coincides with the onset of halo gas heating at the gas-richness threshold scale. This sharp increase serves as a striking reminder of the importance of distinguishing true dwarfs from low-mass galaxies, where the latter are often defined by \mstar $<10^{10}$ \msun, thus including transitional galaxies over a $\sim$0.5 dex mass range.}

\section{Conclusions}
\label{sec:conclusions}
Below is a summary of our results:
\begin{enumerate}
    \item Building on \citet[][P22]{Polimera22}, we have presented a uniquely comprehensive and complete $z\sim0$ AGN inventory, built on the volume- and mass-limited RESOLVE and ECO surveys, which are highly complete into the dwarf galaxy regime. Our census combines AGN identified using three methods: the BPT plot with traditional and alternate BPT demarcations (for Conventional AGN, Composites, and ``S06 Bonus AGN'' based on \citealt{Stasinska06}), the BPT and VO plots together (for SF-AGN, the ``star-forming AGN'' defined by \citetalias{Polimera22} as VO-plot AGN that are {SF based on Ke01/Ka03 BPT demarcations --- note these are usually SF but sometimes Bonus AGN based on S06 demarcations}), and a modification of the \citet[][S18]{Satyapal18} mid-IR color criteria (for WISE AGN). Where available, our inventory also includes archival X-ray and broadline \ha AGN, but this archival information is non-uniform and is not used to define our primary statistics. {(Sections \ref{sec:ch3midiragn}, \ref{sec:ch3opticalagnmethod}, and \ref{sec:archival}; Table~\ref{tab:agntable})}

    \item For BPT/VO AGN, we used data from three independent SDSS emission-line catalogs --- MPA-JHU, NSA, and Portsmouth \citep[][respectively]{Tremonti04, Blanton11, Thomas13} -- crossmatched to RESOLVE+ECO and homogenized following \citetalias{Polimera22}. For WISE AGN, we used custom WISE photometry described herein and validated against prior work including the WXSC (T.\ H.\ Jarrett, 2024, priv. comm.). We have also combined our WISE photometry with custom NUV data \cite[][based on deep photometry and machine-learning, respectively]{Hutchens23, Carr25} to compute new SFRs for the RESOLVE and ECO database; these SFRs are shown to be competitive with state-of-the-art prior work in the Appendix. {(Sections \ref{subsec:ch3emlinedata} and \ref{subsec:ch3midirphot}; Appendix; Tables \ref{tab:irdatatable} and \ref{tab:optdatatable})}

    \item We focus on optical spectroscopy and mid-IR photometry because the SDSS and WISE surveys observed nearly all galaxies in our surveys, though in practice AGN searches are limited by S/N requirements. For the 8352 RESOLVE+ECO galaxies in our complete baryonic-mass-limited parent sample (\mbary $>$ $10^{9.2}$), our S/N cuts reduced the AGN \textit{search} samples to 5443 galaxies for WISE classification, 5325/3857/3388 (MPA-JHU/Portsmouth/NSA) for BPT-only classification, and 2566/1149/1396 (MPA-JHU/Portsmouth/NSA) for BPT+VO classification. {Table~\ref{tab:agnstats}} presents baryonic-mass-limited AGN statistics as percentages of each search sample and of the full parent sample, while {Table~\ref{tab:agnstatsstellar}} presents analogous stellar-mass-limited statistics (\mstar $>$ $10^{8.9}$). \textit{Our work is unusual in providing AGN frequencies normalized relative to the full volume- and mass-limited galaxy population, i.e., true occupation fractions, thus facilit
ating comparison with simulations and putting AGN found in different search samples on a common footing.}

   \item We used the threshold and bimodality mass scales to divide dwarfs, giants, and transitional galaxies based on observed and predicted shifts in gas accretion physics. Applying baryonic mass divisions at $10^{9.7}$ and $10^{10.6}$ \msun, we measured full-population AGN frequencies of $\sim$2--5\% in dwarfs, $\sim$16--27\% in transitional galaxies, and $\sim$20--48\% in giants (Figure~\ref{fig:barchart}), where the ranges reflect systematics between the three crossmatched SDSS catalogs. Roughly equivalent stellar mass divisions at $10^{9.5}$ and $10^{10.5}$ \msun\ yield similar results. \textit{The sharp uptick in AGN frequency for transitional galaxies implies that AGN frequencies for true dwarfs should be distinguished from those for \mstar $<$ $10^{10}$ \msun\ low-mass galaxies \citep[cf.][ and references in the Mezcua erratum]{Salehirad22, Mezcua24, Pucha25}.}  {(Sections \ref{sec:ch3intro}, \ref{sec:optwisestats}, \ref{sec:optwisestats1}, and \ref{sec:optwisestats5})}

    \item Our dwarf AGN frequency exceeds most prior estimates by $\sim$2--5$\times$ when all results are expressed as or converted to full-population percentages, e.g., from X-ray data \citep{Birchall20, Latimer21} and the DESI survey \citep[][using only the BPT diagram]{Pucha25}. \textit{A key driver of our higher detection rate is inclusion of the non-traditional SF-AGN and S06 Bonus AGN categories.} Our full-population dwarf AGN frequency is broadly consistent with AGN occupation fractions in recent simulations \citep{Haidar22} and provides a statistically robust reference point for future simulations. However, the MaNGA IFU results of \cite{Mezcua24} suggest we may be missing a large population of wandering AGN. {(Section~\ref{sec:agnfreq})}

    \item Adding in AGN detections from (non-uniform) archival X-ray/optical data increases the RESOLVE+ECO AGN occupation fractions for dwarfs, transitional galaxies, and giants to $\sim$3--6\%, {$\sim$17--31\%}, and {$\sim$24--50\%}, respectively. {(Section~\ref{sec:archival})}

    \item We found WISE AGN to be rare ($\lesssim$1\% of galaxies at all masses and $\lesssim$0.3\% of dwarfs). Our detected WISE AGN show $\sim$50\% overlap with BPT/VO AGN in massive galaxies but almost no overlap in dwarfs (subject to small number statistics: 7 of our 12 dwarf WISE AGN candidates are potential false positives allowed by our not requiring adequate W3 data to apply the \citetalias{Satyapal18} W2$-$W3 color cut). By comparing our optimized WISE AGN selection (using custom resolution-matched r90-aperture colors and strict S/N cuts) to a more typical AllWISE AGN selection (using catalog colors and common literature S/N cuts), we have argued that AllWISE catalog-based AGN selection may be susceptible to $\sim$60\% contamination while missing $\sim$40\% of true AGN. Our custom photometry reveals a cleaner SF/AGN division that corresponds well with the \citetalias{Satyapal18} W1$-$W2 color cut. {(Section~\ref{sec:ch3midiragn} and Figure~\ref{fig:midiragn})}

    \item BPT/VO AGN demographics are strongly mass dependent. Dwarfs host mostly SF-AGN, Composites, or S06 Bonus AGN, where SF-AGN are up to $\sim$2$\times$ more common. Transitional galaxies host mostly Composites, Bonus AGN, and Conventional AGN, but Conventional AGN are up to $\sim$2$\times$ less common. Giants host mostly Composites and Conventional AGN, where Conventional AGN are up to $\sim$2$\times$ more common. Though systematics between crossmatched catalogs can be large enough to shift the relative abundance of the dominant AGN types, all crossmatched samples agree on the dominant types in each mass range. {Figure~\ref{fig:agnpiechart}} provides a visual summary of these trends, using MPA-JHU crossmatched data and baryonic mass divisions. \textit{We list sample statistics from this figure below, but see Tables \ref{tab:agnstats} and \ref{tab:agnstatsstellar} for NSA/Portsmouth-based results. We stress, as discussed in \citetalias{Polimera22}, that despite sharing 
the same underlying data, crossmatched measurements from these three catalogs yield poor agreement on individual AGN detections, and only moderate agreement on overall statistics.} {(Sections \ref{sec:optwisestats}--\ref{sec:optwisestats4})} \\
    \\
\noindent Dwarfs (3691 galaxies, \mbary $= 10^{9.2}-10^{9.7}$ \msun):
        \subitem SF-AGN - 2.1\%
        \subitem S06 Bonus AGN - 1.2\%
        \subitem Composite - 1.1\%
        \subitem Conventional BPT AGN - 0.4\%
        \subitem WISE AGN - 0.3\% \\
        \\
\noindent Transitional Galaxies (3863 galaxies, \mbary $= 10^{9.7}-10^{10.6}$ \msun):
        \subitem SF-AGN - 0.8\%
        \subitem S06 Bonus AGN - 8.0\%
        \subitem Composite - 9.5\%
        \subitem Conventional BPT AGN - 7.5\%
        \subitem WISE AGN - 0.4\% \\
        \\
\noindent Giants (798 galaxies, \mbary $\geq 10^{10.6}$\msun):
        \subitem SF-AGN - 0.0\%
        \subitem S06 Bonus AGN - 3.9\%
        \subitem Composite - 13.3\%
        \subitem Conventional BPT AGN - 29.2\%
        \subitem WISE AGN - 0.8\% \\
	
\item As seen in {Figure~\ref{fig:agnprop}}, \textit{BPT/VO AGN types form a continuum reflecting gas content, metallicity, and recent star formation.} The continuum runs from SF-AGN (at the high-SF end), to S06 Bonus AGN, then Composites, then Conventional AGN (at the low-SF end). The mass dependence of these properties is consistent with the observation that low metallicity and strong SF shift traditional BPT AGN signatures toward SF-AGN (VO-only AGN) signatures in dwarfs (\citetalias{Polimera22}). \textit{Thus, including SF-AGN and S06 Bonus AGN is essential to recovering the full dwarf AGN inventory.} {(Section~\ref{sec:discmass})}

\item The inherently star-forming nature of S06 Bonus AGN and SF-AGN implies that a subset of these may be primarily ionized by WRs, HMXBs, and/or ULXs, which could affect the inferred dwarf AGN frequency (for example, as in Figure~7 of \citetalias{Polimera22}). We have argued against avoiding all such contamination by raising the \citetalias{Kewley01} demarcations in the VO diagrams (as suggested by \citealt{Law21} and \citealt{Richardson25}, R25), as the \citetalias{Kewley01} demarcations strike a balance between false positives and false negatives. Galaxies with weak AGN (f$_{\rm AGN} < 0.08$) are conversely expected below the \citetalias{Kewley01} line that defines SF-AGN (\citetalias{Polimera22, Richardson25}). We favor multi-diagnostic or Bayesian multi-line modeling approaches for improved classification, but both are unavoidably biased against SF-AGN detection by the unlucky anticorrelation of [OI] S/N with \OvH (\citetalias{Polimera22}), given that \OvH is the most c
ritical line ratio for SF-AGN identification. {(Sections~\ref{subsec:ch3s06bonus}, \ref{subsec:ch3sfagn}, \ref{sec:agnfreq})}

\item As seen in {Figure~\ref{fig:halomass}}, the continuum of BPT/VO AGN types seems to track the transition from cold-accretion dominated halos below \mhalo $\sim 10^{11.4}$\msun (where SF-AGN prevail in dwarf centrals) to shock-heated halos above \mhalo $\sim 10^{12.0}$\msun (where Conventional AGN prevail in giant centrals). Satellites present a more complex picture, with partial quenching (possibly related to ongoing group assembly) creating a mix of AGN types along the continuum. These results suggest that group halo mass and cold gas accretion are potentially important factors in fueling AGN as well as fueling SF. {(Section~\ref{sec:discmass})}

\end{enumerate}

As the frontier of AGN research expands into the dwarf regime, the defining features of AGN become ambiguous. Active BHs need not be in nuclei \citep{Bellovary21,Mezcua24}, and they may include intermediate mass 10$^2$--10$^4$ \msun seed black holes (IMBHs). The astrophysical identities of dwarf AGN candidates such as SF-AGN may include objects in between 10$^5$ \msun AGN and stellar-mass ULXs, such as hyperluminous X-ray sources (HLXs), which represent a mix of stellar-mass and seed-mass BHs \citep{Amato25}. Evidence for seed-mass BHs in nearby galaxies has become more compelling with the growing body of gravitational wave detections of low-redshift, merger-formed `lite' IMBHs with final masses $\sim$200 \msun \citep[e.g.,][]{RuizRocha25}. The most recent example, GW231123 \citep[$z\sim0.39$,][]{LIGOVirgoKAGRA25}, is intriguing because of the high spins of both progenitors, whose close alignment might reflect growth in tandem by sustained accretion \citep{BartosHaiman25}, th
ough many other formation mechanisms have been proposed. If GW231123-like objects and their progenitors can form in metal-poor dwarfs and galaxy outskirts at the present epoch \citep{Paiella25} --- the same environments where \cite{Mezcua24} found wandering SF-AGN and \cite{Law21} found modestly elevated velocity dispersions that seemed sub-AGN level but accompanied line ratios above the \citetalias{Kewley01} line --- then it is interesting to ask whether a subset of SF-AGN may represent neither stellar-mass BHs nor traditional AGN, but instead IMBHs that fill in the BH mass hierarchy. Exploring this exciting possibility and unveiling the true nature of the low-mass AGN inventory will require breaking degeneracies in modeling with improved spectroscopic and multi-wavelength data.

\section*{Acknowledgements}
We would like to express our deepest gratitude to Dr. Thomas Jarrett, whose invaluable contributions and insights greatly enhanced this work. Tragically, Dr. Jarrett passed away unexpectedly before the publication of this paper. He was an amazing person, full of wisdom and humor and generous to all. We miss him dearly and dedicate this work to his memory. We also thank Michelle Cluver, as we know Tom had insightful conversations with her about this work. We thank Julianna Levanti, Adrienne Erickcek, Gerald Cecil, and the anonymous referee for their helpful comments that strengthened this paper. This work was supported by NSF grant AST-2007351. MSP, SJK, and JMB acknowledge support for this work that was provided by NASA through Chandra Award Number GO1-22089A issued by the Chandra X-ray Center, which is operated by the Smithsonian Astrophysical Observatory for and on behalf of the National Aeronautics Space Administration under contract NAS8-03060. CR acknowledges the support
 of the Elon University Japheth E. Rawls Professorship.   

\section*{Data Availability}
All data releases for the RESOLVE and ECO surveys may be found at \url{https://resolve.astro.unc.edu}, both as machine-readable ASCII tables and in a merged searchable database. Machine-readable versions of Tables~\ref{tab:irdatatable}, \ref{tab:optdatatable}, and \ref{tab:agntable} are also available at \dataset[doi:10.5281/zenodo.20945134]{https://doi.org/10.5281/zenodo.20945134}.

\appendix
\section*{Star Formation Rates for the RESOLVE and ECO Surveys}
\begin{figure*}[]
    \centering
    \includegraphics[width = \linewidth]{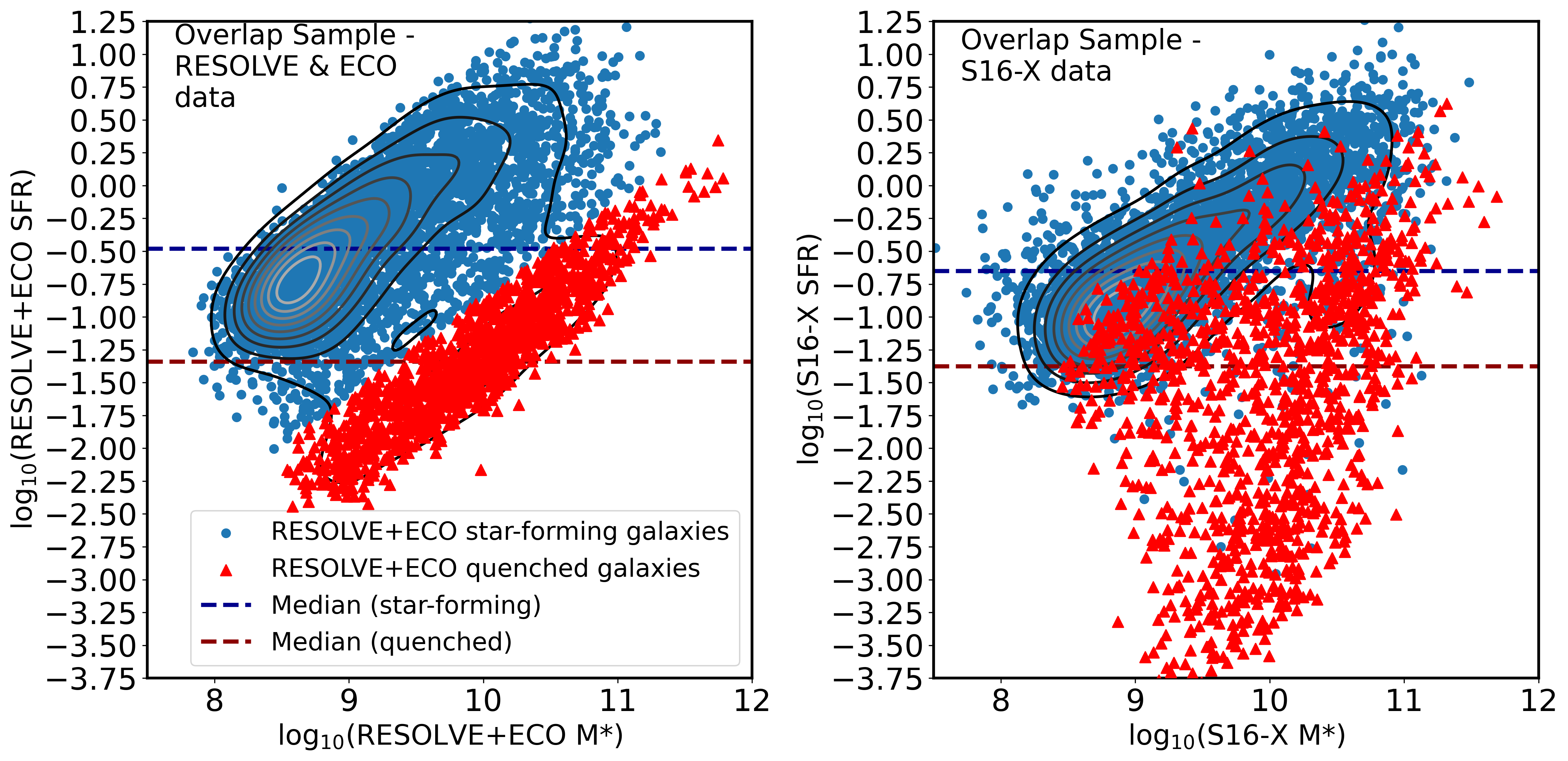}
    \caption{
    Comparison of SFR-\mstar\ relations for the 5535 galaxies with deep NUV data (exposure time $>$1000s) that are found in both RESOLVE/ECO and S16-X; these galaxies are referred to as the \textit{overlap sample}. See Appendix for discussion. In both panels SFRs are a combination of UV and mid-IR SFRs and grayscale lines represent 10th, 20th,... 90th percentile point density contours. Quenched galaxies in the overlap sample are marked with red points, determined following the equation: $\rm{log_{10}}(SFR) < 0.74\rm{log_{10}}(\rm{M_*}) - 8.26$, where the SFR and \mstar are the star formation rate and stellar mass from RESOLVE and ECO.
    }
    \label{fig:sfrmstar_s16res}
\end{figure*}

In this appendix, we compare our star formation rates to those from the master version of the GALEX-SDSS-WISE Legacy Catalog-1\footnote{https://salims.pages.iu.edu/gswlc/} \citep[GSWLC-1, version X;][henceforth S16-X]{Salim16}. 
We note that for the RESOLVE and ECO surveys, we do not have another independent source of SFRs like \ha; the SDSS-based \ha fluxes used for our BPT and VO analysis represent only the central 2--3$\arcsec$ of the galaxies. Therefore, the best way to validate our SFRs is by comparing to S16-X, whose data and methods were relatively similar to ours, albeit there are still some key differences we will discuss below. Among the 14654 unique galaxies in RESOLVE DR4 and ECO DR3 catalogs \citep[][with no mass or volume limits applied]{Hutchens23}, we found 11344 galaxies ($\sim$77\%) in common with S16-X, which form the basis of our comparison. \\

As mentioned in Section \ref{subsec:ch3sfrs}, the RESOLVE and ECO surveys treat galaxies differently with vs.\ without custom photometry derived from deep NUV data, where we define \textit{deep} as having exposure time greater than 1000s. For galaxies with deep custom photometry in either \cite{Eckert15} or \cite{Hutchens23}, we have computed $\rm SFR_{UV}$ from internal and foreground extinction-corrected NUV luminosities (outputs of SED modelling; see Section \ref{subsec:ch3sfrs}). Among the 14654 galaxies in RESOLVE+ECO, 7273 galaxies have deep NUV data. \textit{In the rest of this appendix, when referencing RESOLVE and ECO galaxies, we consider only the 5535 galaxies with deep NUV data that are also present in the S16-X sample, and we call these the {overlap sample}. We use the symbol SFR in this appendix strictly to refer to total UV+mid-IR star formation rates for galaxies with deep NUV data.}\\

Figure \ref{fig:sfrmstar_s16res} shows the logarithmic SFR-\mstar relation for the overlap sample. The left panel shows measurements from the RESOLVE and ECO catalogs and this work. The right panel shows measurements from S16-X, using their unobscured $\rm SFR_{SED}$ and \mstar values computed from SED fitting to GALEX and SDSS data, and their obscured $\rm SFR_{IR}$ computed from the mid-IR luminosity-based SFR calibration of \cite{Kennicutt98} applied to data from the AllWISE Source Catalog\footnote{http://wise2.ipac.caltech.edu/docs/release/allwise/} and the unWISE catalog \citep{Lang16}. To compare total SFRs, we combined the $\rm SFR_{SED}$ and $\rm SFR_{IR}$ from \citetalias{Salim16} using the same equation used for combining our own SFRs, \ref{eqn:sfr} (\citealt{Buat11}; see Section \ref{subsec:ch3sfrs}). \\

\begin{figure}[]
    \centering
    \includegraphics[width = \linewidth]{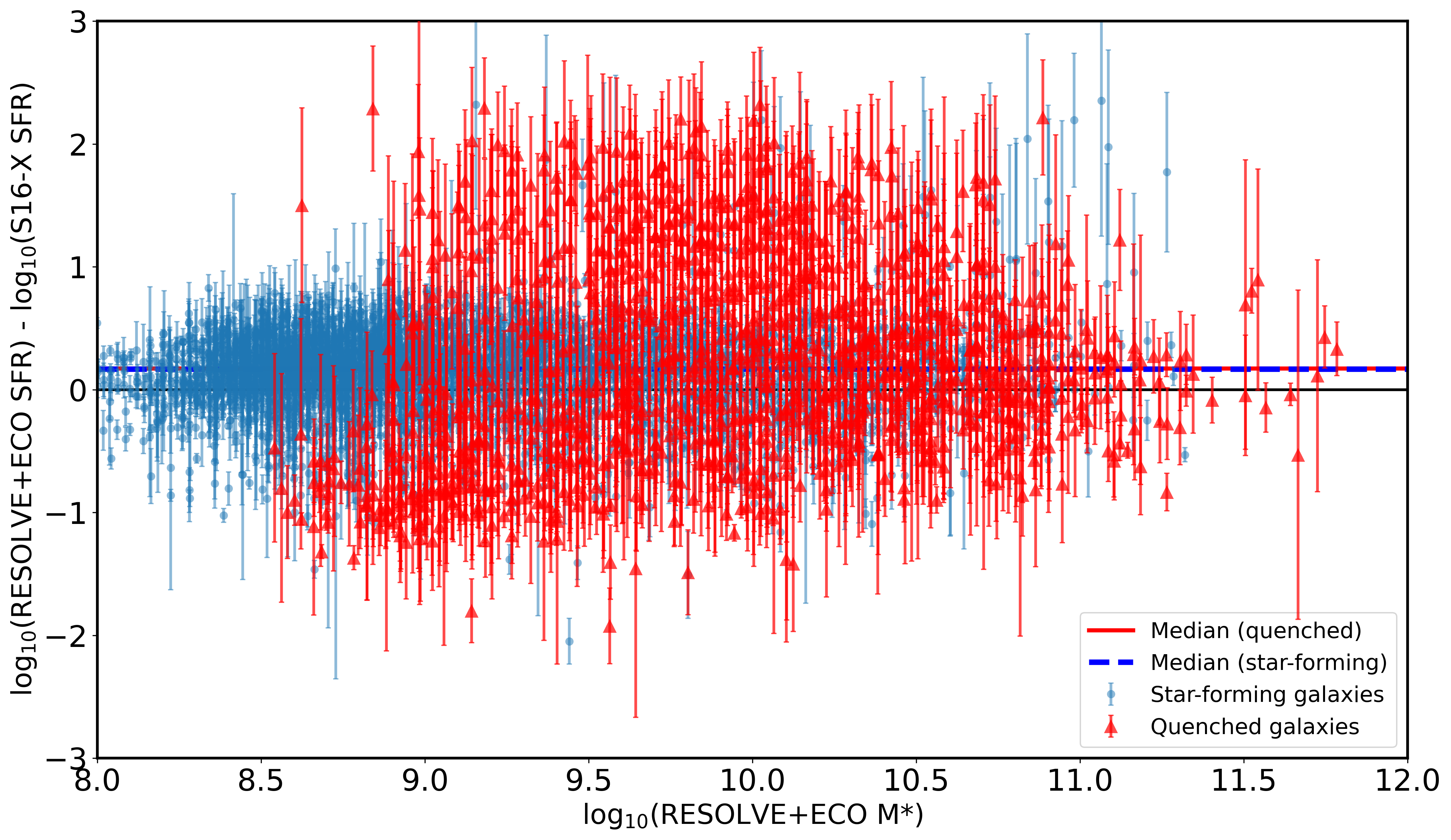}
    \caption{Stellar mass dependence of the differences between the logarithms of total SFR for the S16-X/RESOLVE+ECO overlap sample shown in Figure~\ref{fig:sfrmstar_s16res}, with red and blue again showing quenched and unquenched galaxies. The black line shows an offset of zero. The red and blue lines show the median differences of +0.172 and +0.167 dex for the quenched and unquenched galaxies, respectively, in the sense that RESOLVE+ECO SFRs are higher.
    }
    \label{fig:sfrresidual_s16res}
\end{figure}

For quenched galaxies (red points defined as quenched by the dividing line in Figure~\ref{fig:sfrmstar_s16res}), the SFR distributions are quite different for the two data sets, though the median values of $\rm{log_{10}(SFR)}$ are similar ($-$1.34 for RESOLVE+ECO and $-$1.38 for S16-X). Figure \ref{fig:sfrmstar_s16res} shows that these galaxies form a distinct, tight sequence using RESOLVE+ECO data (left panel) but scatter widely using S16-X data (right panel). We believe that the main reason for this difference is our use of a literature SFR prescription to obtain unobscured $\rm SFR_{UV}$ values, which associates all NUV luminosity to star formation activity irrespective of NUV contributions from the underlying galaxy stellar populations. In contrast, the unobscured $\rm SFR_{SED}$ values in the S16-X catalog come from advanced SED modeling that self-consistently includes the underlying stellar populations. Figure~\ref{fig:sfrresidual_s16res} shows the differences between t
he S16-X and RESOLVE+ECO SFRs as a function of stellar mass for galaxies in common. RESOLVE+ECO SFRs for quenched galaxies in the overlap sample are higher by a median difference of $\sim$0.17 dex, with difference values scattering asymmetrically within $\lesssim$2 dex skewed high around the median. Note that the median difference is larger than the difference between the two medians ($-$1.34 and $-$1.38) mentioned earlier.\\

For star-forming galaxies (blue points in Figure~\ref{fig:sfrmstar_s16res}), the median value of $\rm{log_{10}(SFR)}$ from RESOLVE+ECO data is $-$0.48 while the median from S16-X data is $-$0.65 for the same subsample. In this case, the difference between the two medians (0.17 dex) happens to match the median difference shown in Figure~\ref{fig:sfrresidual_s16res}, and the difference values scatter fairly symmetrically within $\lesssim$1 dex around the median. We believe our SFRs may, however, be superior for these galaxies, as explained below.\\

\begin{figure}[]
    \centering
    \includegraphics[width = \linewidth]{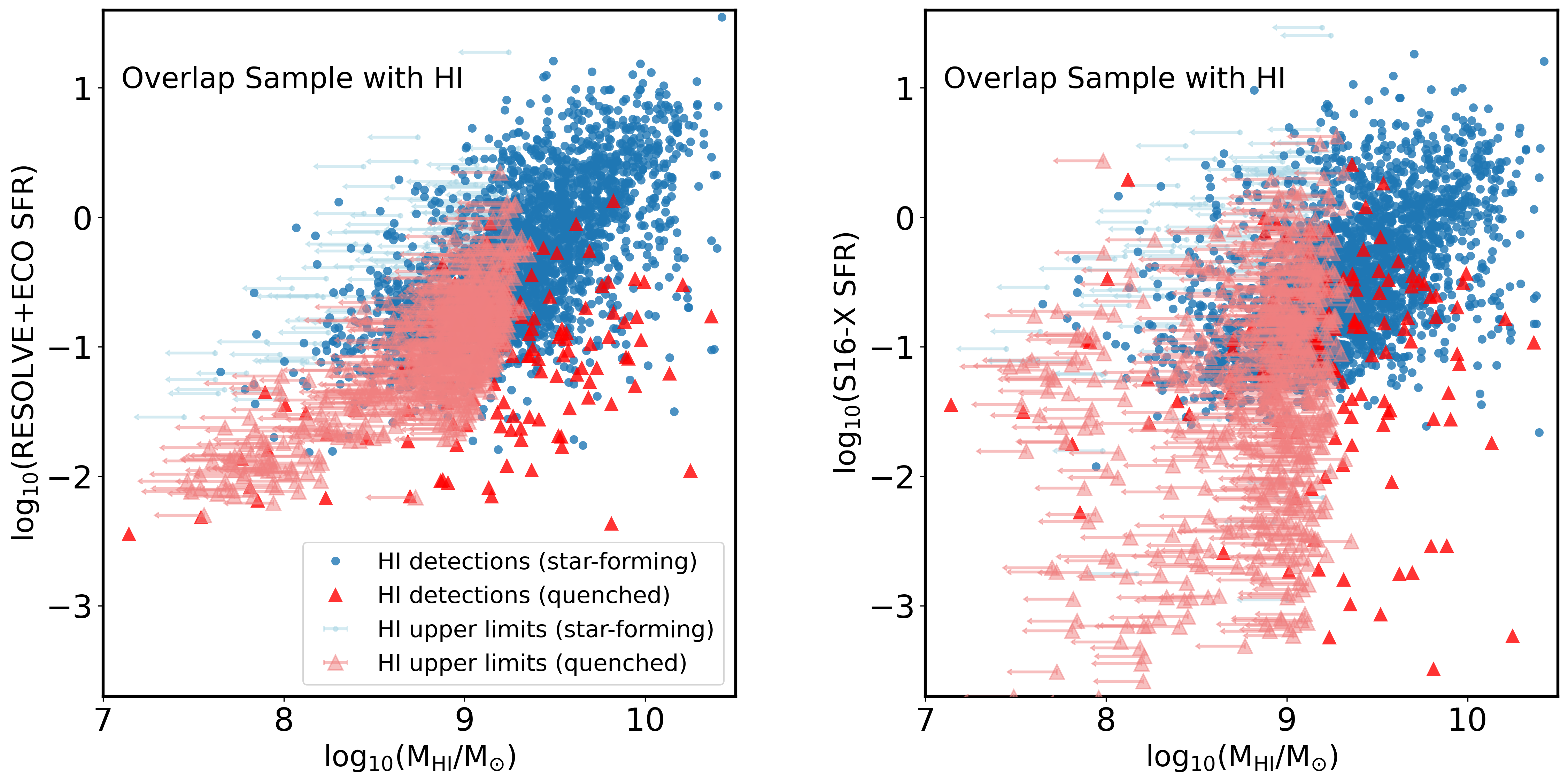}
    \caption{Correlation between $\rm M_{HI}$ \citep{Stark16, Hutchens23} and total SFR either from RESOLVE+ECO (right panel) or from the S16-X catalog \citep[][right panel]{Salim16} for the 3326 galaxies in the overlap sample that have reliable HI mass data (21~cm detections or strong upper limits) in the RESOLVE+ECO database. Red and blue symbols correspond to HI detections for quenched/unquenched galaxies as marked Figure \ref{fig:sfrmstar_s16res}. Light red and light blue arrows correspond to HI upper limits for quenched and unquenched galaxies, respectively.}
    \label{fig:sfrmhi_s16res}
\end{figure}

The atomic gas reservoir is an independent measure of star formation capacity in a galaxy and correlates with long-term star formation history \citep{Kannappan13, Eckert15, Cortese20}. Thus, we used the overlap sample to compare the relations between $\rm M_{HI}$ (atomic gas mass including helium) and the different SFRs from RESOLVE+ECO and S16-X. We used $\rm M_{HI}$ values from \cite{Hutchens23} that were derived from 21cm detections (``mhidet" $>$ 0) or strong upper limits (``mhilim" $>$ 0 and ``mhilim" $\leq$ 0.1 $\times$ \mstar), not values constrained by the photometric gas fractions technique \citep{Kannappan04,Eckert15}. Among the 5535 galaxies in the overlap sample, 3326 galaxies have reliable HI data (detections or strong upper limits). In Figure~\ref{fig:sfrmhi_s16res}, we see that RESOLVE+ECO SFRs correlate more tightly with $\rm M_{HI}$ than do SFRs from S16-X, especially for actively star-forming galaxies (unquenched: blue symbols). \textit{This result suggests 
our SFRs may be superior specifically for star-forming galaxies, though as noted above, they are likely inferior for quenched galaxies as defined by Figure~\ref{fig:sfrmstar_s16res}.} \\

In addition, the 0.17 dex offset between RESOLVE+ECO and S16-X SFRs is not surprising, because we have used reprocessed GALEX NUV photometry designed to recover flux better than the standard GALEX catalog photometry used by S16 (\citealt{Eckert15}; see also \citealt{Kannappan13} and \citealt{Carr25}). NUV fluxes strongly dominate SFRs, so this difference is the most likely explanation for the offset, though it is also true that \cite{Eckert15} used near-infrared data in the SED modeling process while \citetalias{Salim16} did not. Finally, our mid-IR SFRs are also computed differently, but $SFR_{UV}$ dominates $\rm SFR_{IR}$ in the vast majority of galaxies, so this difference cannot be the main driver of the offset. We have also tried using SFRs from the GSWLC-2 catalog from \cite{Salim18} and obtained qualitatively similar results. \\ 

Overall, given that RESOLVE and ECO SFRs are based on high-quality reprocessed GALEX photometry and that RESOLVE and ECO SFRs have a stronger correlation with $\rm M_{HI}$ than do S16-X SFRs, we believe that our SFRs are reliable for the actively star-forming galaxies in the SFR-\mstar main sequence. Deriving more accurate star formation rates for quenched galaxies is beyond the scope of this paper. \\

\bibliography{References}

\end{document}